\documentclass[showpacs,aps,superscriptaddress,letterpaper,nofootinbib]{revtex4}

\usepackage{graphicx}
\usepackage{epsfig}
\usepackage{float}
\usepackage{wasysym}
\usepackage{hyperref}

\newcommand{\be}{\begin{equation}}
\newcommand{\ee}{\end{equation}}
\newcommand{\ba}{\begin{eqnarray}}
\newcommand{\ea}{\end{eqnarray}}

\newcommand{\CP}{\ensuremath{C\!P}} 
\newcommand{\PH}{H}
\newcommand{\ttH}{t\bar{t}\PH}
\newcommand{\bbH}{b\bar{b}\PH}

\newcommand{\tqH}{t{q}\PH}
\newcommand{\tbH}{t{b}\PH}

\newcommand{\Hff}{\PH\!f\!\bar{f}}

\newcommand{\qqbar}{q\bar{q}}

\begin{document}

\begin{flushright}
\vbox{
\begin{tabular}{l}
{TTP16-020}\\
{CERN-TH-2016-135}
\end{tabular}
}
\end{flushright}

\vspace{0.6cm}

\title{Constraining anomalous Higgs boson couplings to the heavy flavor fermions using matrix element techniques}%
\author{Andrei V. Gritsan \thanks{e-mail: gritsan@jhu.edu}}
\affiliation{Department of Physics and  Astronomy, Johns Hopkins University, Baltimore, MD, USA}
\author{Raoul R\"ontsch \thanks{e-mail: raoul.roentsch@kit.edu}}
\affiliation{Karlsruhe Institute of Technology, Karlsruhe, Germany}
\author{Markus Schulze \thanks{e-mail: markus.schulze@cern.ch}}
\affiliation{CERN, European Organization for Nuclear Research, Geneva, Switzerland}
\author{Meng Xiao  \thanks{e-mail: mxiao3@jhu.edu}}
\affiliation{Department of Physics and Astronomy, Johns Hopkins University, Baltimore, MD, USA}

\date{June 9, 2016}

\begin{abstract}
\vspace{2mm}
In this paper we investigate anomalous interactions of the Higgs boson with heavy fermions, employing shapes 
of kinematic distributions. We study the processes $pp \to t\bar{t} + H$, $b\bar{b} + H$, $tq+H$, and $pp \to H\to\tau^+\tau^-$, 
and present applications of event generation, re-weighting techniques for fast simulation of anomalous couplings, 
as well as  matrix element techniques for optimal sensitivity. We extend the MELA technique, which proved to be a powerful 
matrix element tool for Higgs boson discovery and characterization during Run I of the LHC, and implement 
all analysis tools in the JHU generator framework. A next-to-leading order QCD description of the $pp \to t\bar{t} + H$ 
process allows us to investigate the performance of MELA in the presence of extra radiation. Finally, projections for 
LHC measurements through the end of Run III  are presented. 
\end{abstract}


\maketitle

\thispagestyle{empty}


\section{Introduction}
 \label{sect:tth_intro}

The discovery~\cite{Aad:2012tfa,Chatrchyan:2012xdj} of the $H$ boson by the ATLAS and CMS
experiments during Run I of the LHC marked an important milestone in the evolution of our understanding 
of fundamental particle physics. 
One of the most important goals now is a precise understanding of the newly discovered state, 
including its couplings to other particles as well as its $\CP$ nature. Any significant deviation from 
Standard Model (SM) predictions would reveal the existence of new physics in the Higgs sector
and should to be classified according to its anomalous coupling structures. Likewise, in the case 
of a discovery of a new resonance at the LHC, a similar program of investigating properties 
and couplings is required.

Since experimental efforts during Run I mostly focused on the $H$ boson decaying to a pair of vector bosons\footnote{
In this paper, $HVV$ couplings induced by closed fermionic loops are still considered to be 
couplings to vector bosons.},
extensive studies of the $HVV$ couplings and corresponding $C\!P$ properties 
were performed~\cite{Khachatryan:2014kca,Khachatryan:2014jba,Aad:2015mxa,Aad:2015gba}, 
leading to results consistent with the Standard Model nature of the $H$ boson with quantum numbers $J^{PC}=0^{++}$.
However, many generic models of new physics predict deviations which are beyond the current experimental 
precision~\cite{Dawson:2013bba}.
This leaves ample room for anomalous interactions to hide as small modifications of the SM structure.
Moreover, a complete knowledge of the SM Higgs mechanism requires the study of 
the $H$ boson interactions with fermions.  While the observation of $Hgg$ and $H\gamma\gamma$ interactions  
established the $H t\bar{t}$ coupling through a closed loop,  a detailed understanding only arises from 
the observation of various mass hierarchies, as well as the minimal flavor-universal Yukawa interaction 
as predicted by the SM. Hence, it is of paramount importance to investigate possible anomalous coupling patterns 
with which different quarks and leptons may interact with the Higgs field. The most promising approach is the study 
of differential distributions in processes with {\it direct} sensitivity through the associated production of 
the $H$ boson with on-shell fermions, $f^\prime\!\bar{f} + H$, or through the decay $H \to f \bar{f}$.

There has been considerable effort in modeling the $\Hff$ couplings and developing tools for their analysis in associated 
production~\cite{BhupalDev:2007ftb,Agrawal:2012ga,Farina:2012xp,Nishiwaki:2013cma,Biswas:2013xva,Ellis:2013yxa,
Khatibi:2014bsa,Wu:2014dba,Demartin:2014fia,Bramante:2014gda,Kobakhidze:2014gqa,Englert:2014pja,Chang:2014rfa,Yue:2014tya,He:2014xla,
Boudjema:2015nda,Kolodziej:2015qsa,Chen:2015rha,Demartin:2015uha,Buckley:2015vsa,
Mileo:2016mxg,Rindani:2016scj,Cirigliano:2016nyn,Dolan:2016qvg}
and in the $H\to\tau^+\tau^-$ decay~\cite{Desch:2003rw,Berge:2012wm,Harnik:2013aja,Berge:2013jra,Philip:2014el,Berge:2015nua,Askew:2015mda}.
Current experimental analyses have measured the coupling strength of $Hb\bar{b}$ and $Ht\bar{t}$ 
only through closed loops~\cite{Khachatryan:2014jba,Aad:2015gba}.
There have been experimental searches for the $H$ boson production in association 
with a single top quark~\cite{Khachatryan:2015ota}
and with $t\bar{t}$~\cite{Khachatryan:2014qaa,Khachatryan:2015ila,Aad:2015iha,Aad:2016zqi}, 
with strong evidence for the latter in Run~I of the LHC. 
The process $\bbH$ has not been studied with a dedicated analysis so far 
but there is evidence for the $H$ boson decay into 
$b\bar{b}$ pairs~\cite{Chatrchyan:2013zna,Aad:2014xzb,Khachatryan:2015bnx}.
The decay of $H\to\tau^+\tau^-$ is observed when results of CMS~\cite{Chatrchyan:2014nva} and ATLAS~\cite{Aad:2015vsa}
are combined, and searches in this decay channel have been performed for specific scenarios beyond the 
SM~\cite{Aad:2015gha,Khachatryan:2015nba,Khachatryan:2015baw}.
However, an interpretation in terms of generic anomalous couplings has not yet been undertaken.

All these measurements require sophisticated tools for the optimal extraction of statistical information, 
as data remain limited for detailed analyses of the fermion couplings. 
The matrix element approach is one such technique, which has been proven successful in setting
constraints on $HVV$ couplings using Run I data from CMS~\cite{Chatrchyan:2012xdj,Chatrchyan:2012jja,Chatrchyan:2013mxa,
Chatrchyan:2013iaa,Khachatryan:2014iha,Khachatryan:2014ira,Khachatryan:2014kca,Khachatryan:2015mma,Englert:2014pja,Khachatryan:2016tnr} 
and ATLAS~\cite{Aad:2013xqa,Aad:2015mxa,Aad:2016nal}.
In this paper, we focus on applications to Run II of the LHC and extend our earlier developed techniques for 
$HVV$ coupling measurements~\cite{Gao:2010qx,Bolognesi:2012mm,Anderson:2013afp}
to $\Hff$ couplings in $t\bar{t}H$, $b\bar{b}H$, and $t{q}H$ 
production\footnote{Unless otherwise noted, $t{q}H$ refers to all combinations
of $t\bar{q}H$, $\bar{t}{q}H$, $t{q}H$, and $\bar{t}\bar{q}H$ with a quark $q\ne  t$.},
as well as to $H\to\tau^+\tau^-$ decays.
These matrix element techniques allow the optimal analysis of the dynamics in the production and decay processes.
Such techniques have been proposed to enhance signal over background in application to $\ttH$ 
production~\cite{Andersen:2012kn,Artoisenet:2013vfa,Khachatryan:2015ila}, and we employ them to probe anomalous $\Hff$ couplings for the first time.
We define the complete set of kinematic observables and the minimal set of matrix-element-based observables 
necessary to perform the measurements. 
Moreover, using a NLO QCD simulation of $\ttH$ process that includes a fully consistent treatment of production and decays at higher orders, 
we demonstrate the robustness of the matrix element approach with respect to additional radiation and loop corrections.

This paper expands our efforts within the broader framework of the JHUGen (JHU generator) and 
MELA (Matrix Element Likelihood Approach) frameworks~\cite{Gao:2010qx,Bolognesi:2012mm,Anderson:2013afp}.
The rest of the paper is organized as follows. In Section~\ref{sect:tth_couplings} the formalism of anomalous
$H$ boson couplings is discussed. Monte Carlo simulation with the JHU generator is introduced in Section~\ref{sect:tth_mc}.
The matrix elements technique and the MELA framework are discussed in Section~\ref{sect:tth_mela}.
A study of  NLO QCD effects is  presented in Section~\ref{sect:tth_nlo}.
In Section~\ref{sect:tth_measure} we discuss the application of these techniques to LHC measurements
and make projections to the end of Run III of LHC. 
Results are summarized in Section~\ref{sect:tth_summary}.


\section{Parameterization of Higgs boson couplings}
\label{sect:tth_couplings}

We describe the interactions between a spin-zero particle $H$ and two fermions through the amplitude
\begin{eqnarray}
&& {\cal A}(\Hff) = - \frac{m_f}{v}
\bar{\psi}_{f} \left ( \kappa_f  + \mathrm{i} \, \tilde\kappa_f  \gamma_5 \right ) {\psi}_{f}\,,
\label{eq:ampl-spin0-qq}
\end{eqnarray}
where $\bar{\psi}_{f}$ and ${\psi}_{f}$ are the Dirac spinors,
$m_f$ is the fermion mass, and $v$ is the SM Higgs field vacuum expectation value.
In the SM, the couplings\footnote{The coupling convention of Ref.~\cite{Gao:2010qx} corresponds to $\kappa_f =-\rho_{1} $ and $\tilde\kappa_f =\mathrm{i} \rho_{2}$.} 
have the values $\kappa_f=1$ and $\tilde\kappa_f=0$.
Any deviation from these values indicates the presence of physics beyond the SM, which may for example arise through heavy loop-induced fields.
In particular, the $\tilde\kappa_f$ coupling parameterizes the contribution of a $C\!P$-odd pseudoscalar boson, and $C\!P$ violation occurs when both $\kappa_f $ and $\tilde\kappa_f$ are nonzero.

One may equivalently choose to express the couplings through a Lagrangian (up to an unphysical global phase)
\begin{eqnarray} 
&& {\cal L}(\Hff) = - \frac{m_f}{v} \bar{\psi}_{f} \left ( \kappa_f + \mathrm{i} \, \tilde\kappa_f \gamma_5 \right ) \psi_{f} \, H,
  \label{eq:lagrang-spin0-qq}
\end{eqnarray}
which allows a connection to be made between the couplings $\kappa_f$ and $\tilde \kappa_f$ and anomalous operators in an effective field theory.
We assume the couplings to be independent of kinematics, which corresponds to dimension-six operators
in the effective field theory. Higher-dimension operators could easily be considered through $q^2$-dependent
couplings in our framework, where $q$ is the momentum transfer. However, in our study we neglect these
higher-dimension contributions because they are expected to be small.
The hermiticity of the Lagrangian requires $\kappa_f$ and $\tilde \kappa_f$ to be real. 
Nevertheless, in order to consider the broadest range of scenarios, we allow $\kappa_f$ and $\tilde \kappa_f$ to be complex, and trust that, should the unitarity of 
scattering amplitudes be violated as a result, it will be restored in the full theory.
It is convenient to parameterize anomalous couplings through a mixing angle, with $\kappa_f \propto \cos\alpha$ and $\tilde\kappa_f\propto \sin\alpha$. 
Equivalently, we introduce the fractions
\begin{equation}
f_{\CP} = \frac{|\tilde\kappa_f|^2} {|\kappa_f|^2 + |\tilde\kappa_f|^2 }
\,,
~~~\phi_{\CP} = {\rm arg}(\tilde\kappa_f / \kappa_f)
\,,
\label{eq:fa3}
\end{equation}
where the $f_{\CP}$ parameter is conveniently bounded between 0 and 1, is uniquely defined, 
and can be interpreted as the cross section fraction corresponding to the pseudoscalar coupling, 
and therefore is directly related to experimentally observable effects.
It is a convenient counterpart of the $f_{a3}$ parameter defined for the $HVV$ 
couplings~\cite{Anderson:2013afp,Khachatryan:2014kca,Aad:2015mxa}.
While the phase $\phi_{\CP}$ can in general take any value between $0$ and $2\pi$,
it is reasonable to assume that the ratio $\tilde\kappa_f / \kappa_f$ is real, that is $\phi_{\CP} = 0$ or~$\pi$.
However, we do not need to impose this restriction and will also consider other values of $\phi_{\CP}$.
The parameters $f_{\CP}$ and $\phi_{\CP}$ in principle depend on the fermion couplings under consideration
and should be denoted $f^{\Hff}_{\CP}$ and $\phi^{\Hff}_{\CP}$, but in most cases this will be clear from the context. 

The $\tqH$ production also involves the $HWW$ coupling. 
We therefore recall the coupling of the $H$ to two vector bosons~\cite{Anderson:2013afp}\footnote{The coupling convention 
of Ref.~\cite{Anderson:2013afp} corresponds to $a_1 =g_{1} $, $a_3 =g_{2} $, and $a_3 =g_{4} $.}
\begin{equation}
{\cal A}(HVV) =  \frac{1}{v} \left( a_1 m_V^2 \epsilon_1^*  \epsilon_2^* 
+ a_2 f_{\mu \nu}^{*(1)} f^{*(2),\mu\nu} 
+ a_3 f_{\mu \nu}^{*(1)} \tilde{f}^{*(2),\mu\nu}  \right),
\label{eq:coupl-spin0-VV}
\end{equation}
where $\epsilon_i$ is the polarization of the vector boson of mass $m_V$ and momentum $q_i$, 
the field strength tensor is $f^{(i),\mu \nu}=\epsilon_i^{\mu} q_i ^{\nu}-\epsilon_i^{\nu} q_i ^{\mu}$ and its dual 
 $\tilde{f}^{(i),\mu \nu}= 1/2 \, \epsilon^{\mu \nu \rho \sigma} f^{(i)}_{\rho \sigma}$.


\section{Monte Carlo simulation}
\label{sect:tth_mc}

\begin{figure}[t]
\centerline{
\includegraphics[width=0.65\linewidth]{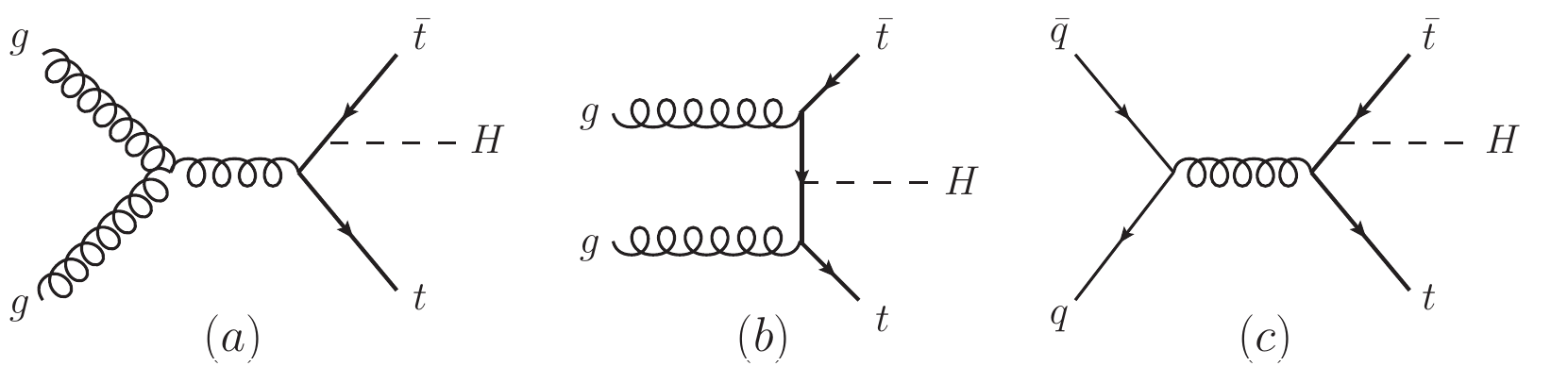}
}
\caption{Representative Feynman diagrams for $\ttH$ production at leading order.\label{fig:feynman-tth}}
\end{figure}

The JHU generator framework~\cite{Gao:2010qx,Bolognesi:2012mm,Anderson:2013afp}
involves both the Monte Carlo generation of unweighted events and the MELA package 
used in the analysis of the $H$ boson couplings. 
For top quark pair production in association with a spin-zero boson $H$ we compute the leading-order 
processes $gg \to t\bar{t}+H$ and $q\bar{q} \to t\bar{t}+H$, followed by spin-correlated top quark decays 
$t \to b W(\to f^\prime\!\bar{f})$ in the narrow-width approximation.
Any leptonic or hadronic decay mode of the top quarks can be described.
Representative Feynman diagrams are shown in Fig.~\ref{fig:feynman-tth}, where we allow for the anomalous 
$Ht\bar{t}$ couplings shown in Eq.~(\ref{eq:ampl-spin0-qq}).
The $H$ boson is considered stable in the respective matrix elements describing production,
and its decay into any possible channel can be introduced subsequently through processing 
the generated events using the JHU generator framework. 

Since hadronic production of $t\bar{t}+H$ final states involves color flow in initial and final states, 
additional jet radiation plays an important role in the description of this process. 
In fact, almost 40\,\% of all $t\bar{t}$ events are accompanied by jets with transverse momentum 
harder than 40\,GeV \cite{Bevilacqua:2015qha}. It is therefore important to study the impact of radiative 
corrections on event kinematics and the matrix element observables.
To this end, we also calculate the next-to-leading order QCD correction to the $pp \to t\bar{t}+H$ process.

The framework for NLO QCD computations is an extension of the TOPAZ code which two of us developed 
for anomalous coupling studies of $t\bar{t}+Z$ final states~\cite{Rontsch:2014cca,Rontsch:2015una}.
We calculate the virtual correction to the $gg$ and $q\bar{q}$ initial states using the numerical 
implementation of $D$-dimensional generalized unitarity techniques~\cite{Ossola:2006us,Ellis:2007br,Giele:2008ve,Ellis:2008ir}.
The real emission corrections involve the partonic processes $gg\to t\bar{t}g$, $q\bar{q} \to t\bar{t}g$,  $qg\to t\bar{t}q$ and  
$\bar{q}g\to t\bar{t}\bar{q}$, which we regularize using the massive dipole subtraction techniques of Refs.~\cite{Catani:1996vz,Catani:2002hc}.
A consistent expansion in the strong coupling constant also requires the computation of the NLO corrections to the top quark decay $t \to b W$ 
and the subsequent $W\to jj$ decay. We account for these contributions in the narrow-width approximation using the implementation
 developed in Ref.~\cite{Melnikov:2009dn}. Non-resonant and off-shell effects are expected to scale parametrically as 
 $\Gamma_t / m_t \approx 1\,\%$ and hence can be safely neglected provided that phase space cuts do not severely 
 constrain the top quark invariant mass. This has been explicitly confirmed in studies for $t\bar{t}+H$ production 
 at LO~\cite{Denner:2014wka} and NLO QCD~\cite{Denner:2015yca}.

We obtain the $pp \to b \bar{b} + H $ process from  $pp \to t \bar{t} + H $ by replacing $m_t \to m_b$ in the matrix elements and phase space (and removing the top quark decay), 
while preserving the five flavor scheme with massless initial state quarks.
Hence, we neglect the newly appearing $t$-channel diagram in $bb\to bb$ reactions which is, however, 
doubly-suppressed by the small $b$ quark parton luminosity. In this way the $H$ boson is always emitted from 
the massive final state quarks only. We believe these approximations are sufficient for studying anomalous 
interactions in our analysis.

\begin{figure}
  \begin{minipage}[t]{0.48\textwidth}
    \includegraphics[width=0.9\linewidth]{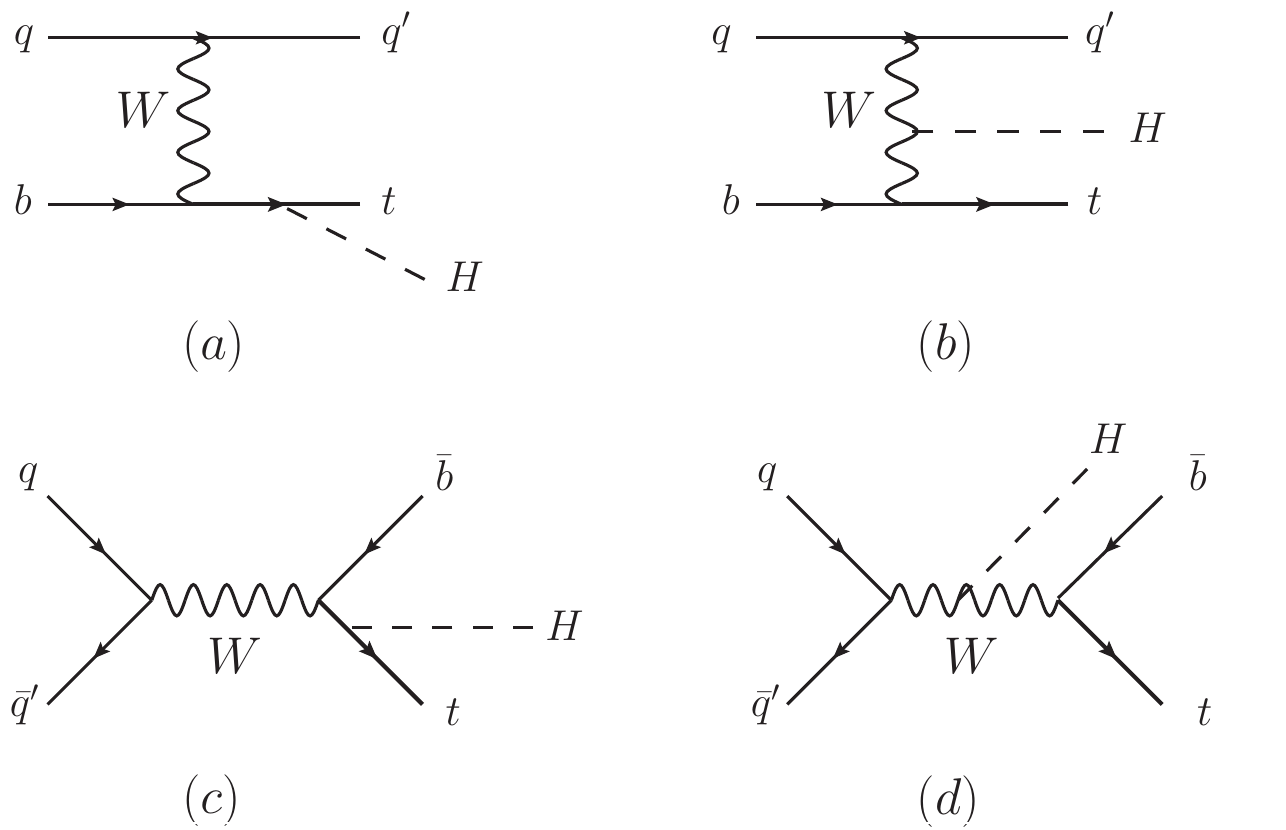}
    \caption{Feynman diagrams describing the single-top production in association with the $H$ boson. 
        The $t$-channel process is shown with the $H$ emitted either from the top quark (a) or from the $W$ boson (b); 
        analogous diagrams in the $s$-channel are shown in (c) and (d). \label{fig:feynman-tqh}}
  \end{minipage}
  \hfill
  \begin{minipage}[t]{0.48\textwidth}
    \includegraphics[width=0.7\linewidth]{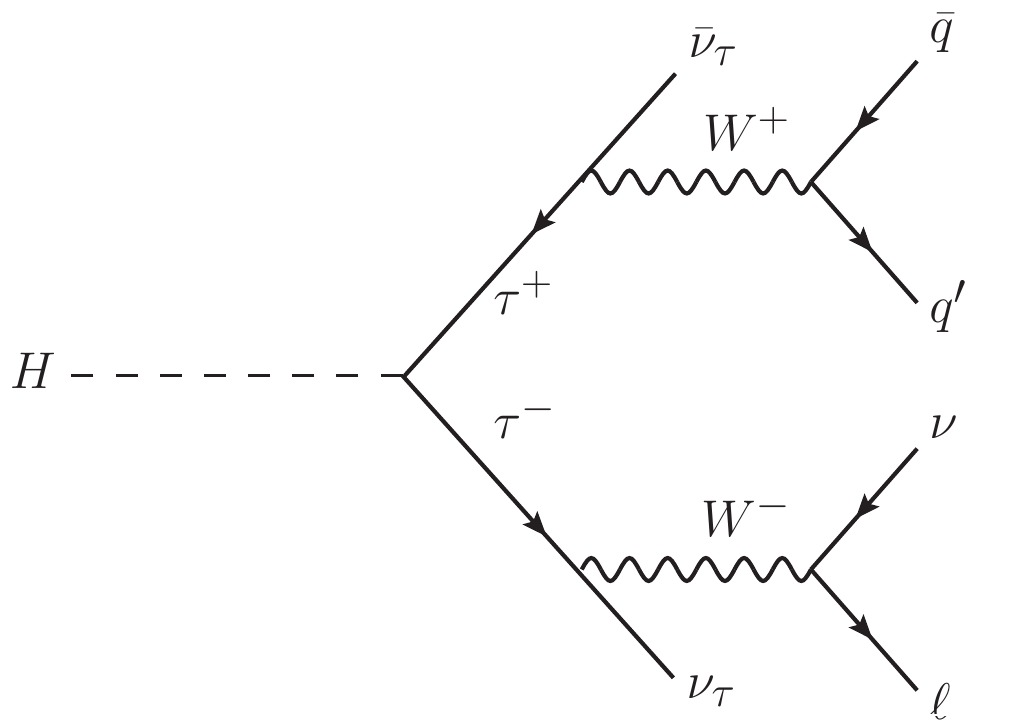} \hspace{0.4cm}
    \caption{Feynman diagram describing the decay $H \to \tau \tau$, with one $\tau$ subsequently decaying leptonically 
$\tau \to \ell \nu_{\tau} \nu$, and the other decaying hadronically $\tau \to q^\prime\bar{q}  \nu_{\tau}$. \label{fig:feynman-htautau}}
  \end{minipage}
\end{figure}

Simulation of the single top quark production process in association with a spin-zero boson relies on the 
partonic processes $qb \to q' t + H$ ($t$-channel process) and $q \bar{q}' \to t \bar{b} + H$ ($s$-channel process).
The former topology is shown in Fig.~\ref{fig:feynman-tqh} (a) and (b), and the latter topology is shown 
in Fig.~\ref{fig:feynman-tqh} (c) and (d). We make use of analytic expressions for the leading-order SM matrix 
elements~\cite{Campbell:2013yla} and extend them to include anomalous couplings, 
keeping the five flavor scheme so that the $H$ boson is never radiated off the initial state $b$-quark.
An extensive comparison of the four and five flavor schemes for this process was performed in Ref.~\cite{Demartin:2015uha}.
Here, we only note that differences between the two schemes are due to missing higher orders in the truncated perturbative series.
Since the perturbative convergence for this process is good, these ambiguities are adequately captured by the scale uncertainty.
Interestingly, in contrast to $\ttH$ and $\bbH$ production, $\tqH$ production includes not only $\Hff$ coupling 
but also $HWW$ coupling (depicted in Fig.~\ref{fig:feynman-tqh} (b) and (d)). 
The interference between these diagrams is destructive and leads to a strongly suppressed production rate 
in the SM~\cite{Maltoni:2001hu}. Therefore any new physics modification of either the $\Hff$ or $HWW$ coupling 
may spoil this suppression and lead to a substantially enhanced production rate and altered kinematics. 
We therefore include anomalous $HWW$ couplings, following Eq.~(\ref{eq:coupl-spin0-VV}).

The study of spin-zero $H$ boson anomalous couplings to tau leptons relies on the matrix element $H \to \tau^+ \tau^-$ 
with subsequent spin-correlated decays $\tau \to \mu \, \nu_\tau \, \nu_\mu $ or $\tau \to q^\prime\bar{q} \, \nu_\tau$. 
These decay chains are illustrated in Fig.~\ref{fig:feynman-htautau}. This decay mode supplements the existing 
$H \to VV$ decays ($V=W,Z,\gamma,g$) within the JHU generator framework such that any $H$ boson production 
process can be interfaced with this decay.
An option for stable $\tau$ leptons allows one to study $H \to \mu^+ \mu^-$ or $H \to e^+ e^-$ decays as well.
Moreover, the tau decay chain encompasses the same structure as the top quark decay, enabling the future study 
of fully spin-correlated decay $pp \to X \to t\bar{t} \to b\bar{b} W(\to f^\prime\bar{f})W(\to f^\prime\bar{f})$, where $X$ 
is any massive spin-zero state. In this work, we will only focus on anomalous coupling studies in $pp \to H \to \tau^+ \tau^-$.
In the current implementation, the form factors for hadronic tau decay are not implemented in the generator and 
instead the inclusive tau decay is simulated. Below we illustrate the re-weighting technique to obtain the $H \to \tau^+\tau^-$ 
process with anomalous couplings using SM simulation of hadronic $\tau$ decays with hadronic form factors 
by the TAUOLA package~\cite{tauola}.

The generation of unweighted events for $H$ boson production in association with heavy-flavor quarks
is performed at leading order (LO),  complemented with parton shower generated by PYTHIA8~\cite{Sjostrand:2006za,Sjostrand:2007gs}. 
The $H$ boson decay is simulated independently from its production. In all cases, the Les Houches Event (LHE) 
file format~\cite{Alwall:2007mw} is used to interface the JHU generator program. 
We also generate weighted events at NLO in QCD for $\ttH$ production to investigate the impact of radiative corrections, as discussed above.
The simulation of the SM processes $\ttH$ and $\bbH$ has been checked against the NLO QCD production simulation 
by POWHEG~\cite{Frixione:2007vw,Hartanto:2015uka,Jager:2015hka}, pseudoscalar $\ttH$ production at NLO QCD has been checked against Ref.~\cite{Frederix:2011zi}, 
while the $H \to \tau^+\tau^-$ decay is validated with TAUOLA. The background $t\bar{t}VV$ samples in this study are generated with MadGraph~\cite{Alwall:2014hca}.

In the following, we focus on the LHC energy of $\sqrt{s}=13$ TeV  and use the following input parameters throughout
\begin{eqnarray}
& m_H = 125.0\,\mathrm{GeV}, 
\quad\quad
& m_t = 173.2\,\mathrm{GeV}, 
\nonumber \\
& m_Z = 91.19\,\mathrm{GeV}, 
\quad\quad
& m_W = 80.39\,\mathrm{GeV},
\nonumber \\
& m_b = 4.2\,\mathrm{GeV}, 
\quad\quad
& m_\tau = 1.8\,\mathrm{GeV}, 
\nonumber \\
& G_\mathrm{F} = 1.16639 \times 10^{-5} \, \mathrm{GeV}^{-2},
\end{eqnarray}
as well as the NNPDF3.0 parton distribution functions~\cite{Ball:2014uwa}.
We summarize the processes relevant to our study of $H$ boson $C\!P$ properties in heavy flavor fermion interactions, 
discussed above, in Table~\ref{table:mc_production}. 
We also show their SM production cross sections and the order in perturbative QCD to which they are simulated.
For each process shown, we provide the matrix elements through the MELA library. 
One direct application of MELA is kinematic discriminants for the optimal analysis,
as discussed in Section~\ref{sect:tth_mela}.
This technique also allows one to re-weight an existing Monte Carlo sample to any model with anomalous couplings without the need for additional CPU-consuming simulation. 
This is particularly important for the LHC experiments where modeling ATLAS and CMS detector response sometimes
requires months of wall-clock time. 
A successful application of this procedure has been presented in Ref.~\cite{Khachatryan:2014kca}.

\begin{table}[t]
\begin{center}
\renewcommand{\arraystretch}{1.5}
\begin{tabular}{|c|c|c|}
\hline
Process & SM cross section  & Simulation\\
        & or branching &  \\
\hline\hline
$\ttH$          & 509 fb & LO + parton shower events \\ 
                  &                &      NLO QCD weighted events            \\
$\bbH$          & 512 fb & LO + parton shower \\ 
$\tqH$ ($t$-channel)          & 73 fb & LO + parton shower events\\ 
$\tbH$ ($s$-channel)          & 3 fb & LO + parton shower events\\ 
$H\to \tau^+\tau^-$    &       6.3\%    & LO + parton shower events\\
\hline
\end{tabular}
\caption{
Summary of MC processes with anomalous $\Hff$ couplings in the production and decay
implemented in the JHU generator and the MELA package. 
The cross sections are listed without systematic uncertainties 
at the $\sqrt{s}=13$\, TeV LHC for a SM Higgs boson mass of $125$\,GeV~\cite{Heinemeyer:2013tqa,Demartin:2015uha}.
\label{table:mc_production}
 }
\end{center}
\end{table}


\section{Matrix element technique}
 \label{sect:tth_mela}

The matrix elements, or multivariate per-event likelihoods, maximize the amount of information 
that can be extracted from a given event. These techniques  were used for example in top and bottom quark, as well as electroweak boson 
measurements, and proved to be powerful tools for the $H$ boson discovery and characterization during Run I of the LHC
on both CMS~\cite{Chatrchyan:2012xdj,Chatrchyan:2012jja,Chatrchyan:2013mxa,Khachatryan:2014iha,Khachatryan:2015cwa,Khachatryan:2014kca,Khachatryan:2015mma,Khachatryan:2015ila} 
and ATLAS~\cite{Aad:2013xqa,Aad:2015mxa,Aad:2016nal} experiments.
As part of the latter development, we investigated application 
of these techniques to the production and decay processes involving $H$ boson 
coupling to vector bosons in Refs.~\cite{Gao:2010qx,Bolognesi:2012mm,Anderson:2013afp}.
Here we extend the MELA technique to the processes involving $H$ boson coupling to heavy flavor fermions. 

We take the $gg(\qqbar)\to\ttH$ processes as an example to define a complete set of kinematic observables
following the full sequence of the process, similar to the $HVV$ production and decay kinematics discussed 
in Refs.~\cite{Gao:2010qx,Anderson:2013afp}. These observables are equivalent to the more familiar 
observables defined in the laboratory frame, as shown in Appendix~\ref{sec:appendixA}, but provide 
a more intuitive insight into the production and decay dynamics. We then define the complete set of matrix element discriminants, following
Refs.~\cite{Bolognesi:2012mm,Anderson:2013afp}, in application to the processes involving heavy fermion couplings
to the $H$ boson. 


\subsection{Kinematics in the $H$ boson production and decay}
\label{sect:tth_mela_angles}

The processes $gg(\qqbar)\to\ttH$, $\tqH$, or $\bbH$ with subsequent decay of the top quarks and the $H$ boson
can be characterized by the four-momenta of the decay products, such as leptons and quark jets. In the case of one final state 
neutrino, its momentum can be deduced from a kinematic fit using mass constraints and
utilizing the missing transverse energy information. 
In the following description we consider the $\ttH$ production in its center-of-mass frame.
Both longitudinal and transverse momenta of the $\ttH$ system can be parameterized separately. They are driven by QCD effects, 
either parton distribution functions of the proton for rapidity or additional jet radiation for transverse momentum. 

\begin{figure}[t]
\centerline{
\setlength{\epsfxsize}{0.37\linewidth}\leavevmode\epsfbox{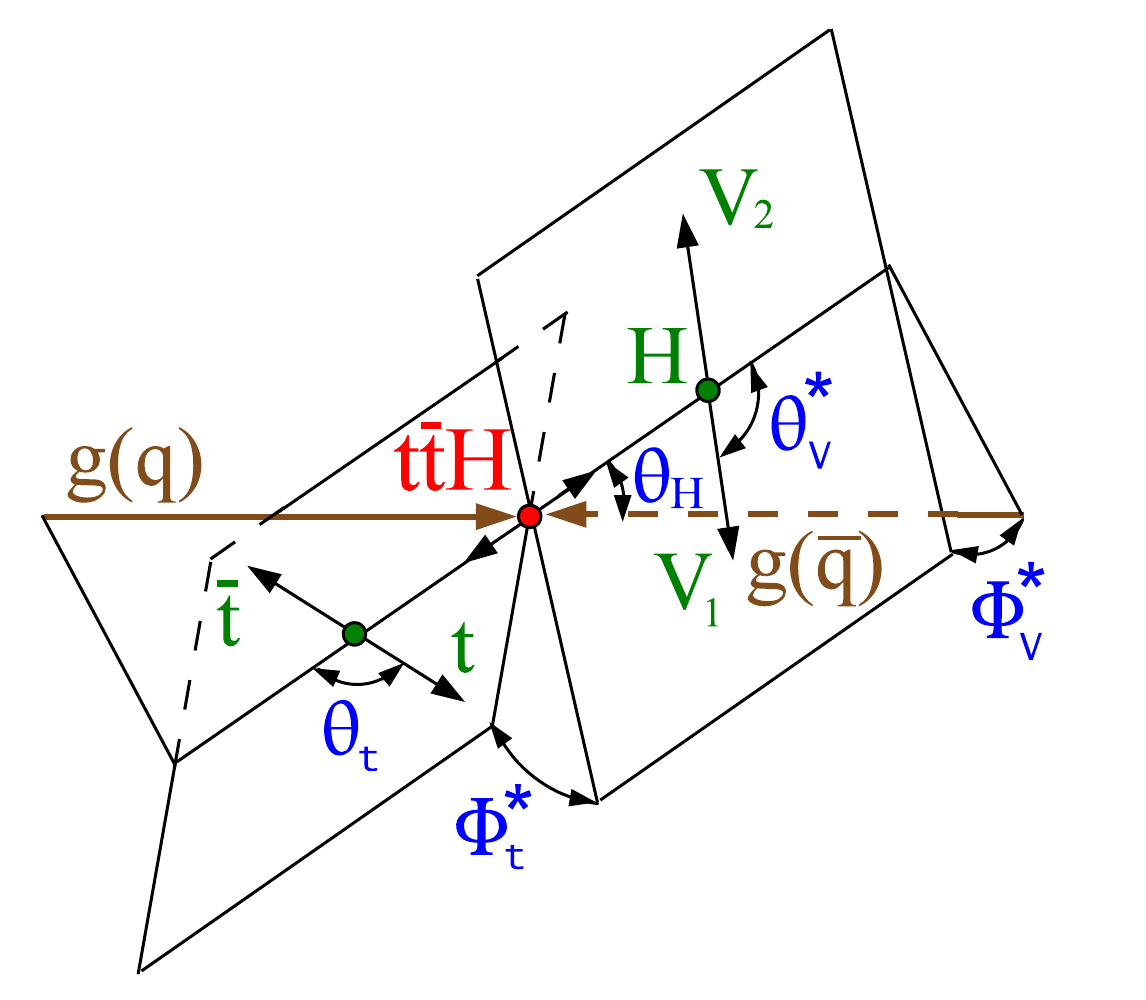}
\setlength{\epsfxsize}{0.37\linewidth}\leavevmode\epsfbox{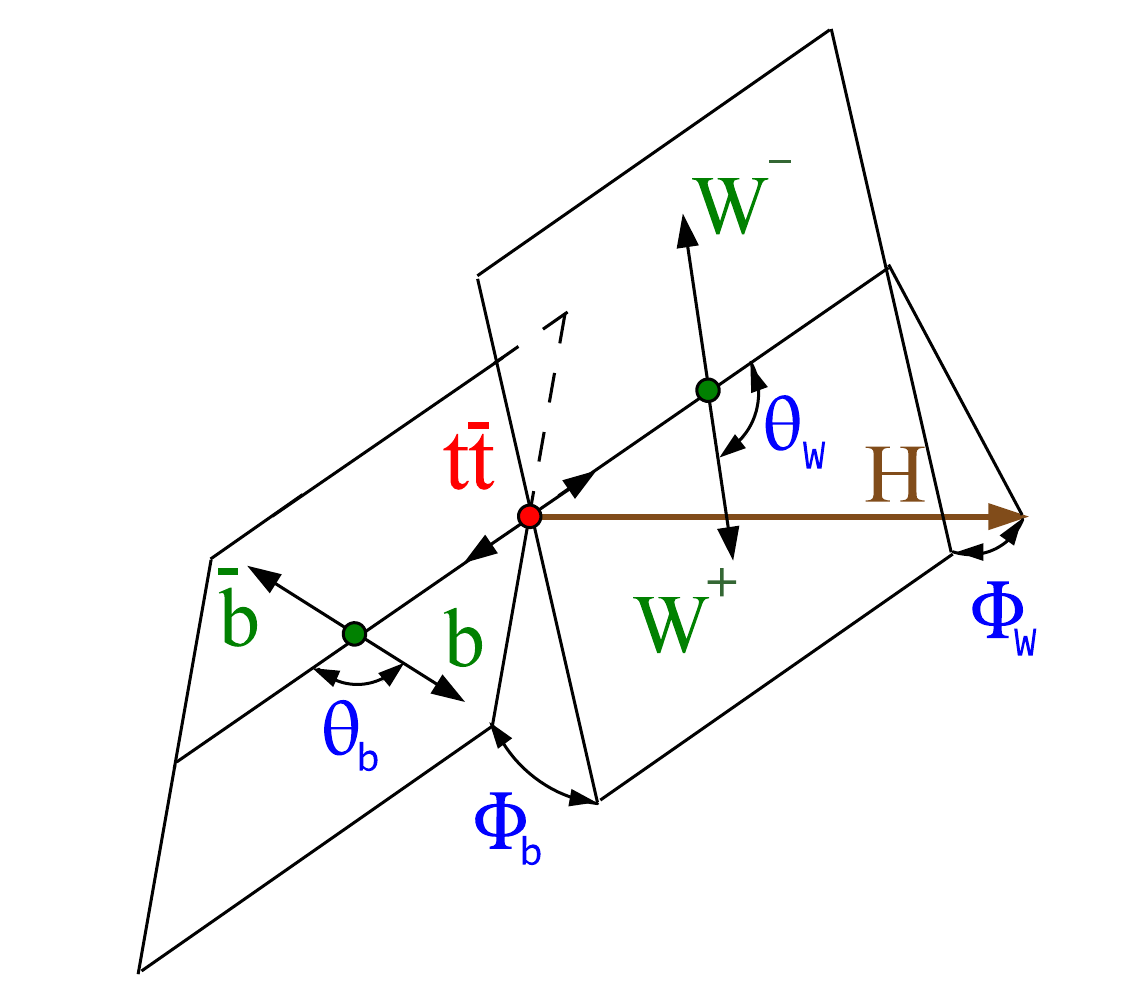}
\setlength{\epsfxsize}{0.26\linewidth}\leavevmode\epsfbox{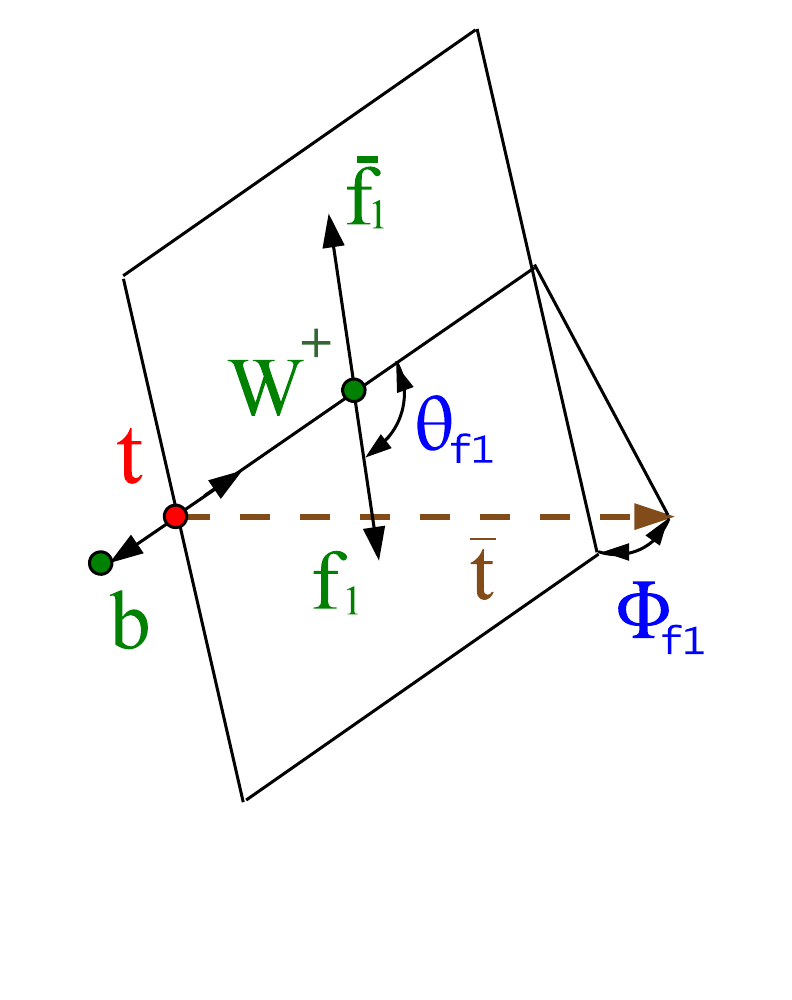}
}
\caption{
The definition of observable in the sequential process of production and decay of $\ttH$, see text for details. 
Each angle is defined in the respective reference frame of the decaying system. 
}
\label{fig:decay}
\end{figure}

Similar to the description of the $H$ boson production and decay with couplings to vector 
bosons~\cite{Gao:2010qx,Bolognesi:2012mm,Anderson:2013afp}, it is convenient to describe the complete kinematics 
of the process by a set of angles and invariant masses, which we generically denote as $\vec\Omega$,
following the sequential processes. The definition of observables in the process 
$gg(\qqbar)\to\ttH\to(W^+b)(W^-\bar{b})H\to(f_1^\prime\bar{f}_1b) (\bar{f}_2f_2^\prime\bar{b})(VV)$
is shown in Fig.~\ref{fig:decay}. 
The following set of angles and invariant masses is defined~as
\begin{itemize}
\item $m_{\ttH}$: invariant mass of the $\ttH$ system;
\item $\theta_H$: angle between the $H$ boson direction and the incoming partons in the $\ttH$ frame;
\item $\theta_{V}^{*}$: angle of the $H\to VV (f\bar{f})$ decay with respect to the opposite $t\bar{t}$ direction in the $H$ frame; 
\item $\Phi^*_{V}$: angle between the production plane, defined by incoming partons and $H$, and $H\to VV (f\bar{f})$ decay plane;
\item $\theta_{t}$: angle between the top quark direction and the opposite Higgs direction in the $t\bar{t}$ frame;
\item $\Phi_{t}^{*}$: angle between the decay planes of the $t\bar{t}$ system and $H\to VV (f\bar{f})$ in the $\ttH$ frame;
\item $m_{tt}$: invariant mass of the $t\bar{t}$ system;
\item $\theta_{W}$: angle between $W^+$ and opposite of the $b\bar{b}$ system in the $W^+W^-$ frame;
\item $\Phi_{W}$: angle between the production $(b\bar{b})(W^+W^-)H$ plane and the plane of the $W^+W^-$ system in the $t\bar{t}$ frame;
\item $\theta_{b}$: angle between the $b$ quark and opposite of the $W^+W^-$ system in the $b\bar{b}$ frame;
\item $\Phi_{b}$: angle between the planes of the $b\bar{b}$ and $W^+W^-$ systems in the $t\bar{t}$ frame;
\item $m_{Wb1}$ or $m_{Wb2}$: invariant mass of the $W^+b$ or  $W^-\bar{b}$ system;
\item $\theta_{f1}$ or $\theta_{f2}$: angles between fermion direction and opposite of the $b$ or $\bar{b}$ quark 
          in the $W^+$ or $W^-$ frame;
\item $\Phi_{f1}$ or $\Phi_{f2}$:  angle between the $W^+$ or $W^-$ decay plane
          and the $\bar{t}W^+b$ or $tW^-\bar{b}$ plane in the $t$ or $\bar{t}$ quark frame;
\item $m_{f1\bar{f}1}$ or $m_{f2\bar{f}2}$: invariant mass of the $f_1\bar{f}_1$ or  $f_2\bar{f}_2$  system.
\end{itemize}
The decay of the $H$ boson with angles $\theta_{V}^{*}$ and $\Phi^*_{V}$ is shown only for illustration, 
their distribution is flat for a spin-zero $H$ boson production due to the lack of spin correlations between the production and decay processes. 
Their complete description is discussed in Ref.~\cite{Gao:2010qx} in terms of the two equivalent angles
$\theta^*$ and $\Phi_{1}$.
The grouping of the $W^+W^-$ and $b\bar{b}$ systems, as opposed to $W^+b$ and $W^-\bar{b}$,
is motivated by enhanced spin-correlation effects visible with the corresponding observables. 
The complete multidimensional distribution retains full information with either approach. 

\begin{figure}[ht]
\centerline{
\setlength{\epsfxsize}{0.32\linewidth}\leavevmode\epsfbox{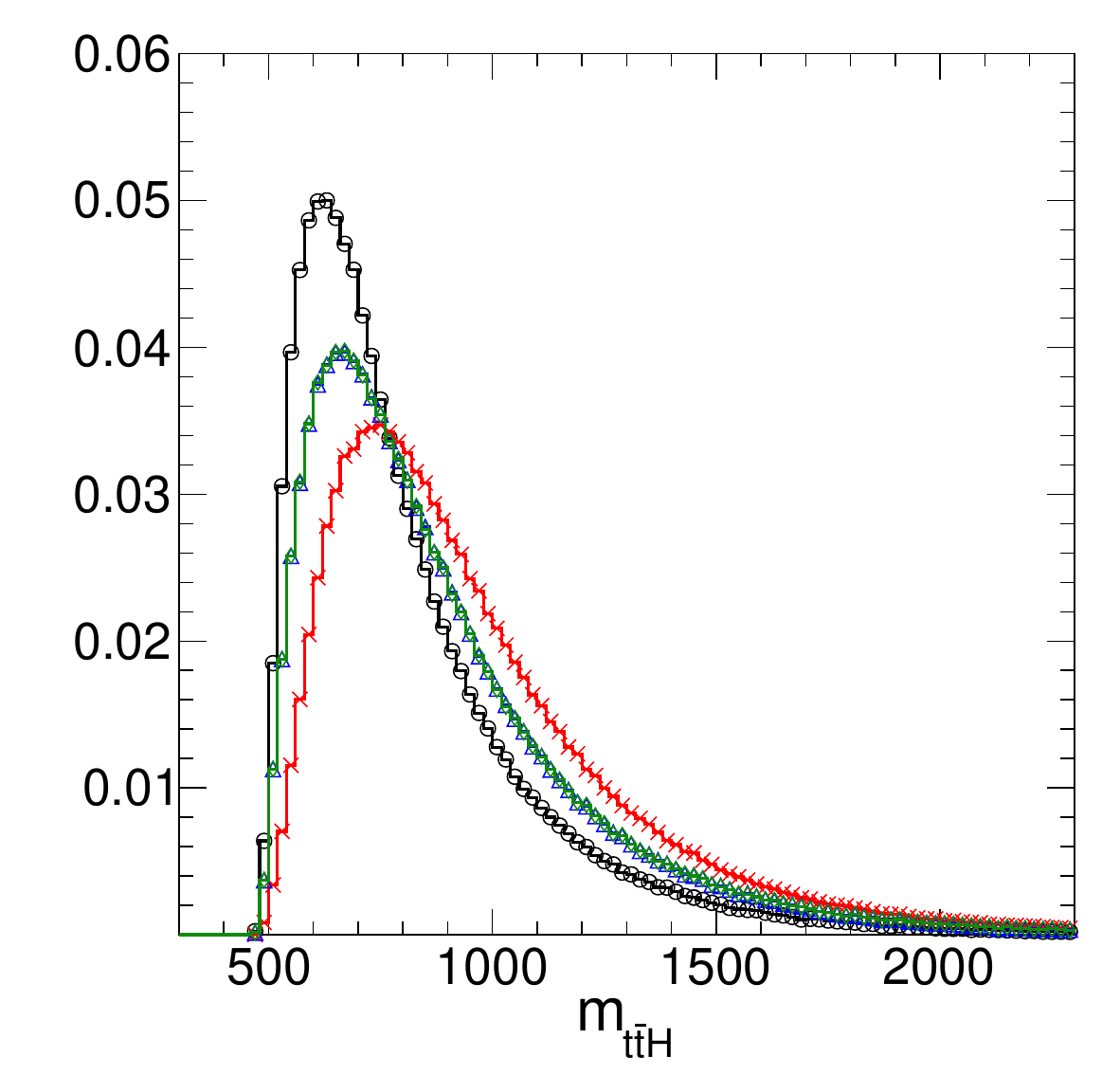}
\setlength{\epsfxsize}{0.32\linewidth}\leavevmode\epsfbox{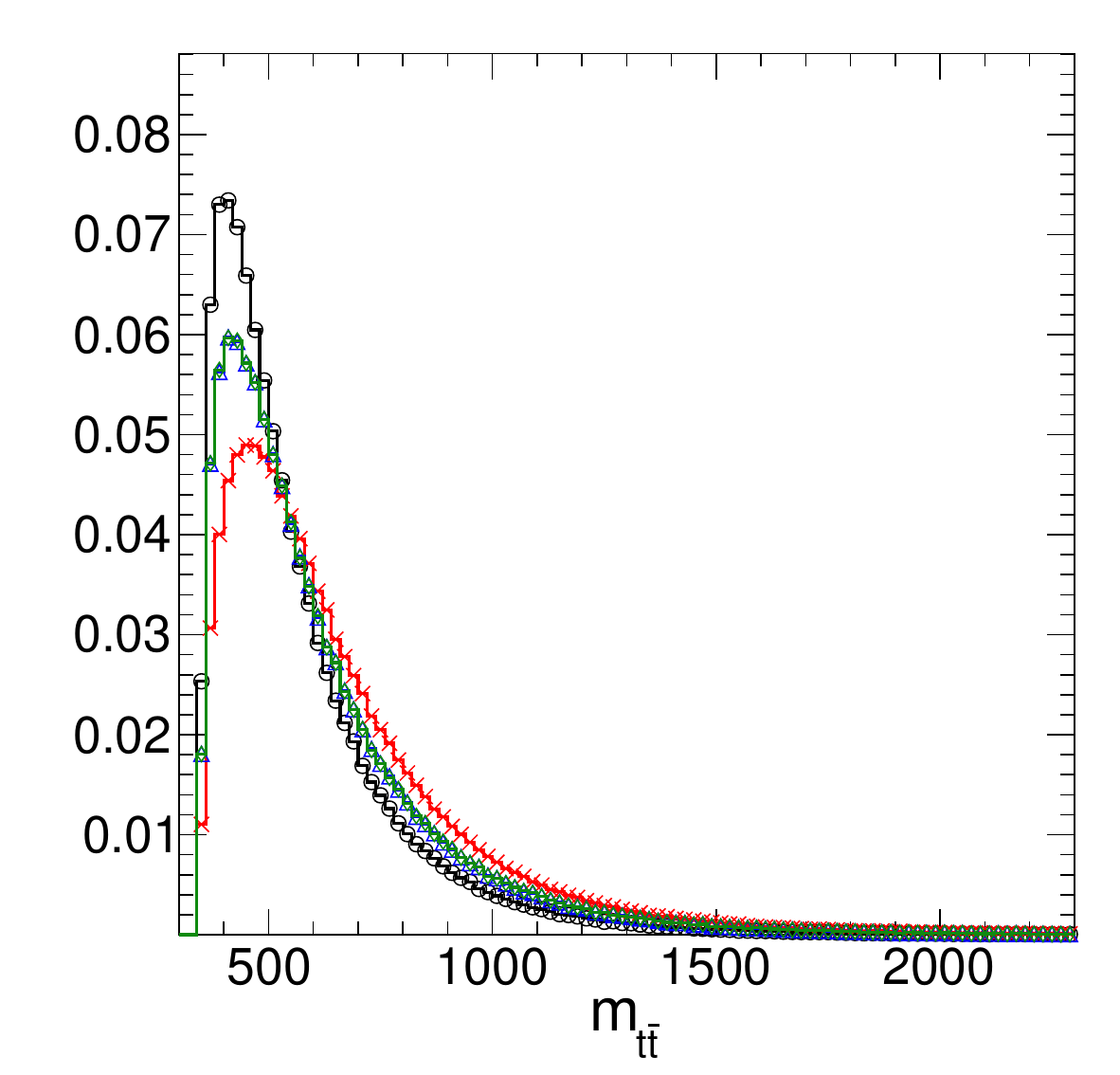}
\setlength{\epsfxsize}{0.32\linewidth}\leavevmode\epsfbox{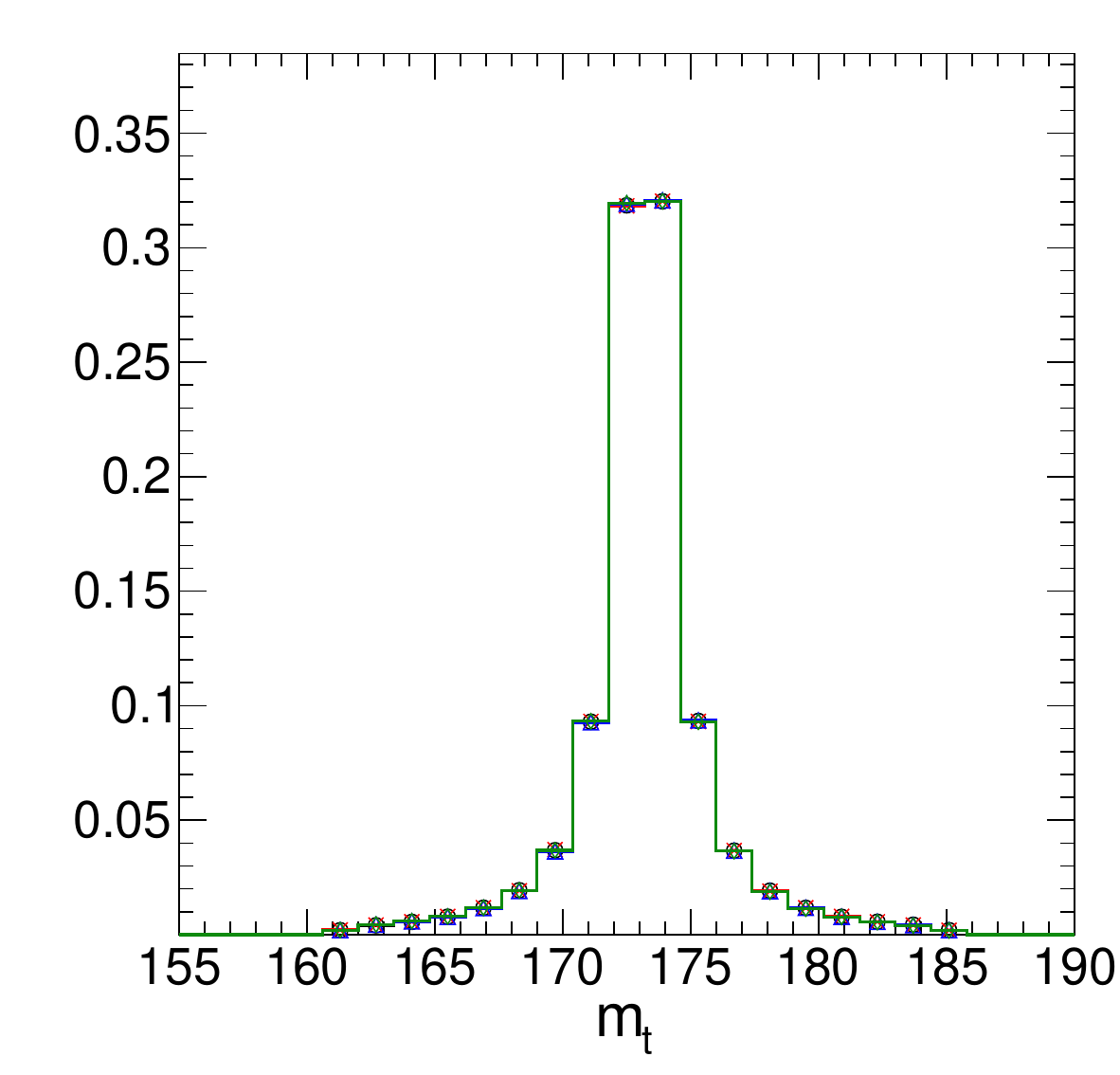}
}
\centerline{
\setlength{\epsfxsize}{0.32\linewidth}\leavevmode\epsfbox{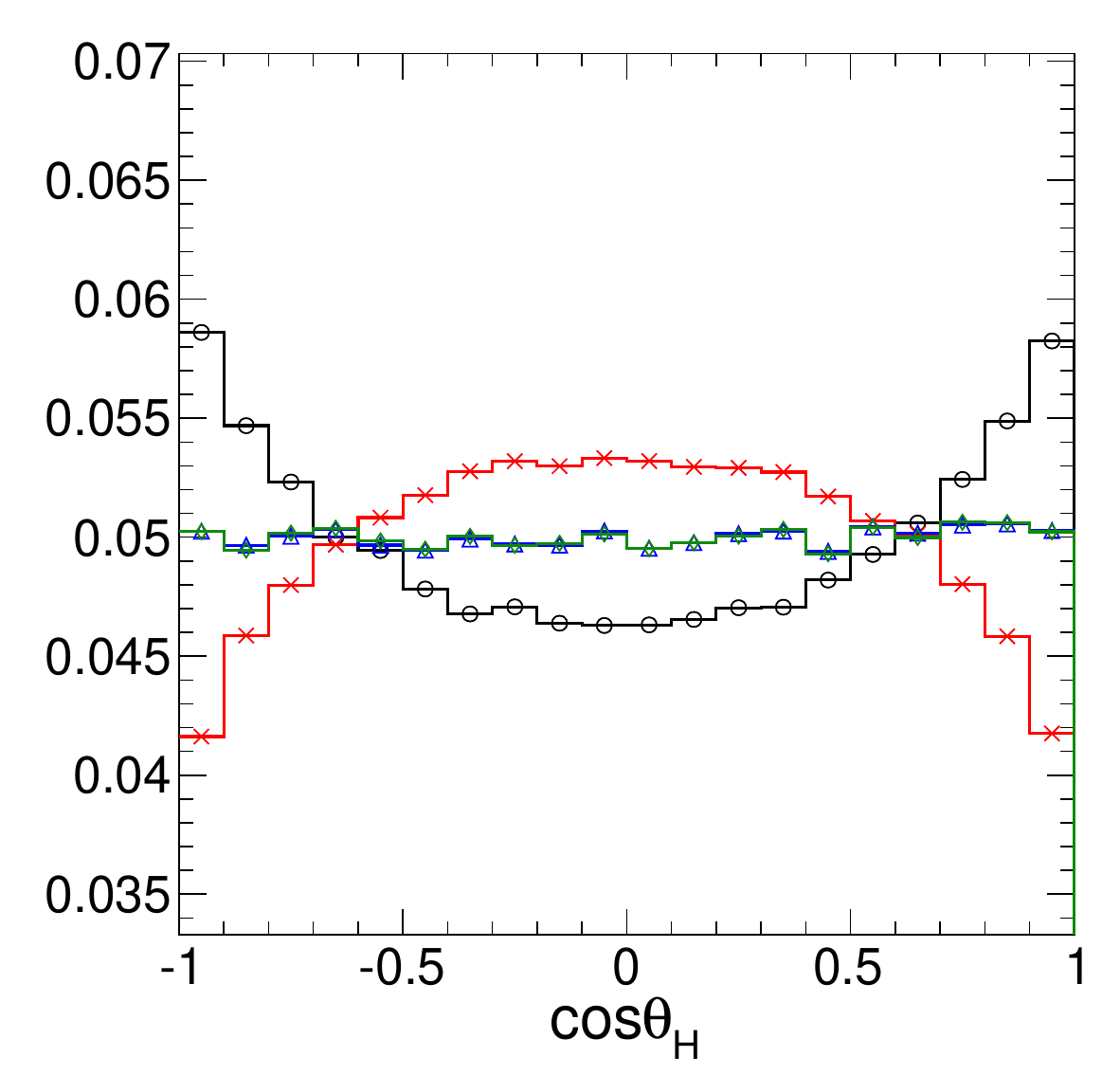}
\setlength{\epsfxsize}{0.32\linewidth}\leavevmode\epsfbox{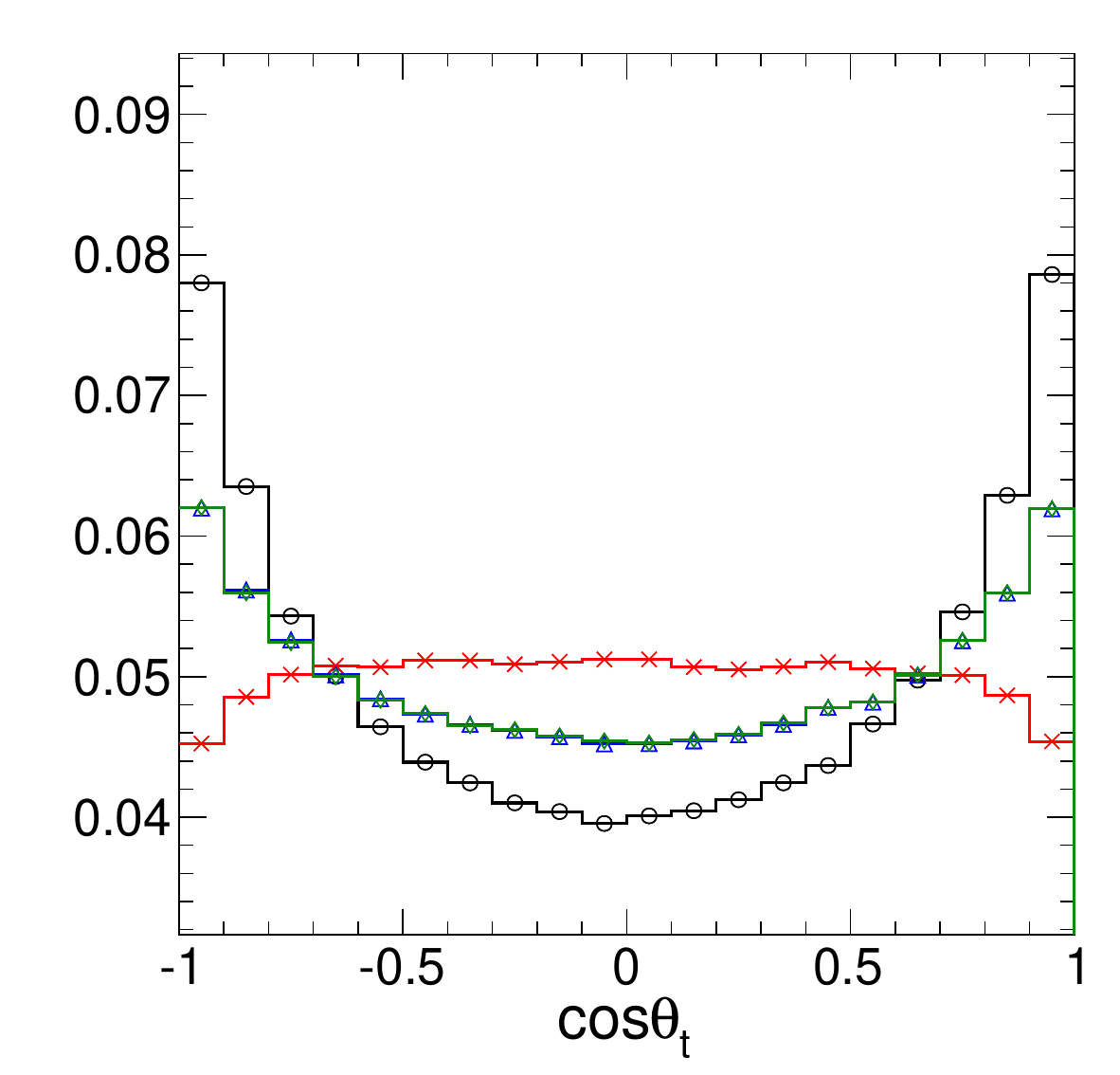}
\setlength{\epsfxsize}{0.32\linewidth}\leavevmode\epsfbox{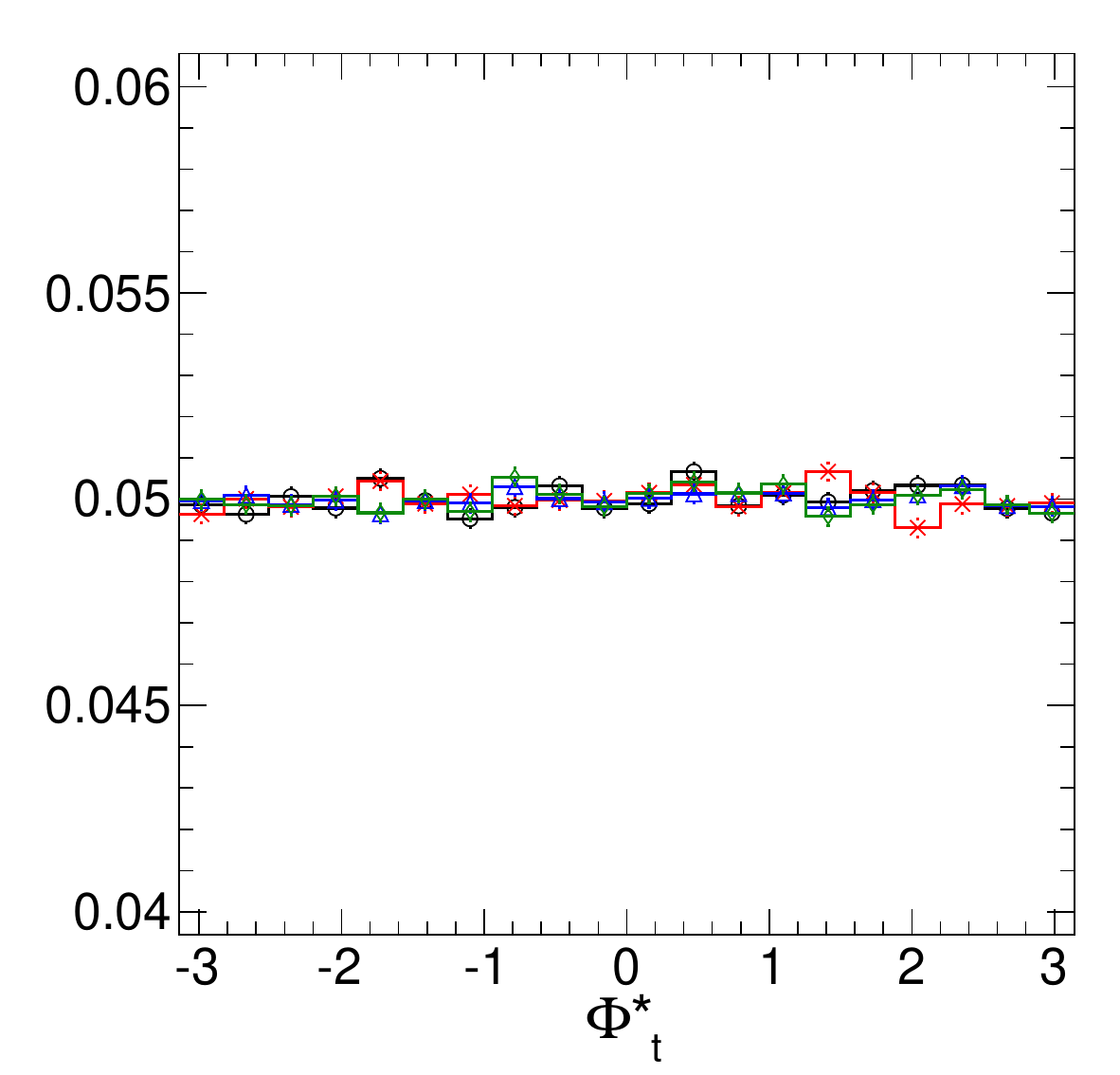}
}
\centerline{
\setlength{\epsfxsize}{0.32\linewidth}\leavevmode\epsfbox{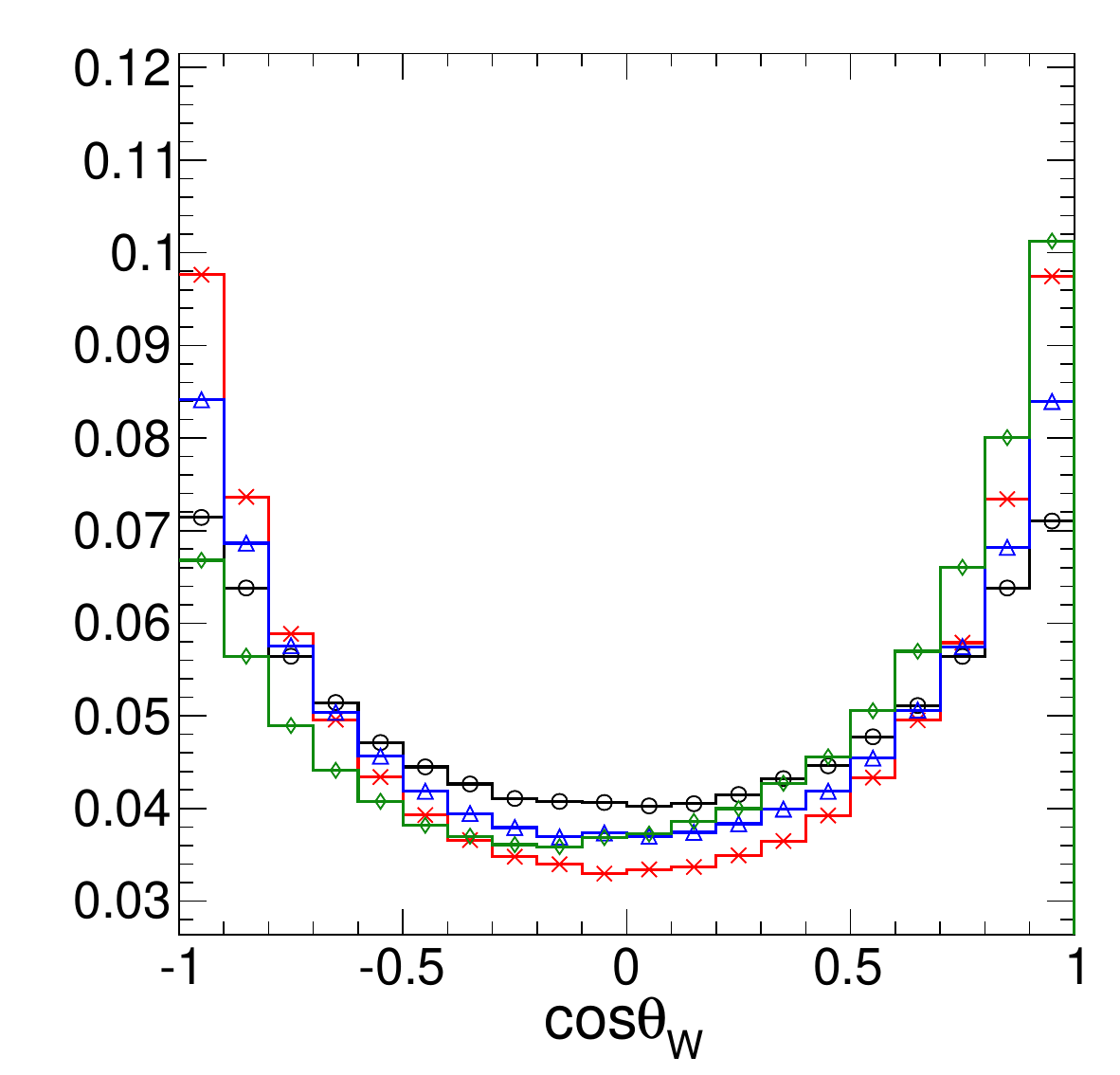}
\setlength{\epsfxsize}{0.32\linewidth}\leavevmode\epsfbox{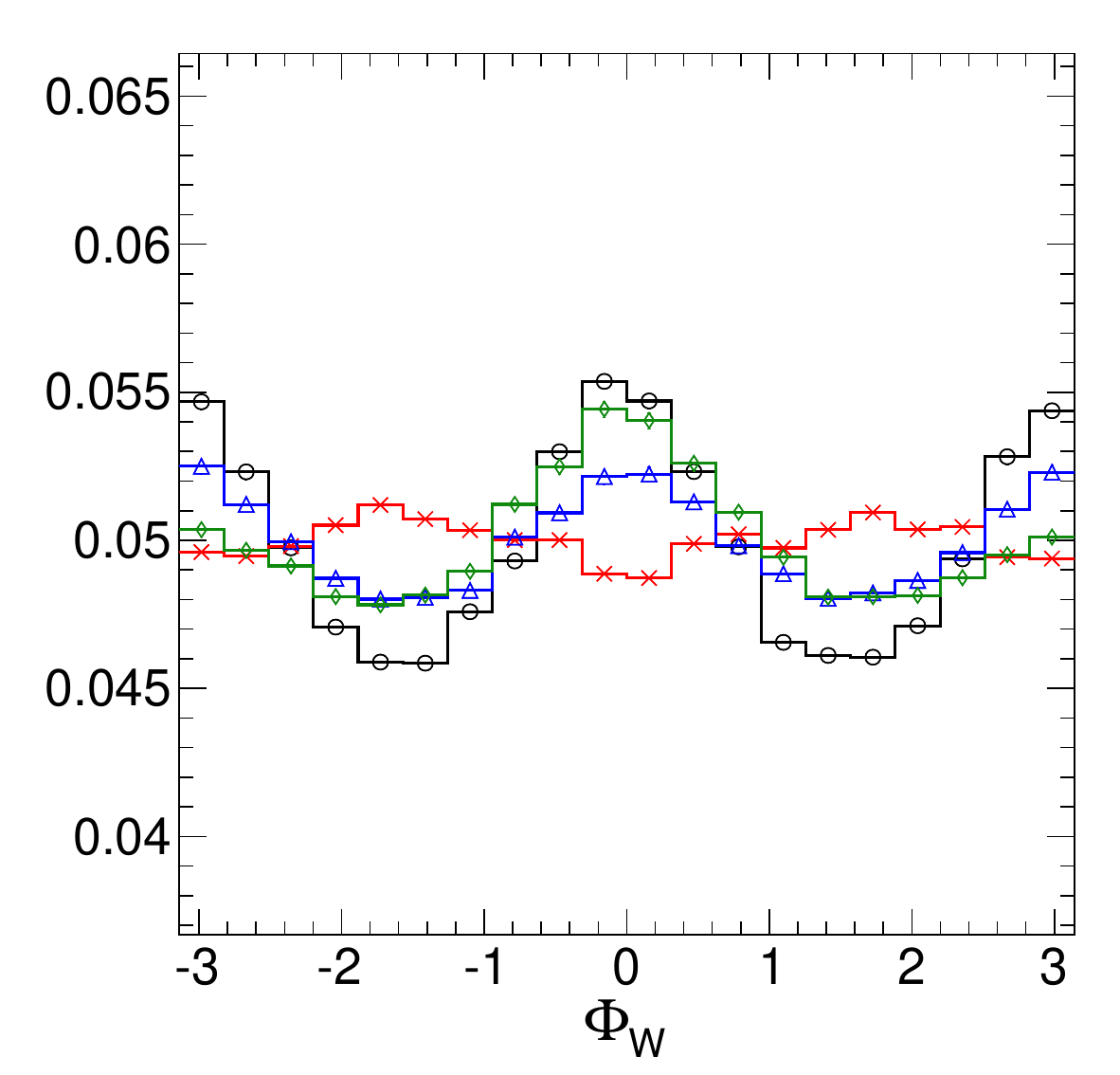}
\setlength{\epsfxsize}{0.32\linewidth}\leavevmode\epsfbox{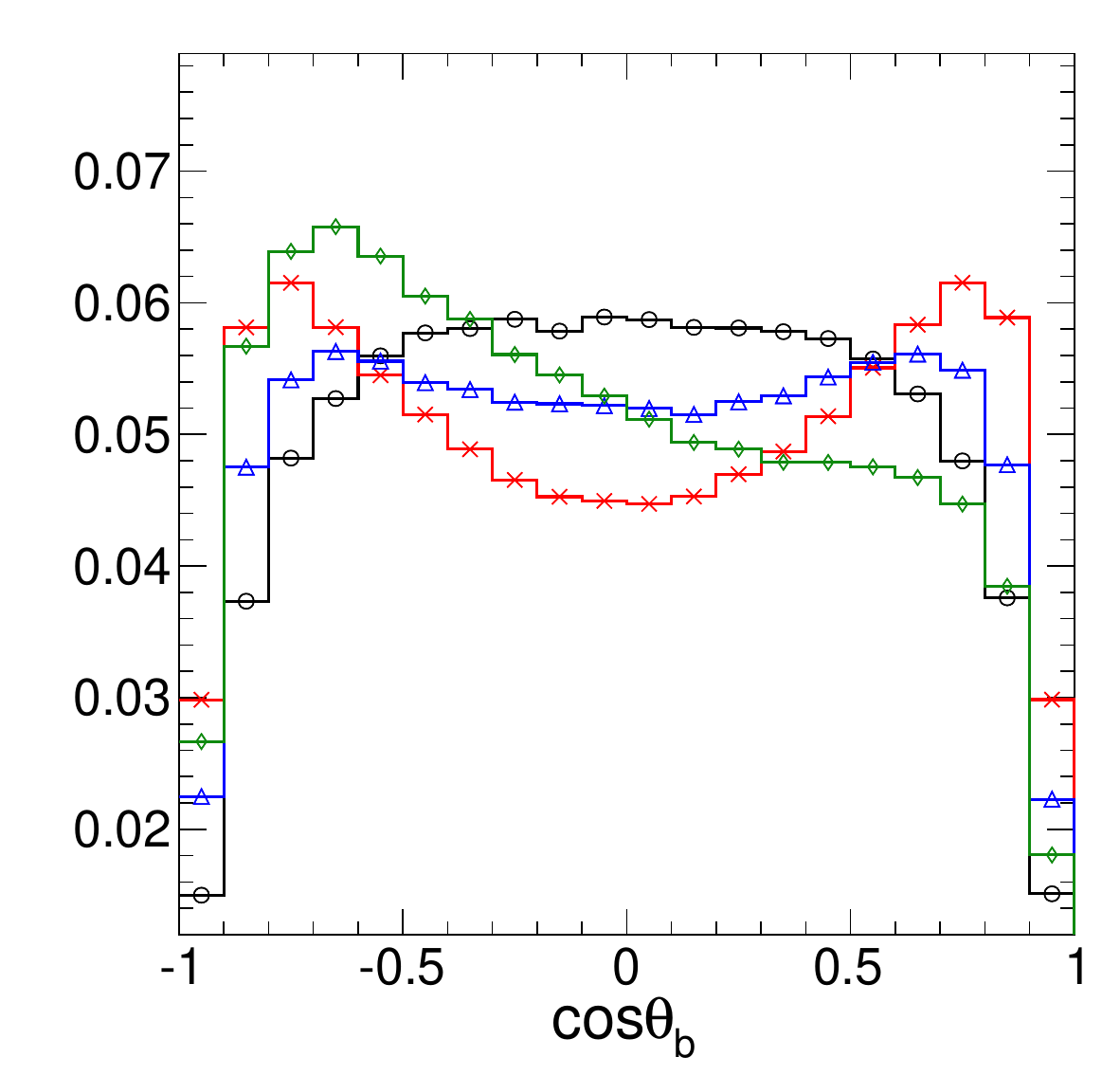}
}
\centerline{
\setlength{\epsfxsize}{0.32\linewidth}\leavevmode\epsfbox{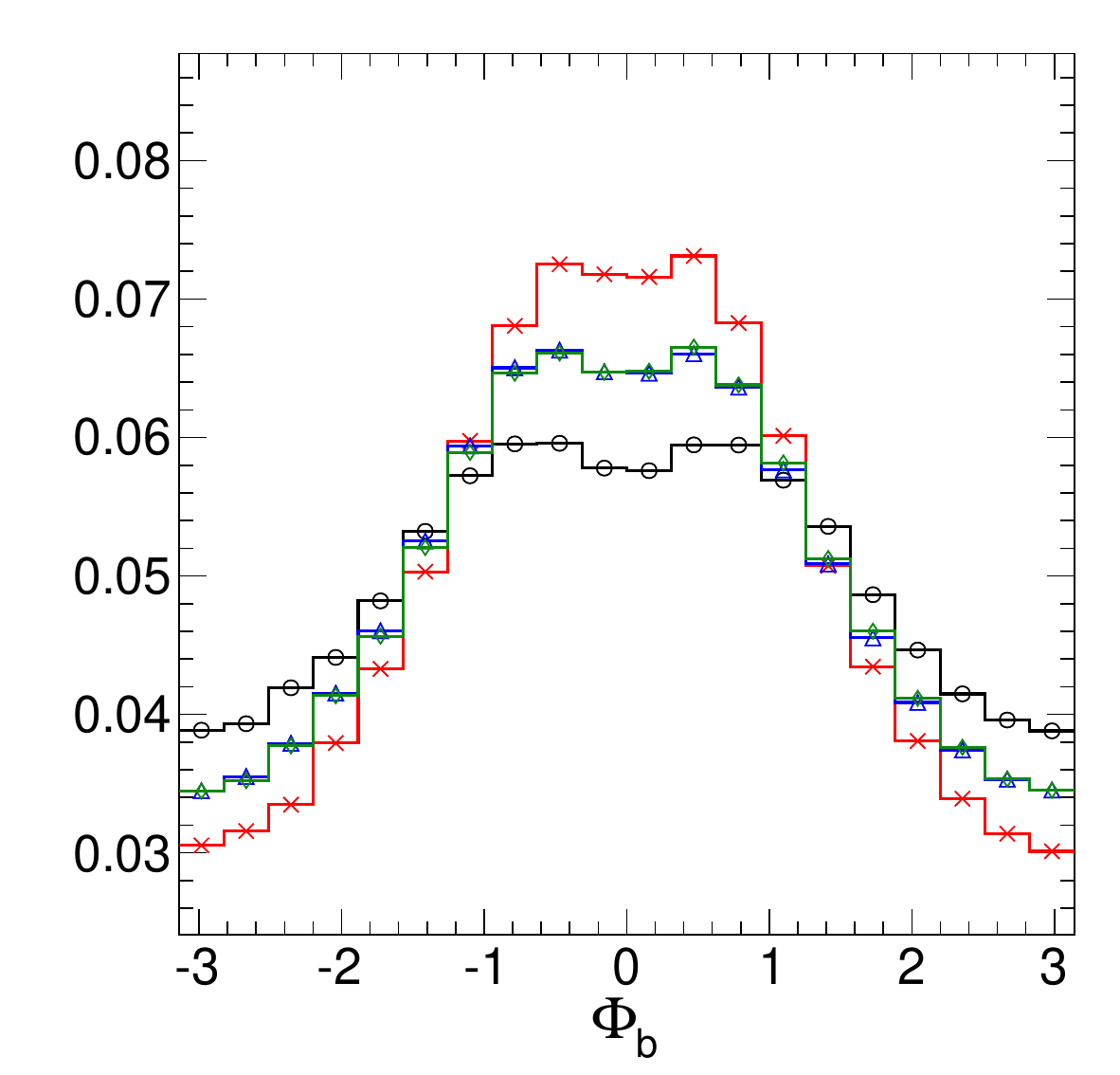}
\setlength{\epsfxsize}{0.32\linewidth}\leavevmode\epsfbox{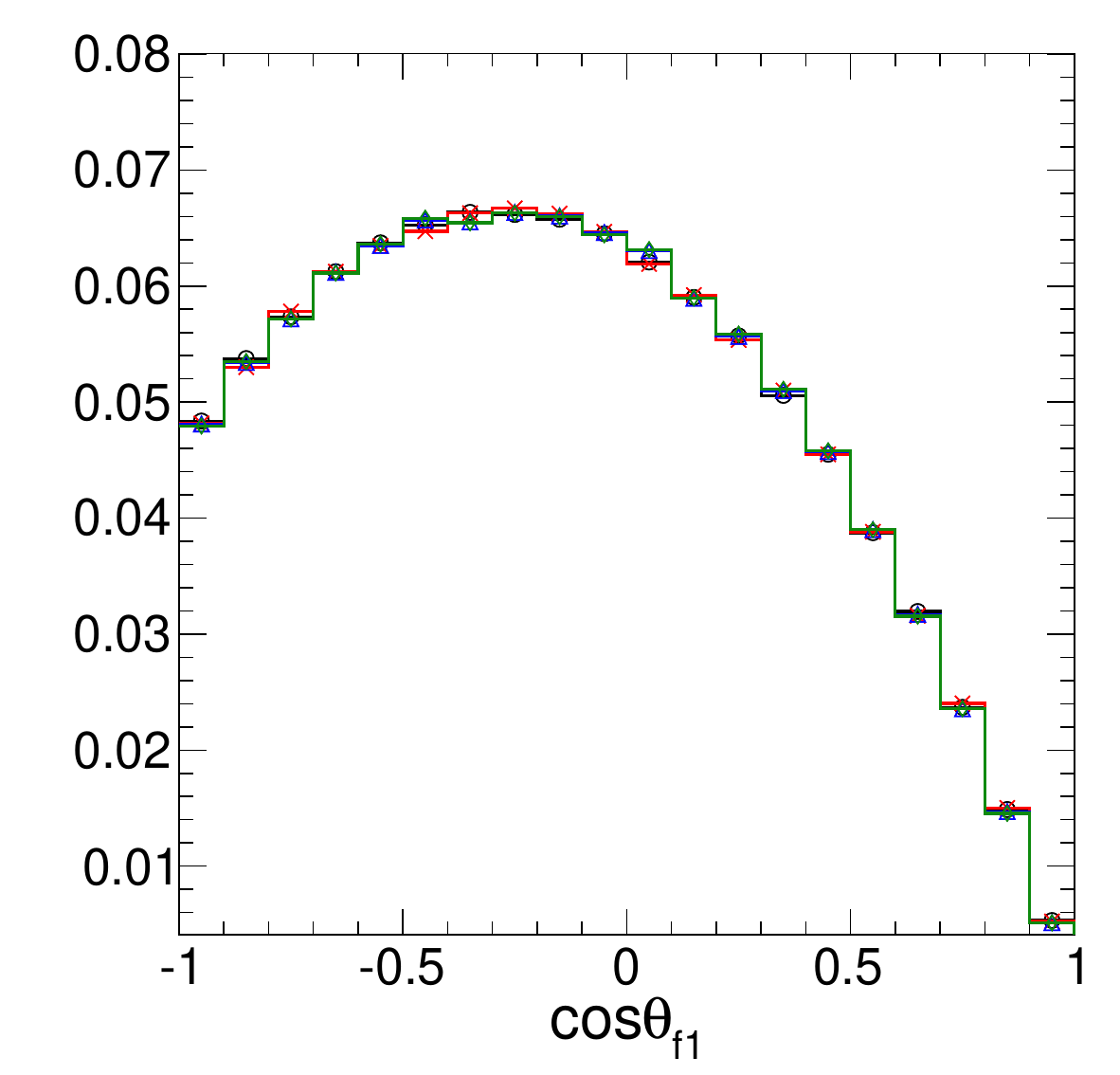}
\setlength{\epsfxsize}{0.32\linewidth}\leavevmode\epsfbox{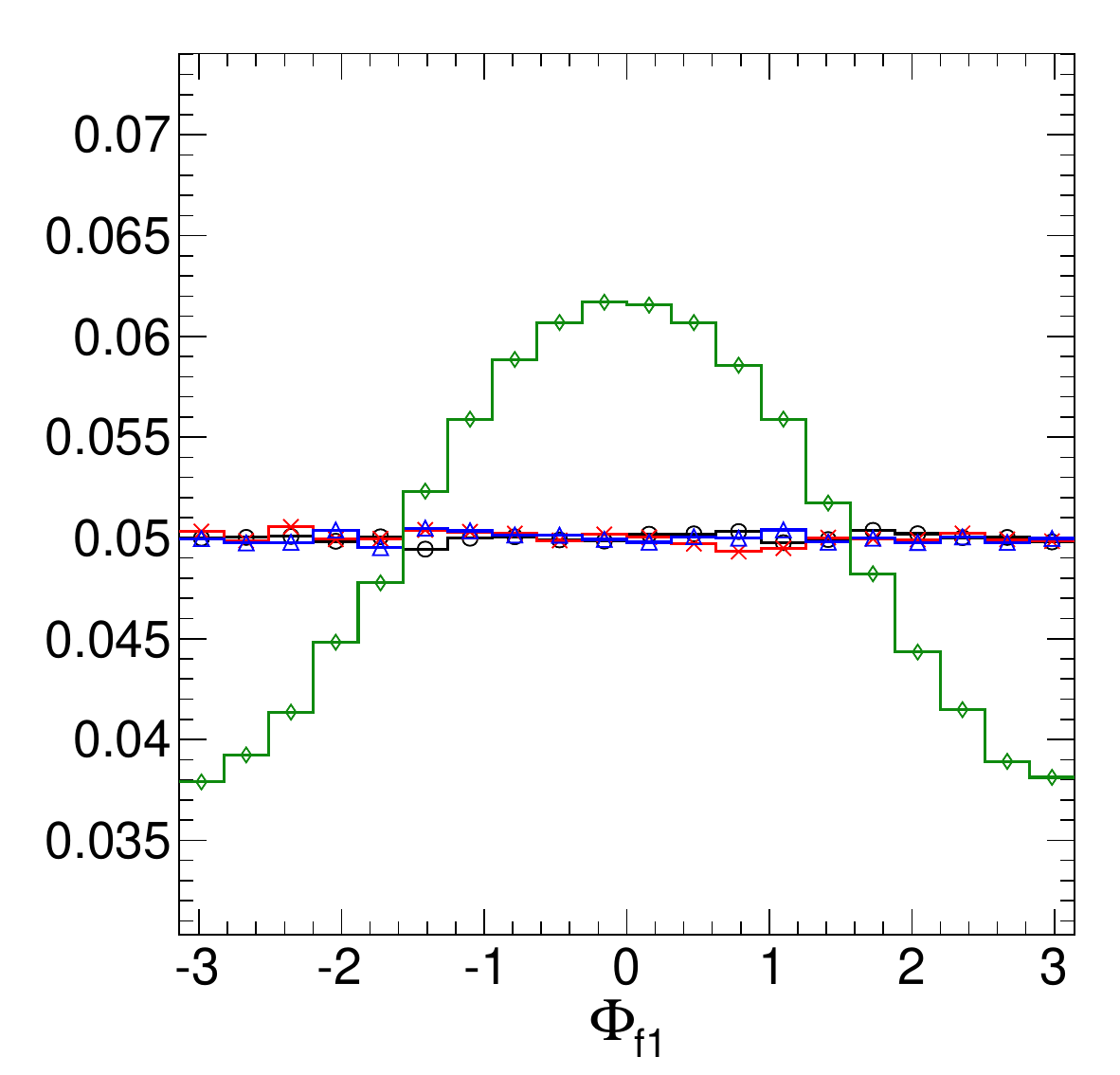}
}
\caption{
The normalized angular and mass distributions in the process $pp \to\ttH$ corresponding to four
scenarios of anomalous $\ttH$ couplings:
$f_{\CP}=0$ (SM $0^{+}$, black circles), $f_{\CP}=1$ (pseudoscalar $0^{-}$, red crosses),
$f_{\CP}=0.28$ with $\phi_{\CP}=0$ (blue triangles) and $\phi_{\CP}=\pi/2$ (green diamonds). 
The LHC $pp$ energy of 13 TeV and $H$ boson mass of 125 GeV are used in simulation.
See text for definition of all observables. 
}
\label{fig:angular}
\end{figure}

Figure~\ref{fig:angular} shows the non-trivial angular distributions in the process $pp\to\ttH$ 
corresponding to four scenarios of anomalous $\ttH$ couplings: pure scalar, pseudoscalar, and 
two mixed scenarios with $f_{\CP}=0.28$  (corresponding to the equal cross-section of scalar and 
pseudoscalar processes) and different phases. 
Most angular observables exhibit a clear difference between the scalar and pseudoscalar processes.
Only observables appearing in sequential decay of the top quarks are sensitive to $\phi_{\CP}$.
As noted earlier, these observables together with the boost of the $\ttH$ system
are equivalent to other observables defined in the laboratory frame, 
as shown in Appendix~\ref{sec:appendixA}, but provide complete kinematics required as input to the
matrix element tools and emphasize particular features in the process. 

The description of observables $\vec\Omega$ in the other processes $pp \to\tqH$ and $\bbH$ 
follows by analogy, with only a subset of observables available due to lack of sequential decay of 
at least one associated quark. 

\clearpage


\subsection{Matrix element likelihood approach}
\label{sect:tth_mela_mela}

With the kinematics of a process reflected in the complete set of observables $\vec\Omega$, 
one could in principle analyze the data in this multi-dimensional space. However, this often becomes
impractical with a large number of observables, as illustrated in Section~\ref{sect:tth_mela_angles},
when parameterization of probabilities and detector effects in such a multi-dimensional space becomes
difficult. Reducing the number of observables is a possible approach, but essential information may be lost. 
Machine learning techniques can approximate optimal functions that can depend on a large number of inputs,
but those also require training and therefore perform only as well as training goes. These techniques 
are typically targeted to discriminate between certain categories of events and are not optimal for dealing
with quantum mechanical interference effects which become essential in physics processes, 
though for possible solutions see Ref.~\cite{Cranmer:2015bka}.

The matrix element techniques are the methods based on multivariate per-event likelihoods 
prepared using a phenomenological calculation for the process of interest. They may employ the
same calculation as used in the Monte Carlo event generators or may be reformulated to 
represent analytical distributions of observables of interest, such as $\vec\Omega$. 
Such matrix elements, if used properly, are guaranteed to retain full information about the event.
The difficulty in using matrix element methods often comes from non-trivial detector effects which 
alter multivariate likelihoods. This issue is greatly simplified by utilizing ratios of the matrix 
elements in which certain detector effects cancel, most importantly variation of reconstruction efficiency 
as a function of observables. Resolution effects may be introduced with transfer functions, or neglected
when their effect on performance is small. Missing degrees of freedom, such as neutrinos, may be either
constrained from the global event information, as we illustrate below, or integrated out in the matrix
element calculation. 
In the end, either machine learning or matrix element techniques could be used in the analysis of the data, 
but in either approach, it is ultimately the matrix elements which guide us in maximizing the amount 
of information, as they are also used in machine training through Monte Carlo. 

The basic idea of the MELA technique is to project kinematics on the minimal set of discriminants calculated 
as ratios of the matrix elements.  It has already proved to be a powerful tool for the $H$ boson discovery 
and characterization during Run I of the LHC as applied primarily to the $H$ boson coupling to the vector bosons. 
For a simple discrimination of two hypotheses, the Neyman-Pearson lemma~\cite{Neyman:1933} guarantees that the ratio
of probabilities ${\cal P}$ for the two hypotheses provides an optimal discrimination power. However, for a continuous set of
hypotheses with an arbitrary quantum-mechanical mixture several discriminants are required for an optimal measurement
of their relative contributions. For example, probability for interference of two contributions could be presented as
%
\begin{equation}
{\cal P}_{\rm sig}(\vec{x}_{i}; f_{\CP}, \phi_{\CP}) =  
(1-f_{\CP}) \, {\cal P}_{0^+}(\vec{x}_{i}) + f_{\CP} \, {\cal P}_{0^-}(\vec{x}_{i}) 
+ \sqrt{f_{\CP}(1-f_{\CP})} \, \left( {\cal P}_{\rm int}(\vec{x}_{i}) \cos\phi_{\CP}+ {\cal P}^\perp_{\rm int}(\vec{x}_{i}) \sin\phi_{\CP}\right)  
\,,
\label{eq:fractions-P}
\end{equation}
where ${\cal P}_{\rm int}$ and ${\cal P}^\perp_{\rm int}$ describe quantum mechanical interference of  $J^P=0^+$ and $0^-$ terms.
One could apply the Neyman-Pearson lemma to each pair of points in the parameter space of $(f_{\CP}, \phi_{\CP})$,
but this would require continuous, and therefore infinite, set of probability ratios. However, an equivalent information is 
contained in a linear combination of only three probability ratios, which can be treated as three independent observables. 
For $H$ boson physics at proton or lepton colliders, such discriminants are introduced in Ref.~\cite{Anderson:2013afp} as
\begin{eqnarray}
\label{eq:kd0m-mela}
&& {\cal D}_{0-} = \frac{{\cal P}_{\rm 0^+}(\vec\Omega) }{{\cal P}_{\rm 0^+}(\vec\Omega)  +{\cal P}_{0^-} (\vec\Omega) } \,, \\
\label{eq:kdcp-mela}
&& {\cal D}_{\CP} = \frac{{\cal P}_{\rm int}(\vec\Omega) }{{\cal P}_{\rm 0^+}(\vec\Omega)  +{\cal P}_{0^-} (\vec\Omega) } \,, \\
\label{eq:kdcpperp-mela}
&& {\cal D}^\perp_{\CP} = \frac{{\cal P}^\perp_{\rm int}(\vec\Omega) }{{\cal P}_{\rm 0^+}(\vec\Omega)  +{\cal P}_{0^-} (\vec\Omega) } \,,
\end{eqnarray}
which become the optimal discriminants for the process with four contributions in Eq.~(\ref{eq:fractions-P}).

In Eq.~(\ref{eq:fractions-P}), $\vec{x}_{i}$ is a set of observables describing the process,
which may be $\vec\Omega$ when calculating the discriminants, or may be discriminants
themselves when performing the analysis later. The number of discriminants can also
be reduced by dropping ${\cal D}^\perp_{\CP}$ assuming $\sin\phi_{\CP}=0$, 
which is the case for real $\kappa_f$ and $\tilde\kappa_f$ in Eq.~(\ref{eq:lagrang-spin0-qq}).
On the other hand, with additional contributing amplitudes the number of observables grows. 
For example in the presence of background the ${\cal D}_{\rm bkg}$ discriminant is introduced
which can also be supplemented by the interference discriminant if there is quantum mechanical
interference between the signal and background processes.
The corresponding two discriminants are defined as 
\begin{eqnarray}
\label{eq:kdbkg-mela}
&& {\cal D}_{\rm bkg} = \frac{{\cal P}_{\rm 0^+}(\vec\Omega) }{{\cal P}_{\rm 0^+}(\vec\Omega)  +{\cal P}_{\rm bkg} (\vec\Omega) } \,, \\
\label{eq:intbkg-mela}
&& {\cal D}^{\rm int}_{\rm bkg} = \frac{{\cal P}_{\rm int}^{\rm bkg} (\vec\Omega) }{{\cal P}_{\rm 0^+}(\vec\Omega)  +{\cal P}_{\rm bkg} (\vec\Omega) } \,. %
\end{eqnarray}
Calculating a discriminant analogous to ${\cal D}_{\rm bkg}$ for the pseudoscalar signal hypothesis is not
necessary as a combination of Eqs.~(\ref{eq:kd0m-mela}) and~(\ref{eq:kdbkg-mela}) carries the needed information. 
The number of discriminants grows with the number of free components in the model; for example the 
background may interfere with different signal components and those may require different observables. 
However, typically there is a limited set of interference discriminants which become of practical interest, 
as we illustrate below. 

The probabilities $\cal{P}$ in Eqs.~(\ref{eq:kd0m-mela}--\ref{eq:intbkg-mela})
are the physical cross sections given by the product of parton distribution functions convoluted 
with the partonic cross sections that are proportional to the squared matrix elements.
The latter depend on the full event kinematics as measured in the experiment or simulated by a Monte Carlo generator.
They are computed at LO and do not include detector effects. However, as we illustrate in the following studies, 
they remain nearly optimal even after higher order or detector effects are introduced.
The probabilities $\cal{P}$ in Eq.~(\ref{eq:fractions-P}) may be treated as templates of the limited number of 
optimal discriminants when the analysis is performed. These templates are obtained from numerical simulation 
of the processes accounting for parton showering and detector effects. In the following analysis we limit the 
maximum number of discriminants to three, which we find to be both practical and close to optimal. 

The complete set of optimal discriminants in Eqs.~(\ref{eq:kd0m-mela}--\ref{eq:intbkg-mela})
was introduced earlier in experimental analysis of $HVV$ processes with LHC data by the
CMS~\cite{Chatrchyan:2012xdj,Chatrchyan:2012jja,Chatrchyan:2013mxa,Khachatryan:2014iha,Khachatryan:2015cwa,Khachatryan:2014kca,Khachatryan:2015mma} 
and ATLAS~\cite{Aad:2013xqa,Aad:2015mxa,Aad:2016nal} experiments,
and phenomenological studies supporting this development~\cite{Gao:2010qx,Bolognesi:2012mm,Anderson:2013afp}.
For example, it was shown that the complete set $x_i=\{ {\cal D}_{0-}, {\cal D}_{\CP}, {\cal D}_{\rm bkg} \}$
was optimal for the measurement of $f_{a3}$, a parameter equivalent to $f_{\CP}$, for the real $HVV$ couplings. 
A subset of equivalent observables was also introduced independently in earlier work
on different topics~\cite{Atwood:1991ka,Davier:1992nw,Diehl:1993br}. 
Here we apply this formalism to the measurement of the $H$ boson anomalous couplings to the 
heavy flavor fermions for the first time. 


\begin{figure}[t]
{\includegraphics[width=0.32\linewidth]{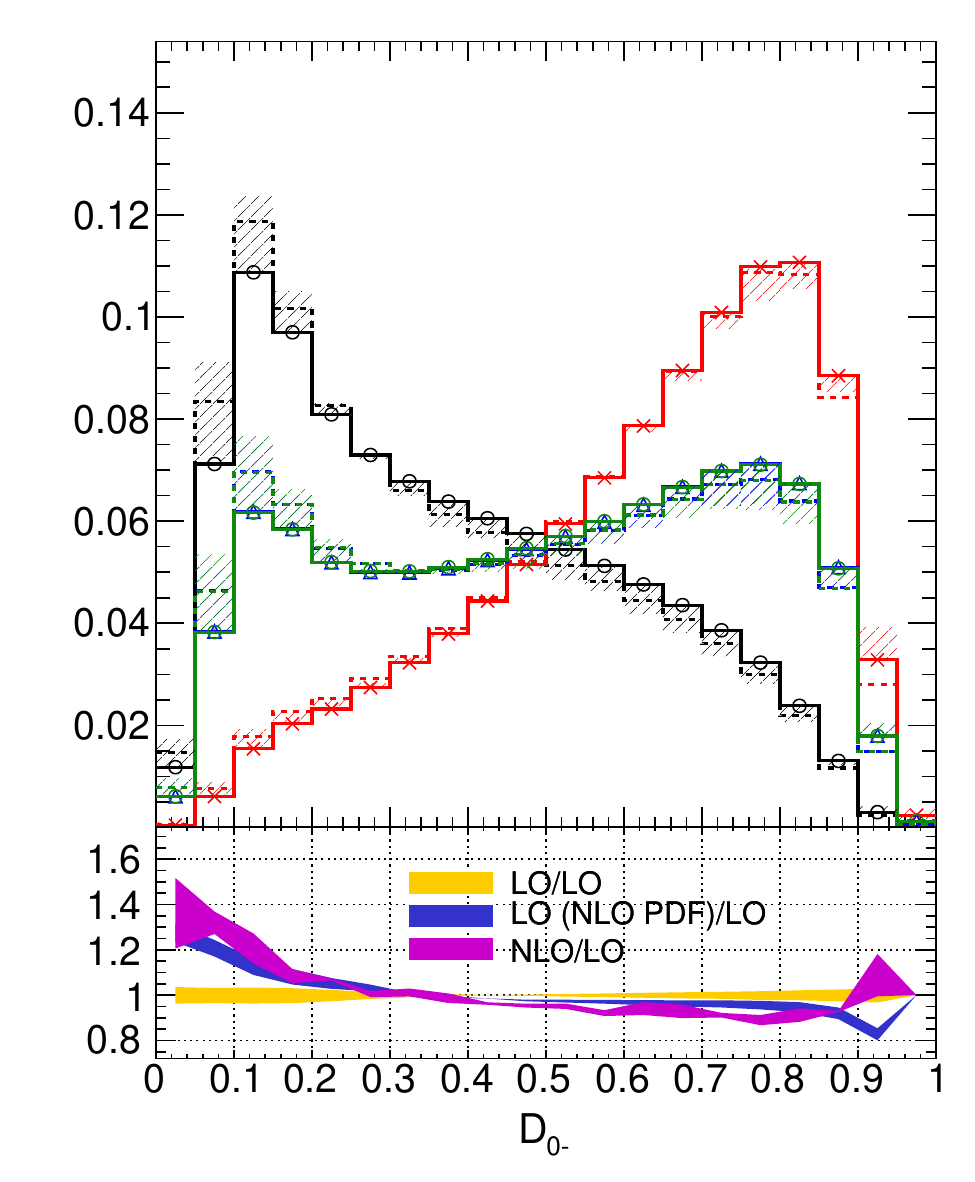}}
{\includegraphics[width=0.32\linewidth]{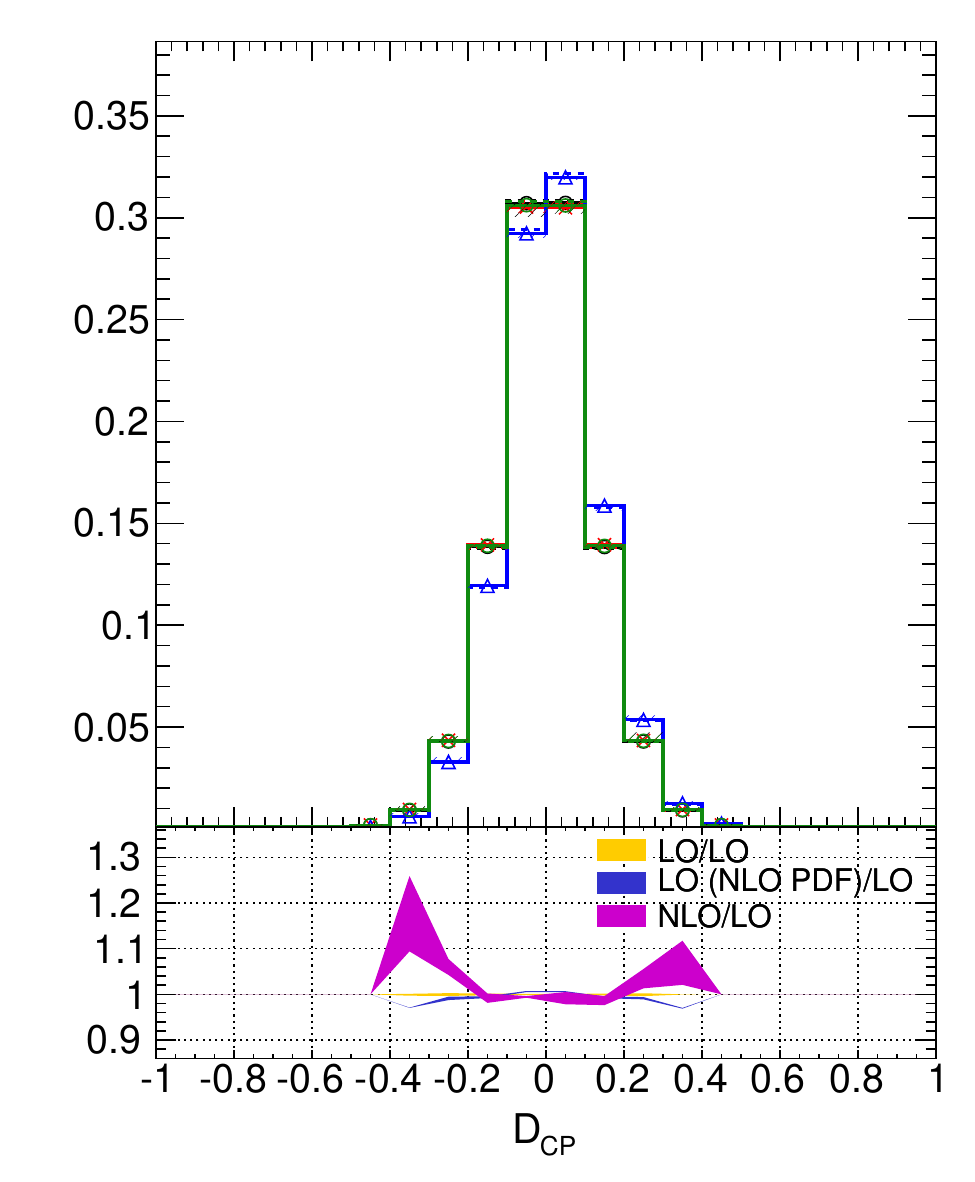}}
{\includegraphics[width=0.32\linewidth]{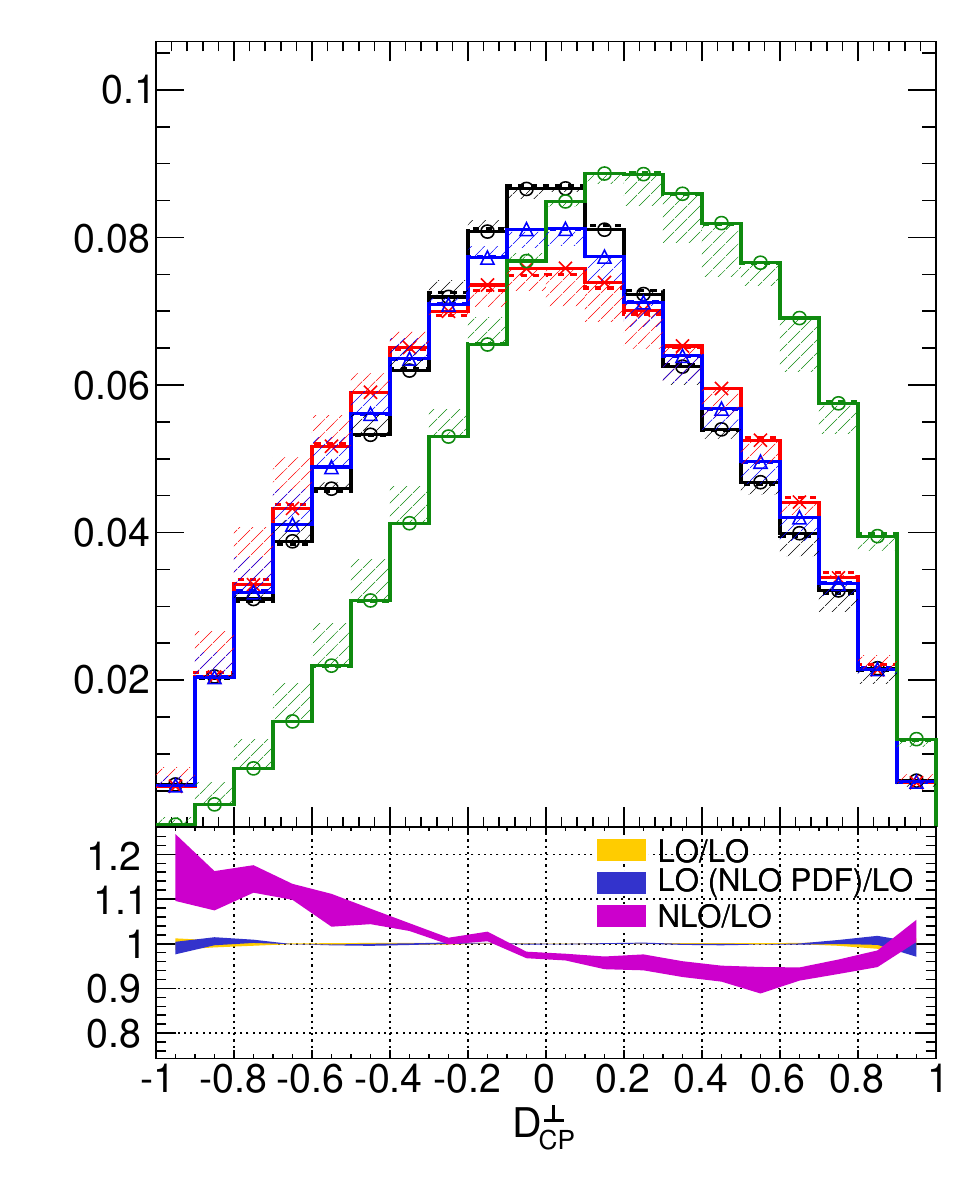}}
\caption{
Top: the $D_{0-}$ (left), ${\cal D}_{\CP}$ (middle), and ${\cal D}_{\CP}^\perp$ (right)
discriminant distributions for the $\ttH$ process for the $H$ boson models 
$J^P=0^{+}$ (red crosses), $0^{-}$ (black circles), 
$f_{\CP}=0.28$ with $\phi_{\CP}=0$ (blue triangles) 
and $\phi_{\CP}=\pi/2$ (green diamonds). 
Solid histogram shows distributions generated at LO in QCD. The hatched region covers
the range between LO and NLO distributions. The dashed histograms shows distributions 
generated with LO matrix element and NLO PDFs. 
Bottom: ratios of distributions for $f_{\CP}\cos(\phi_{\CP})=0.28$ with the ranges corresponding to the QCD scale variations,
where the denominator is the distribution generated at LO without considering scale variations. 
The LO/LO ratio is centered around 1 and its width is the effect of scale variation. 
The LO(NLO PDF)/LO ratio includes distribution generated with LO matrix element and NLO PDFs. 
The NLO/LO ratio includes distribution generated at NLO.
}
\label{fig:D0m}
\end{figure}

\subsection{Application to the $\ttH$ process}
\label{sect:tth_mela_ttH}

The large number of observables $\vec\Omega$ defined in Section~\ref{sect:tth_mela_angles}
for the $\ttH$ process can be compressed in a compact form with only three discriminants 
$x_i=\{ {\cal D}_{0-}, {\cal D}_{\CP}, {\cal D}_{\CP}^\perp \}$ as defined in Eqs.~(\ref{eq:kd0m-mela}--\ref{eq:kdcpperp-mela}), 
which is sensitive to the measurement of anomalous $\ttH$ couplings.
The distributions for the discriminants are shown in Fig.~\ref{fig:D0m} for $J^P=0^{+}$, 
$0^{-}$, and mixed states. The nominal studies presented here are based on LO in QCD calculations. 
Variations due to NLO matrix elements, PDFs, and QCD scale uncertainties are also 
shown in Fig.~\ref{fig:D0m} and are discussed in more detail in Section~\ref{sect:tth_nlo}. 

As one can see from both the discriminant definitions and distributions in Fig.~\ref{fig:D0m}, 
the ${\cal D}_{0-}$ is sensitive to the relative size of $\CP$ contributions, 
while ${\cal D}_{\CP}$ and ${\cal D}_{\CP}^{\perp}$ are sensitive to $\CP$ mixture
leading to forward-backward asymmetry in the presence of both $\CP$ amplitudes
for the real and complex ratio of couplings $\kappa_f / \tilde\kappa_f$, respectively. 
It is interesting to observe that the asymmetry is not strongly pronounced in the case of 
real couplings even when using the full top decay chain information. 
The asymmetry is more pronounced in the case of 
complex couplings, as seen in the ${\cal D}_{\CP}^{\perp}$ distribution, which can be
traced to the $\Phi_{f1}$ distribution in Fig.~\ref{fig:angular}. 
The asymmetry in both ${\cal D}_{\CP}$ and ${\cal D}_{\CP}^{\perp}$ disappears when top 
decay information is not used in the matrix element, which reflects the fact that spin correlations
in the $t\bar{t}$ system decay are essential for observing effects sensitive to $\CP$ mixture. 

At the moment of discovery of the $\ttH$ process, precision will be limited by statistics and the 
${\cal D}_{0-}$ discriminant will provide the most information about the $\CP$ components 
in the process. As smaller anomalous contributions get tested, the importance of the interference
discriminant will grow, but ultimately the full information is contained in the complete set
of discriminants. 


\begin{figure}[t]
{\includegraphics[width=0.32\linewidth]{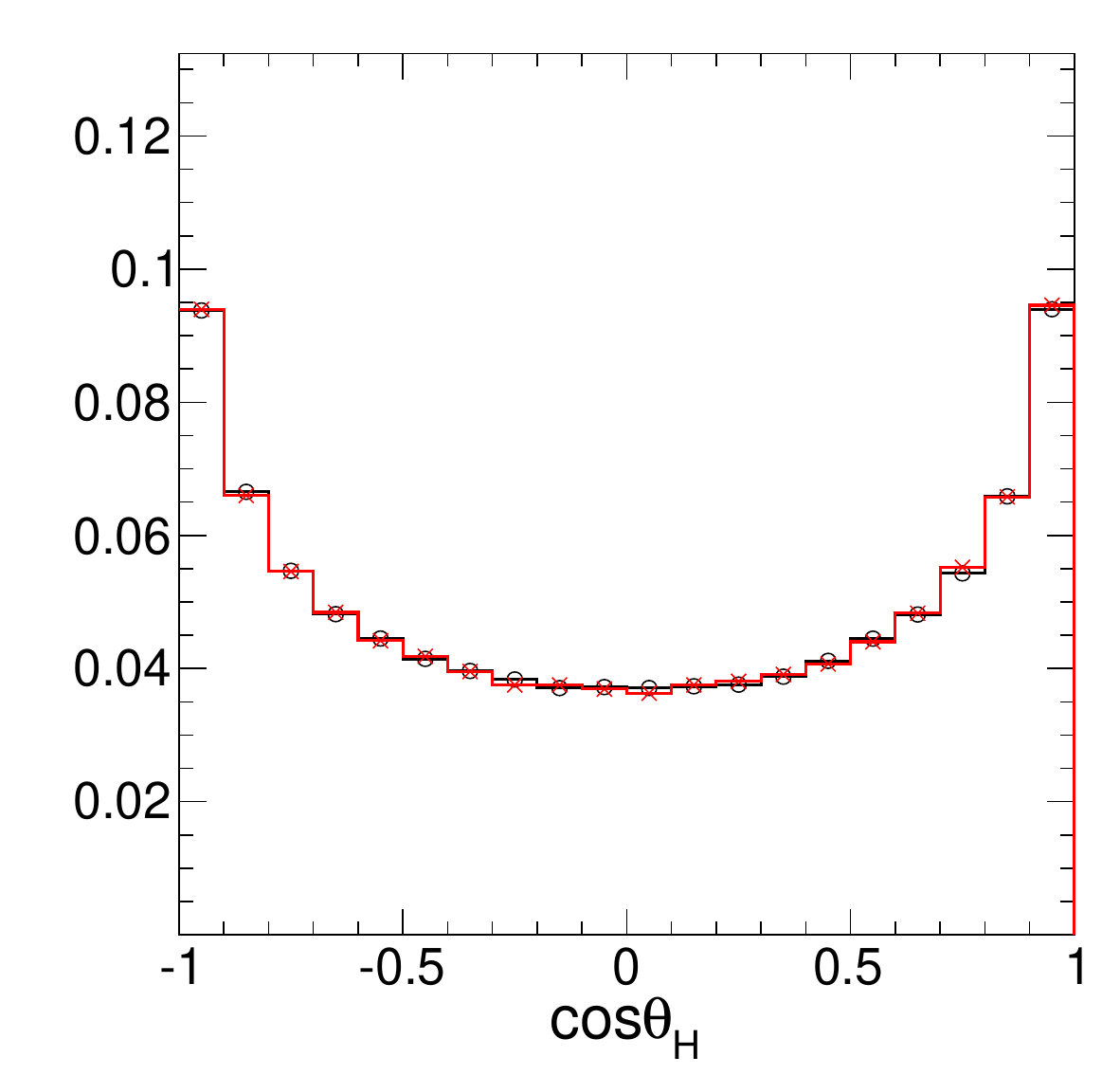}}
{\includegraphics[width=0.32\linewidth]{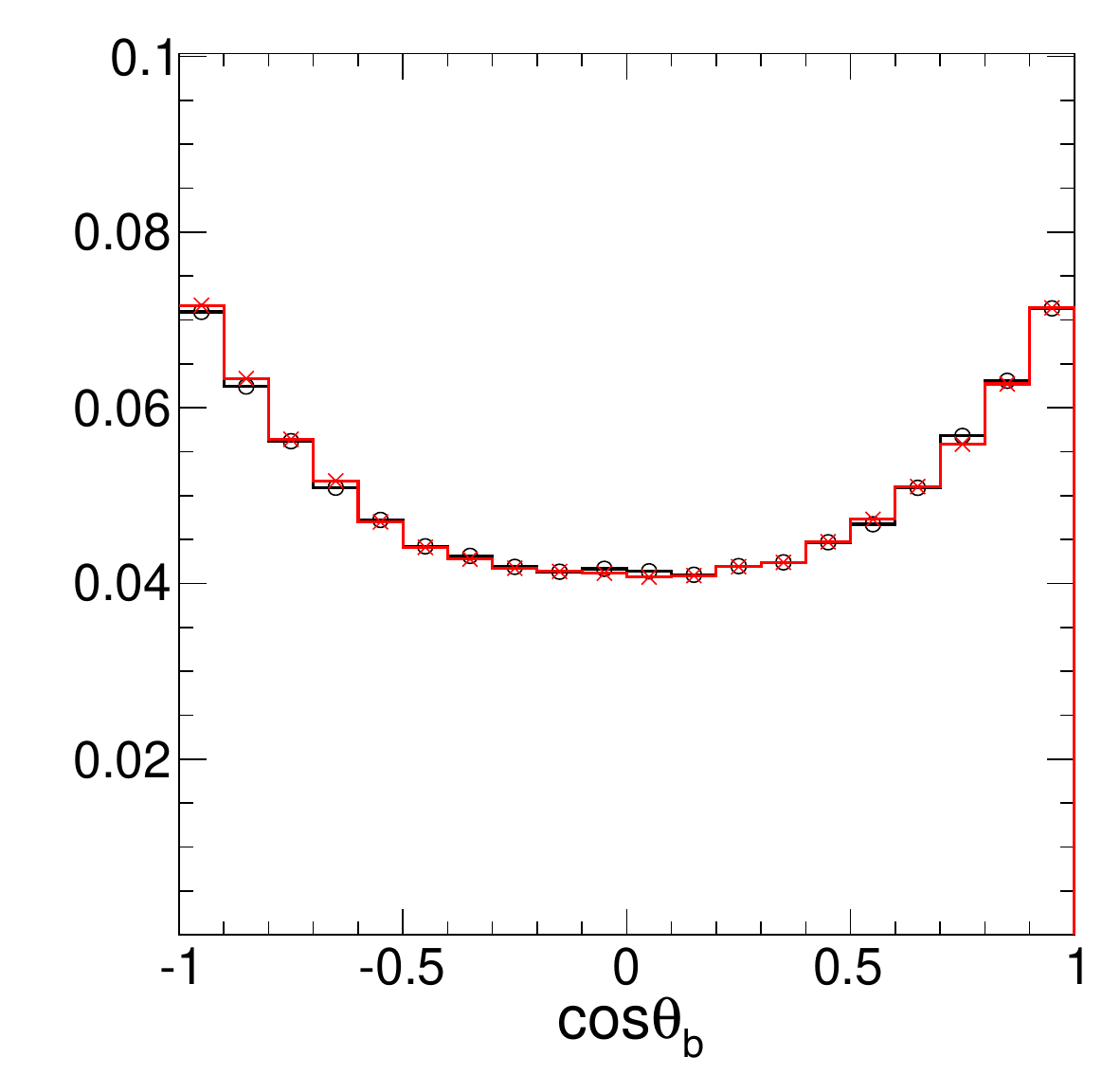}}
{\includegraphics[width=0.32\linewidth]{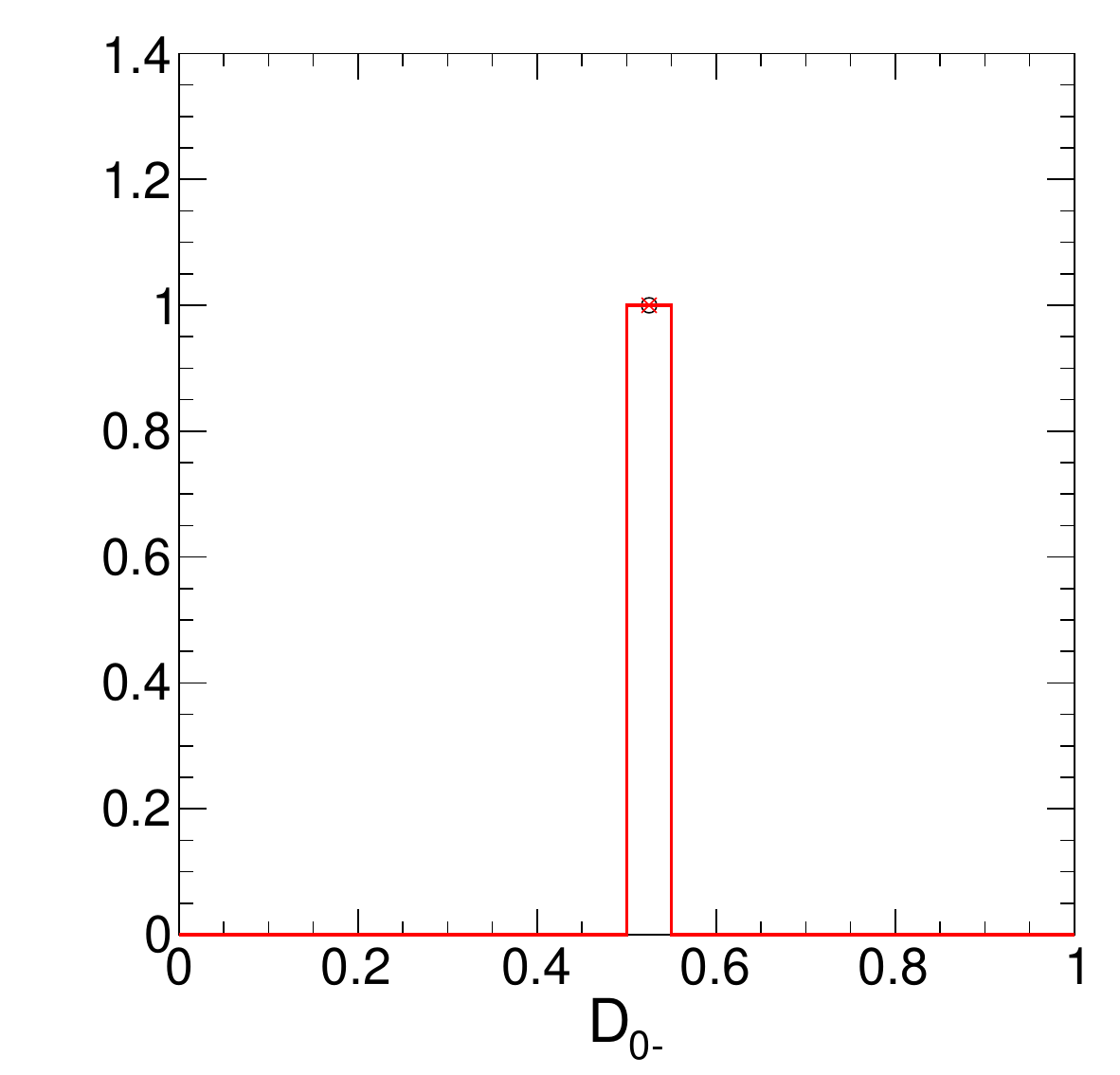}}
\caption{
Distributions for the $\bbH$ process: $\cos\theta_H$ (left), $\cos\theta_b$ (middle),
and $D_{0-}$ discriminant (right) for $J^P=0^{+}$ (black circles) and $0^{-}$ (red crosses).
}
\label{fig:D0m_bbH}
\end{figure}

\subsection{Application to the $\bbH$ process}
\label{sect:tth_mela_bbH}

The $\bbH$ and $\ttH$ processes are very similar with the main difference being  the heavy quark mass which, in fact, 
has a significant impact on the sensitivity of kinematic shapes to the $\Hff$ couplings.
This is because shape sensitivity arises from the mixing of left- and right-handed helicities at the matrix element level.
Therefore, this effect becomes proportional to the mass of the associated quark and becomes 
essentially invisible in the $\bbH$ process. 
In Fig.~\ref{fig:D0m_bbH}, we plot the angular distributions as well as the matrix element discriminant for the $\bbH$ process,
analogous to $\ttH$ process distributions shown in Figs.~\ref{fig:angular} and~\ref{fig:D0m}.
The different $\CP$ states have almost identical distributions, as follows from the helicity flip suppression
discussed above. Therefore, we conclude that it will be very challenging to probe the $\CP$ nature of $Hb\bar{b}$ coupling 
through shape analyses in the $\bbH$ production mode. 
 

\subsection{Application to the $\tqH$ process}
\label{sect:tth_mela_tqH}

\begin{figure}[htb]
	{\includegraphics[width=0.32\linewidth]{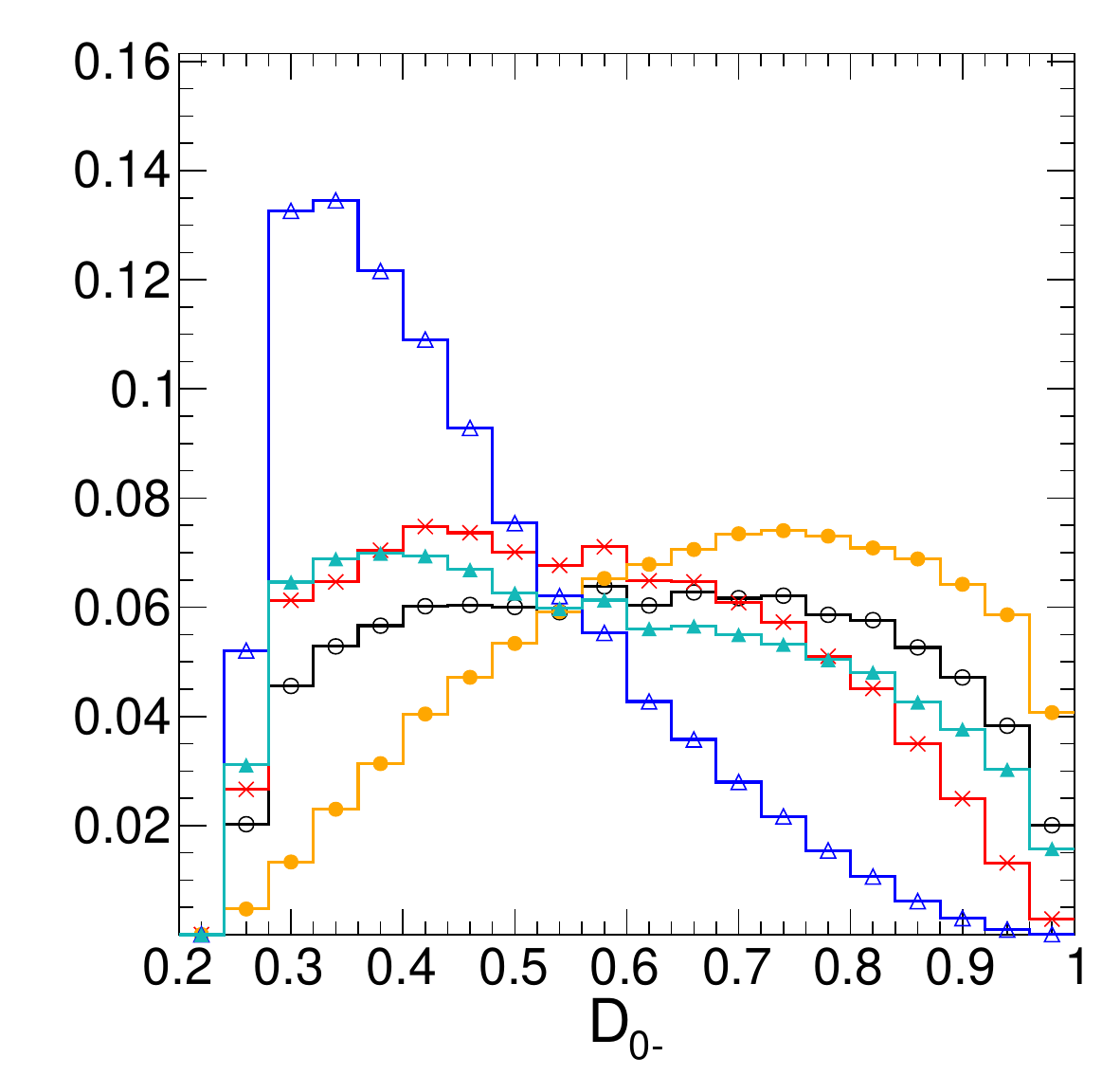}}
	{\includegraphics[width=0.32\linewidth]{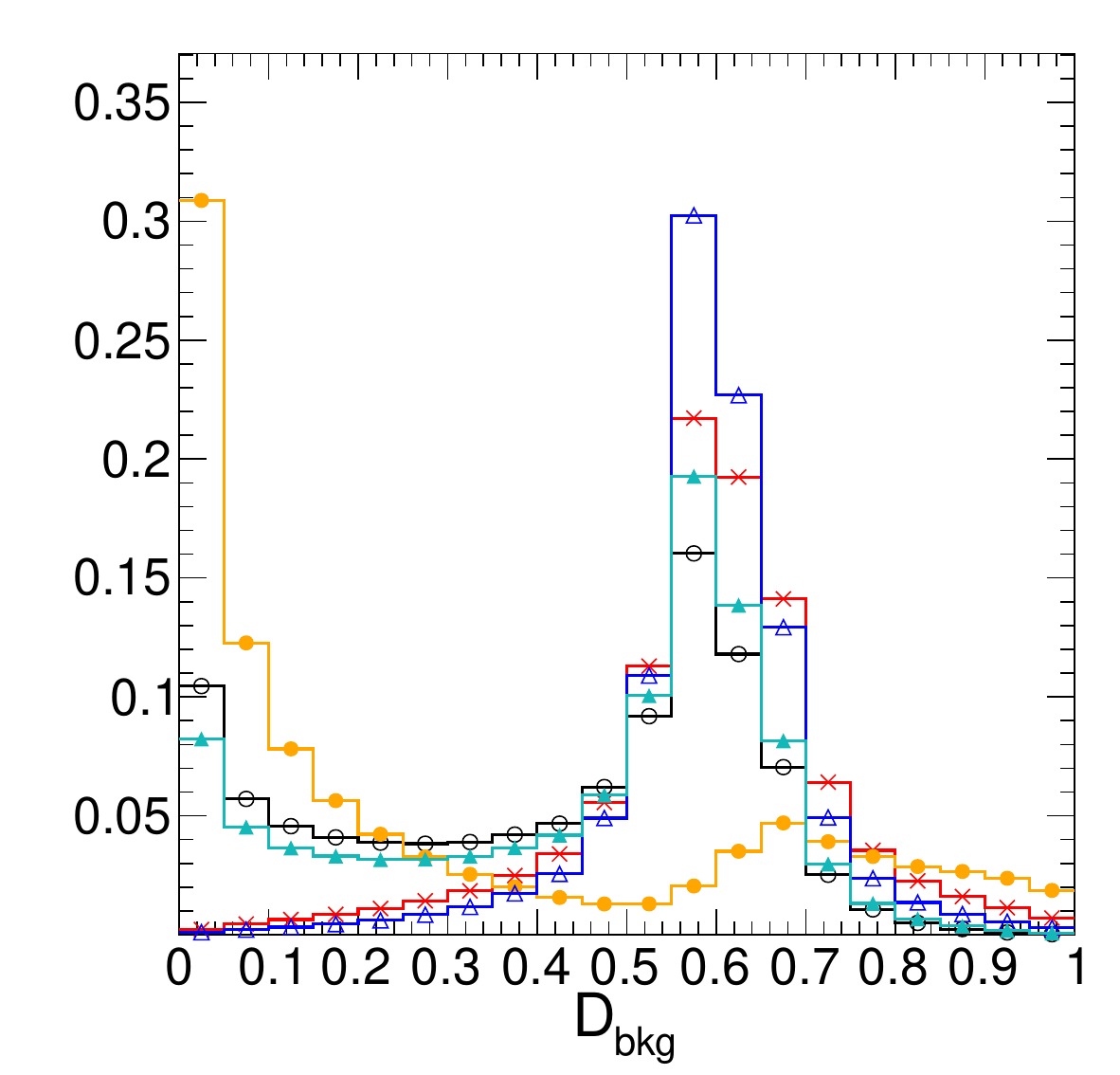}}
	{\includegraphics[width=0.32\linewidth]{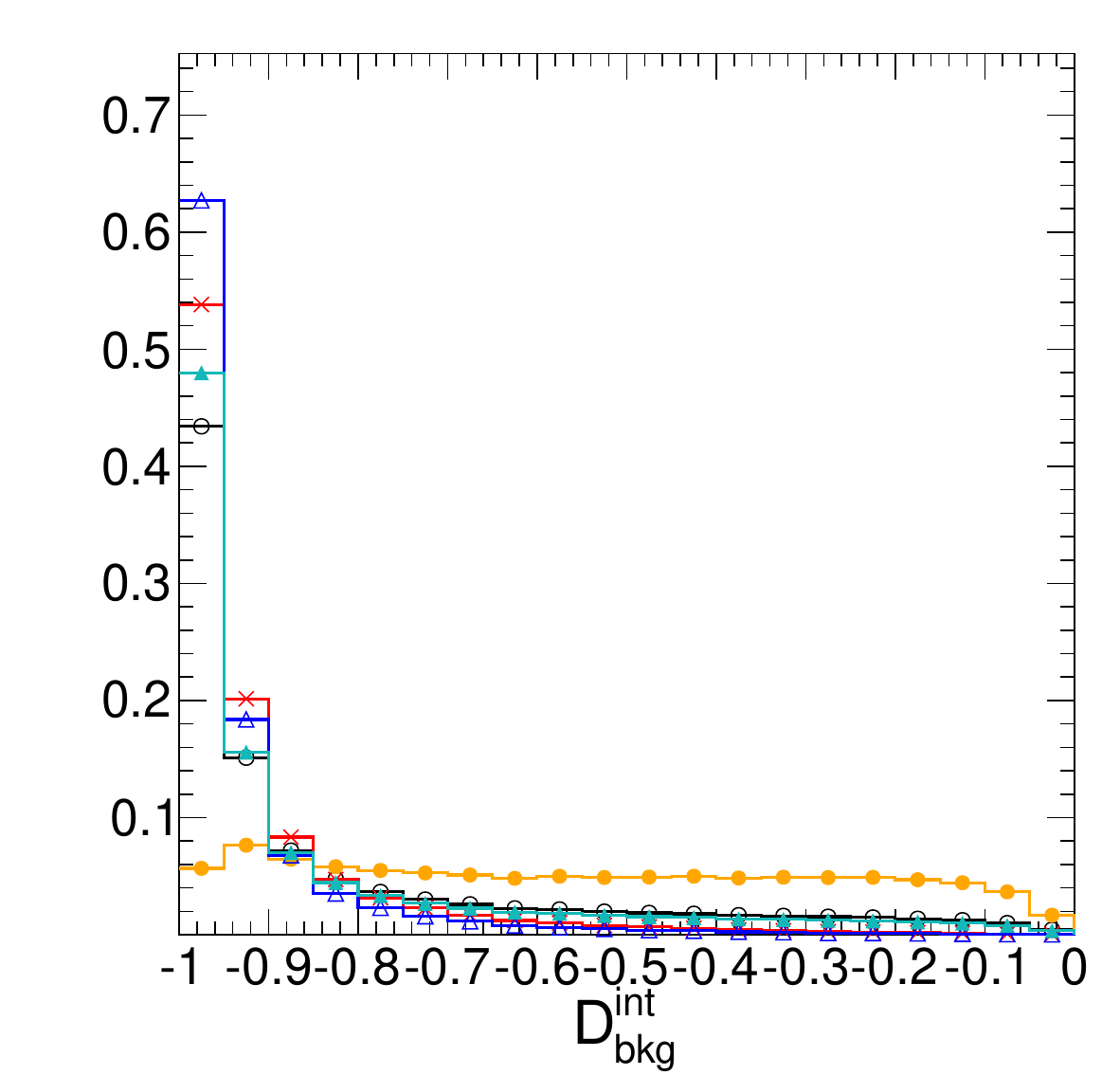}}
	\caption{
Distributions of ${\cal D}_{0-}$ (left), ${\cal D}_{\rm bkg}$ (middle), and  ${\cal D}^{\rm int}_{\rm bkg}$ (right) discriminants 
for the $\tqH$ production process, where five contributions are considered: 
$\Hff$ coupling as $J^P=0^{+}$ signal (red crosses) or as $0^{-}$ signal (blue triangles),
$HVV$-induced process as background (black circles), sum of $\Hff$ and $HVV$ processes, including interference, 
with $0^{+}$ $\Hff$-induced signal (orange closed circles) or $0^{-}$ $\Hff$-induced signal (cyan triangles).
	}
	\label{fig:D0m_tHq}
	\end{figure}

The $\tqH$ production process features both fermion and vector boson couplings of the $H$ boson, as shown 
in Fig.~\ref{fig:feynman-tqh}. 
Interference between the $\Hff$ and $HVV$-induced diagrams in Fig.~\ref{fig:feynman-tqh} is destructive
in the SM, but any deviation in either size or sign of either contribution could lead to a significant
change in observations. Therefore, in this paper we illustrate the approach where two parameters of interest are
determined: relative size of the $\Hff$ and $HVV$ contributions, including their relative phase, and the relative
size of the anomalous $\Hff$ coupling. 
In this context, contributions from the $HVV$ process could be considered as background and we consider 
only SM-like $HVV$ coupling with $a_1\ne0$ in Eq.~(\ref{eq:coupl-spin0-VV}),
while the signal process with the $\Hff$ coupling is allowed to have an arbitrary anomalous contribution. 
Therefore, the three discriminants ${\cal D}_{0-}$, ${\cal D}_{\rm bkg}$, and  ${\cal D}^{\rm int}_{\rm bkg}$ as defined in 
Eqs.~(\ref{eq:kd0m-mela}, \ref{eq:kdbkg-mela}, \ref{eq:intbkg-mela}) provide the most relevant information for
this analysis. Their distributions are shown in  Fig.~\ref{fig:D0m_tHq}. 

There is a clear difference between distributions for the alternative hypotheses, such as between 
$J^P=0^+$ and $0-$ $\Hff$-induced signal in ${\cal D}_{0-}$, or between $J^P=0^+$ signal and $HVV$-induced process 
in ${\cal D}_{\rm bkg}$. It is important to stress that destructive interference between the $J^P=0^+$ signal 
and $HVV$-induced processes with SM couplings leads to distributions which are very different from the direct sum 
of the two distributions, as shown in Fig.~\ref{fig:D0m_tHq}. In particular, the ${\cal D}^{\rm int}_{\rm bkg}$ discriminant 
shape is significantly distorted due to effect of interference, while the other two discriminants also exhibit sizable
differences as well. This feature leads to strong separation power between different hypotheses even with small
number of events in analysis, as we show below. 

\subsection{Application to the $H\to\tau^+\tau^-$ process}
\label{sect:tth_mela_HTT}

\begin{figure}[htb]
{\includegraphics[width=0.32\linewidth]{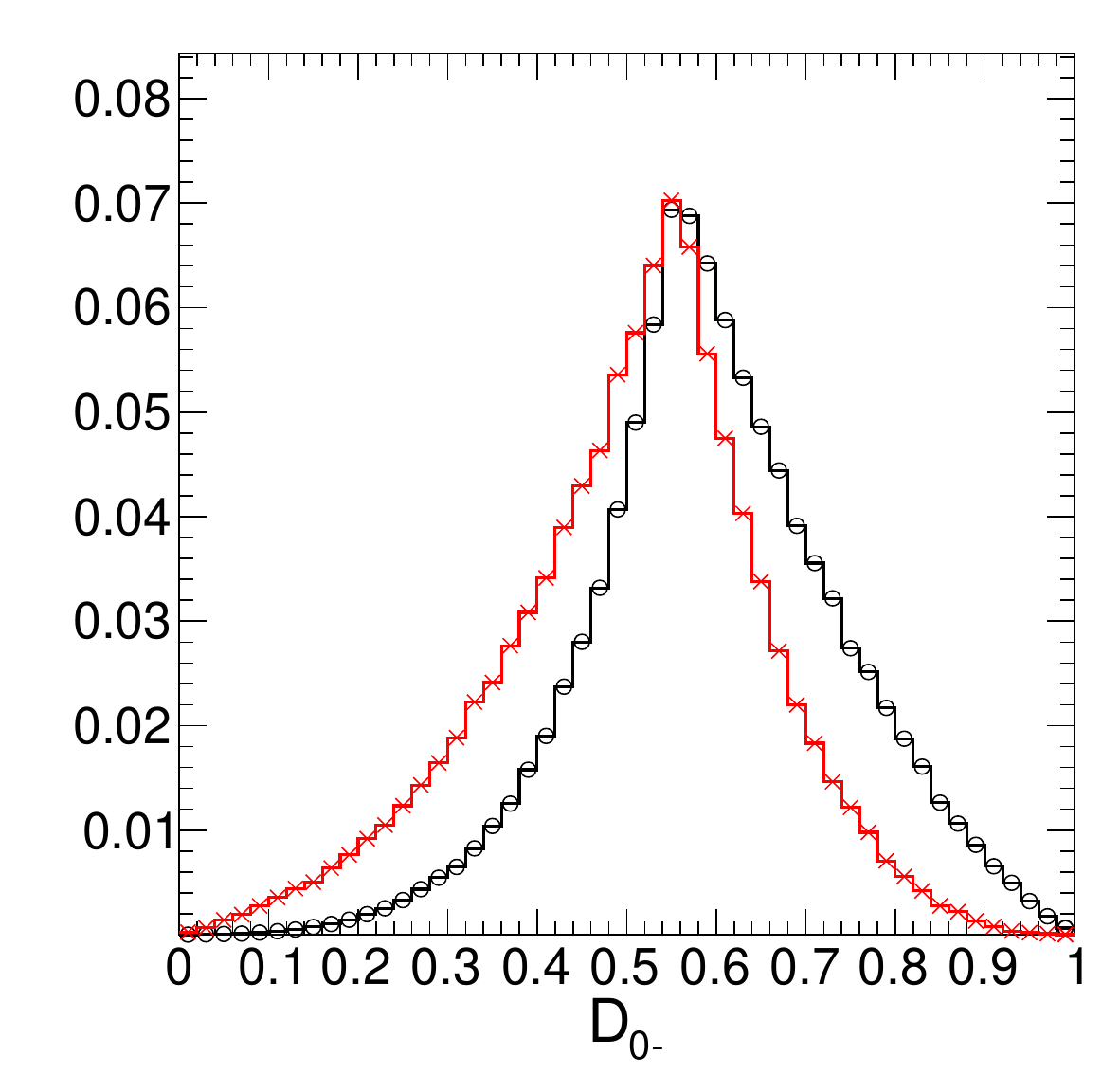}}
\caption{
Distributions of the $D_{0-}$ discriminant for $J^P=0^{+}$ (black circles) and $0^{-}$ (red crosses) models
in the ideal process $H\rightarrow \tau^+\tau^- \rightarrow (\ell\nu\nu) (\ell\nu\nu)$ simulated by JHUGen
and assuming all final state particles are reconstructed. 
}
\label{fig:D0m_htt_lep}
\end{figure}

In the $H\rightarrow \tau^+\tau^-$ process, it is possible to define the full sequential decay kinematics
and construct the matrix elements using information about all final state particles. 
This is illustrated for the process $H\rightarrow \tau^+\tau^- \rightarrow (\ell\nu\nu) (\ell\nu\nu)$ 
with the $D_{0-}$ discriminant in Fig.~\ref{fig:D0m_htt_lep}. Even though there is a strong separation
power between the $J^P=0^{+}$ and $0^{-}$ models in this case, there is little practical application because
reconstruction of the four neutrinos is not possible. Therefore, only limited information can be retained in
reconstructed observables and we use the matrix elements for the purpose of MC re-weighting techniques below.
 
\begin{figure}[htb]
{\includegraphics[width=0.32\linewidth]{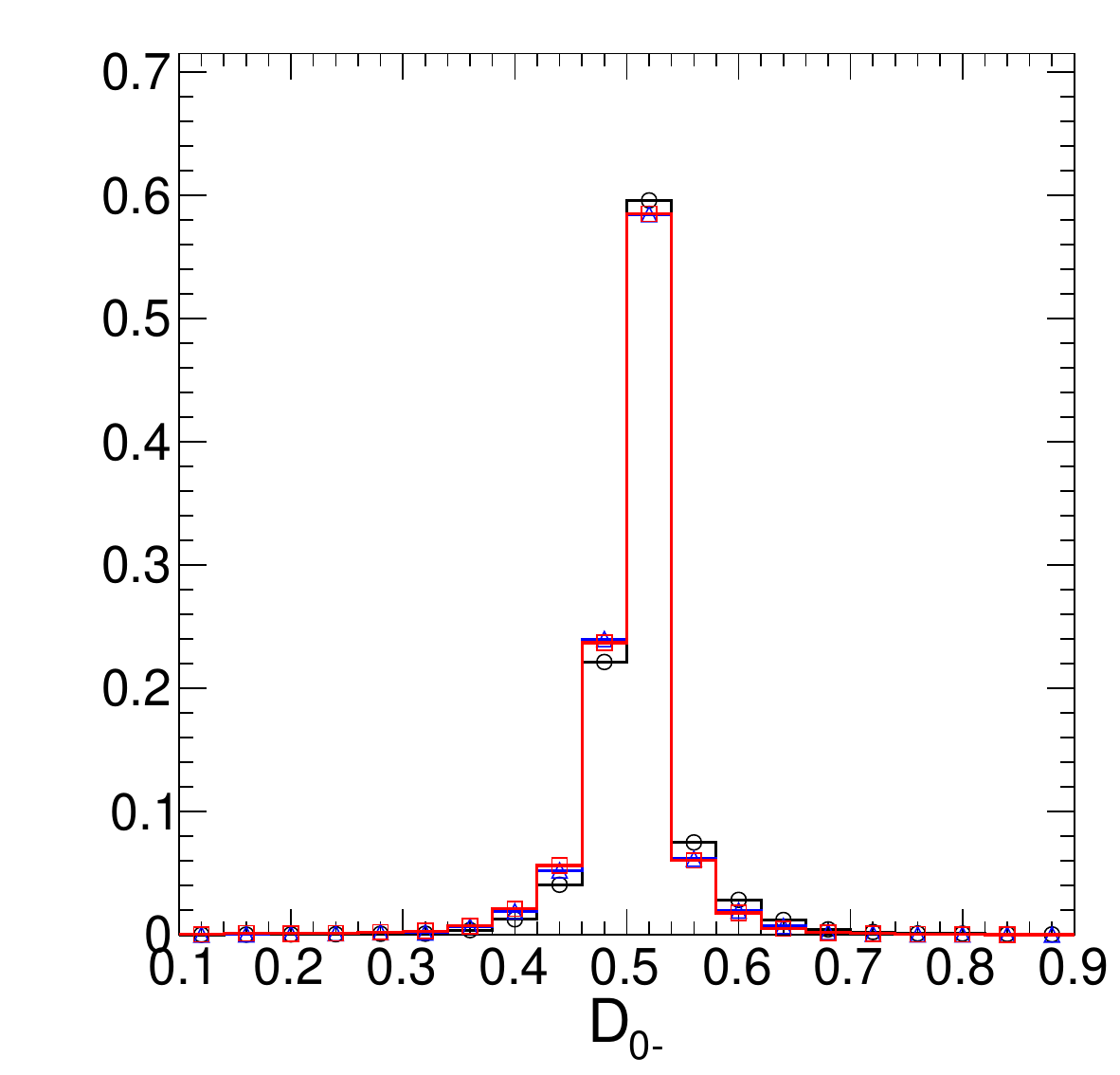}}
{\includegraphics[width=0.32\linewidth]{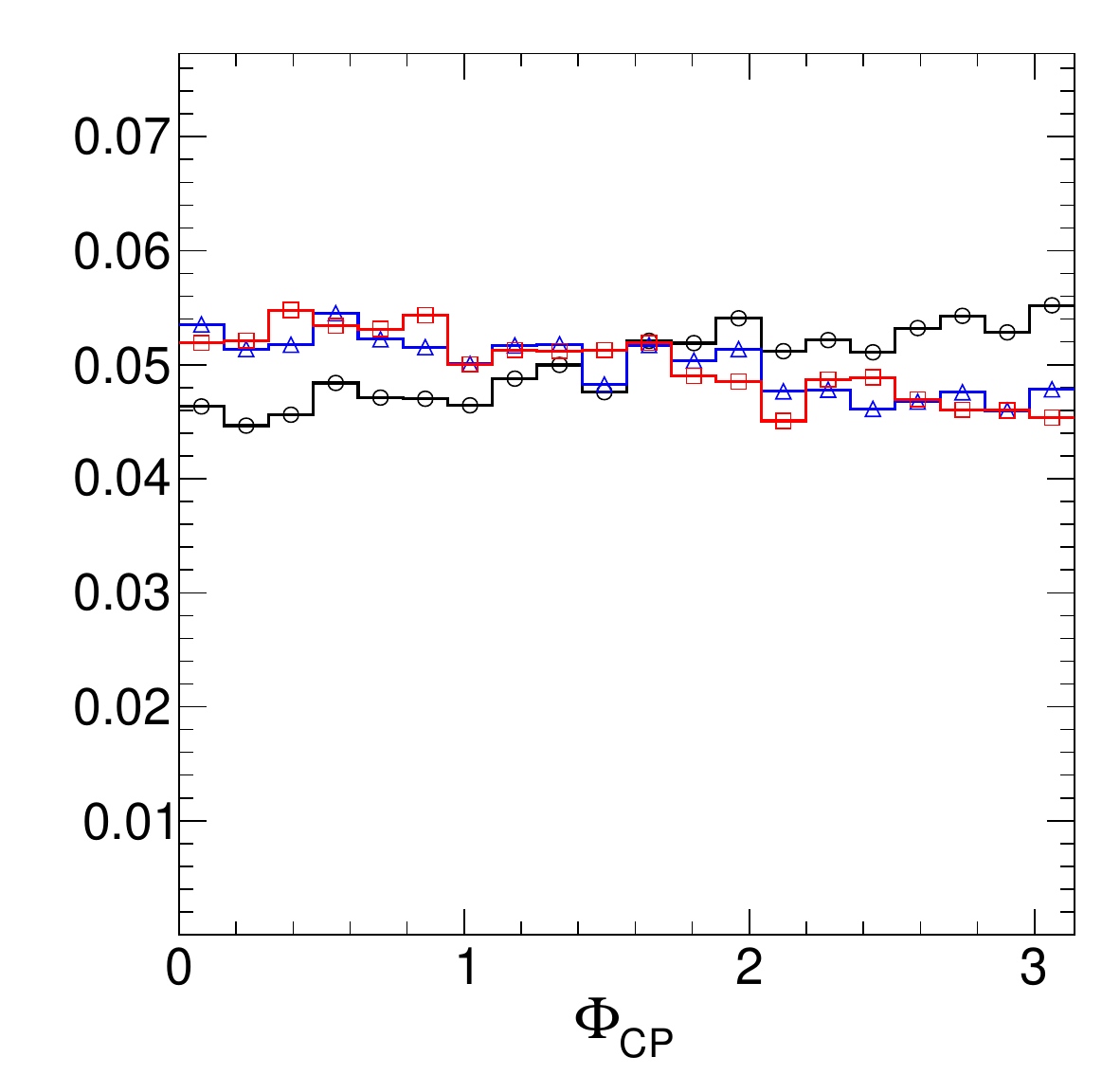}}
\caption{
The distributions of $D_{0-}$ (left) and $\Phi_{CP}$ (right) for $H\rightarrow \tau\tau \rightarrow (X\nu) (X^\prime\nu)$ simulation, 
where $X$ is a hadronic state. The $J^P=0^+$ (black circles) and $0^-$ (red squares) distributions are generated
with TAUOLA, including hadronic form-factors, and the $0^-$ (blue triangles) distributions are obtained from the 
$0^+$ simulation using MELA re-weighting. 
}
\label{fig:D0m_htt}
\end{figure}

In the case of hadronic $\tau$ decay, we provide the matrix element for the $H\rightarrow \tau^+\tau^- \rightarrow (X\nu) (X'\nu)$
process, where $X$ could be any hadronic particle decayed from $\tau$, e.g. $\pi$, $\rho$, $a_1$. Figure~\ref{fig:D0m_htt} shows 
the $D_{0-}$ discriminant constructed using this matrix element, in a hadronic final state. The events are decayed through TAUOLA, 
including hadronic form-factors for particle hadronization, for the $J^P=0^+$ and $0^-$ models. 
In addition, the $J^P=0^+$ events are re-weighted to the $0^-$ model using the MELA weights, which allow us to 
create any model with arbitrary anomalous couplings. The $D_{0-}$ discriminant can be compared to other observables
proposed for analysis of the $H\rightarrow \tau^+\tau^-$ decay, for example $\Phi_{\CP}$~\cite{Philip:2014el}, defined as
\begin{equation}
  \Phi_{\CP} = \mathrm{acos}(\vec{n}_X \cdot \vec{n}_{X'}) \,;~{\rm where}~~
  \vec{n}_{X} = \frac{\vec{q}_{X} \times \vec{q}_{\tau-}}{\left|\vec{q}_{X} \times \vec{q}_{\tau-}\right|}\,,
\end{equation}
$\vec{q}_{\tau^-}$ and $\vec{q}_{X,X'}$ are the 3-momentum of $\tau^-$ and $X$ or $X^\prime$ in the $H$ boson rest frame. 
The two observables, $D_{0-}$ and $\Phi_{CP}$, carry similar information for analysis of anomalous couplings, but the
$D_{0-}$ discriminant is somewhat more powerful.

\section{NLO QCD study of kinematic discriminants}
\label{sect:tth_nlo}

Let us now discuss the effects of higher order QCD corrections on the modeling and performance
of anomalous coupling discrimination. As described in Section~\ref{sect:tth_mc}, 
we compute the NLO QCD predictions for $pp \to \ttH$ production followed by the spin-correlated 
top quark decays at NLO QCD in the narrow-width approximation.
Neglecting QCD corrections in the description of the $pp \to t\bar{t}+H$ process constitutes the 
dominant theoretical uncertainty on its cross section. We find that residual scale uncertainty on the 
total cross section is reduced from 21\,\% to 9\,\% when going from LO to NLO in QCD.
The corrections on shapes of basic kinematic distributions are up to $\pm 10\,\%$.

In the earlier work~\cite{Anderson:2013afp}, we studied the impact of NLO in QCD effects in production on the
anomalous coupling discrimination in decay $H \to VV$. However, the production and decay processes 
carry no spin correlation and additional radiation from the production stage is largely decoupled from 
the color-neutral $H \to VV$ decay dynamics. Hence, it is straightforward to use LO matrix elements to 
characterize $HVV$ couplings, even in the presence of initial-state radiation.
This is in contrast to the $pp \to t\bar{t} \, (\to b\bar{b}WW ) + H(\to VV/f\bar{f})$ process where initial and final state 
particles radiate color charges and the top quarks exhibit spin correlations, all of which affect the
studied dynamics. 

A fully consistent extension of the matrix element method beyond the LO requires both event generation and 
matrix element discriminants at higher orders. The main complication stems from kinematic configurations where 
hadronic activity is clustered by a jet algorithm. Commonly-used jet algorithms combine soft and collinear radiation 
in subsequent $2 \to 1$ recombination steps. Hence, the resulting jet either acquires some invariant mass which 
does not correspond to the fundamental parton mass, or the jet violates global momentum conservation.
This feature prohibits the use of jet momenta in a LO matrix element, which has on-shellness (of quarks and gluons) 
and momentum conservation built-in from first principles.  A systematic solution to this issue at NLO QCD is part of 
active research and first elegant solutions have been presented 
in Refs.~\cite{Alwall:2010cq,Campbell:2012cz,Campbell:2013hz,Martini:2015fsa}, 
where modified jet algorithms are proposed to map resolved and unresolved parton configurations onto 
their proper matrix elements. These approaches have promising prospects for future measurements, 
but they require the use of new jet algorithms that are currently not used in experimental analyses. 
Moreover, only solutions for either colorless final states or colorless initial states have been presented 
in the literature. A fully developed application to e.g. top quark pair production at NLO QCD is not yet available.
A variation away from the exact NLO treatment has also been presented in Ref.~\cite{Soper:2014rya}, where additional radiation is 
included through a parton shower approximation. This approach allows to include multiple emissions 
and has been applied to Higgs boson physics in Refs.~\cite{Soper:2011cr,Englert:2015dlp}.

In this paper, we take a pragmatic and more simplistic approach. We retain leading order matrix elements in the discriminants 
of Eqs.~(\ref{eq:kd0m-mela}--\ref{eq:intbkg-mela}) and probe them 
with events from leading {\it and} next-to-leading order simulation, and also compare those to variations due to 
PDFs, QCD scales, and parton showering. The mismatch between the LO discriminants and NLO simulation does
not formally allow us to claim optimal discrimination power by virtue of the Neyman-Pearson lemma, where
constructed likelihoods should be interpreted as fundamental probabilities. However, we demonstrate that
NLO corrections to the shapes of kinematic distributions in the $pp \to \ttH$ process are small and 
sometimes indistinguishable when compared to other associated uncertainties. Therefore, the LO discriminants 
${\cal D}$ maintain their discriminating power beyond the well-defined leading-order, and we can continue 
to use them as robust and powerful tools for anomalous coupling studies.

In Fig.~\ref{fig:D0m}, we compare the impact of LO versus NLO events probing the LO discriminants ${\cal D}$.
The solid histograms show the distribution for LO events, whereas the hashed bands indicate the shift due to NLO corrections. 
We note that the general shapes of the various distributions are maintained and only minimally distorted. The separation
power between the extreme $J^P=0^+$ and $0^-$ hypotheses is largely unaffected by the presence of higher order corrections. 
The most powerful discriminating observable ${\cal D}_{0-}$ receives very small corrections in range within the bulk 
of the distributions, as shown in the lower pane of Fig.~\ref{fig:D0m}. Moreover, most of this correction appears already with 
the PDF variations before NLO corrections at the matrix element level. Hence, the bulk of corrections that we observe stems 
only from different input parameters and PDFs. 
The width of the bands in all lower panes of Fig.~\ref{fig:D0m} corresponds to scale variations by a factor of two around 
the central scale $\mu=m_t+ m_H/2$. Studies presented in Fig.~\ref{fig:D0m} do not include parton showering.
However, as we show in Section~\ref{sect:study_tth} and Fig.~\ref{fig:D0m_reco_ttH}, inclusion of parton showering in
LO simulation brings this simulation even closer to NLO modeling with parton showering. 

We therefore conclude that discrimination power of the MELA approach is guaranteed even when higher-order corrections 
are considered  in the $pp \to \ttH$ process and additional jets are present in the event sample. Soft and collinear radiation, 
which generates massive jet momenta, can be handled in the matrix element approach and does not spoil the discrimination 
power. These higher-order effects are within the uncertainties of the PDF, scale choice, and parton showering. 


\section{Application to $C\!P$ parity measurements in $\ttH$, $\tqH$, and $\bbH$}
\label{sect:tth_measure}

In this section, we estimate the potential for $C\!P$ measurements in the $\ttH$, $\tqH$, and $\bbH$
processes on LHC with 300\,fb$^{-1}$ of proton-proton collision data collected at 13\,TeV center-of-mass energy. 
This is the integrated luminosity expected by the end of Run III of LHC in about seven years. 
Projections to other luminosity scenarios are usually trivial extensions as long as uncertainties 
remain limited by statistics. While there is a strong evidence for the $\ttH$ production in Run I of LHC,
none of these processes have been firmly established yet. However, we rely on experimental studies 
of these processes in 
Refs.~\cite{Khachatryan:2014qaa,Khachatryan:2015ila,Khachatryan:2015ota,Aad:2015iha,Aad:2016zqi} 
for realistic event reconstruction projections. 

As the first observation, following Section~\ref{sect:tth_mela_bbH}, we conclude that 
it will not be possible to measure $C\!P$ in the $\bbH$ production process in the LHC program.  
For the $\ttH$ and $\tqH$ processes, we consider the $H \to \gamma \gamma$ decay mode 
to tag the $H$ boson, as a clean signature with sizable branching fraction. 
We also consider the $H \to ZZ\to 4\ell$ final state in the $\ttH$ process for comparison,
but its contribution is small due to the small branching fraction.
We use the hadronic decay of one top quark final state so that the full kinematics can be reconstructed.
In the $\ttH$ case, the other top quark is reconstructed in the leptonic channel. 
Inclusion of other final states of either $H$ boson or top would only enhance expected precision, 
but the  decays we consider are representative of the typical analyses of these processes. 

In this study, the $\ttH$, $\bbH$, and $\tqH$ processes with SM or anomalous couplings are generated with 
the JHU generator. The only non-negligible background that we need to consider is SM $t\bar{t} \gamma \gamma$ 
production as a background to the $\ttH$ study with the photon decay of the $H$ boson, which is simulated with MadGraph. 
The MC samples are interfaced to PYTHIA8 for parton shower and hadronization. 
In order to model detector effects, lepton and photon $p_T$ are smeared with 1\% and 4\% resolution. 
The jets are reconstructed in a cone of $R=0.5$ using anti-$k_T$ algorithm and their energy is smeared by 20\%. 

The event selection criteria follow those of the  LHC analyses~\cite{Khachatryan:2014qaa}.
We require the leptons, photons, and jets to have $p_T > 5$, 10, and 30 GeV, and $|\eta| < 2.4$, 2.4, and 4.7, respectively. 
Jets within $\Delta R<0.2$ of the leptons or photons are removed. 
In the $\ttH$ analysis, an event should have at least four jets and a $b$-tagged jet. 
The $b$-tagging efficiency (62\%) and fake rate for the light quark jets (6\%) follow experimental study~\cite{Khachatryan:2014qaa}.
To fully reconstruct the semileptonic decay of the $t\bar{t}$ system, we use the constraint fit from Ref.~\cite{constraintfit}. 
The four-momenta of four jets, MET, and one lepton are used in the kinematic fit with the masses of the top quarks
and the $W$ bosons as constraints. If more than four jets are reconstructed, the combination that gives the best 
$\chi^2$ is selected. The four-momenta of all decay products of the $t\bar{t}$ system are obtained from this fit
and are used in the further analysis.
In the $\tqH$ analysis, exactly four jets and a $b$-tagged jet are required in order to remove hadronic $\ttH$ events. 
The combination of three jets with the mass closest to the top is treated as the top decay product in this process. 
The required number of reconstructed leptons and photons depends on the studied final state.
If required, the leading photon should have $p_T>33$ GeV and $p_T/m_{\gamma\gamma}>0.5$.
In the $H \to 4\ell$ channel, two pairs of opposite sign and same flavor leptons should have invariant mass 
greater than 40 and 12 GeV. The invariant mass of the $H$ boson candidate is required to be between 100 and 140 GeV. 

In the case of the $\tqH$ process with $H \to \gamma \gamma$, the main other contribution is 
cross-feed from the $\ttH$ process with the same $H$ boson decay. 
The $\ttH$ process with the $4\ell$ decay of the $H$ boson has negligible  background,
while with the $\gamma\gamma$ decay the dominant background is the SM $t\bar{t}\gamma\gamma$ production.
The expected number of events of signal and background events at 300\,fb$^{-1}$ is shown in Table~\ref{table:nevents_ttH}.
We would like to note that these expected yields are quoted for the SM scenario where destructive interference 
between the $\Hff$ and $HVV$-induced $\tqH$ processes leads to a small number of expected events.
However, this interference may become constructive with the non-SM couplings. 
The cross section for $t\bar{t}\gamma\gamma$ processes suggests background yield to be smaller
than the signal. However, the LHC studies with data-driven methods suggest larger background~\cite{Khachatryan:2014qaa}.
Therefore, we conservatively set the $t\bar{t}\gamma\gamma$ background yield to be twice the signal
in the invariant mass window specified above. 

\begin{table}[h]
\begin{center}
\caption{
Number of events expected in the SM for signal and other contributions in the study of $\Hff$ couplings
discussed in text with 300\,fb$^{-1}$ at 13 TeV.
}
\begin{tabular}{|l|c|l|c|}
\hline
~signal process & signal yield & ~other process & other yield \\
\hline
~$\ttH$, $H \to \gamma \gamma$~ & 50.3 & ~$t\bar{t}\gamma\gamma$~ & 100.6  \\
~$\ttH$, $H \to 4\ell$~ & 4.3 & ~negligible~ & 0 \\
~$\tqH$, $H \to \gamma \gamma$~ & 3.2 & ~$\ttH$, $H \to \gamma \gamma$~ & 10.2 \\
\hline
\end{tabular}
\label{table:nevents_ttH}
\end{center}
\end{table}


\subsection{Study of the $\ttH$ process}\label{sect:study_tth}

\begin{figure}[ht]
\centerline{
\includegraphics[width=0.38\linewidth]{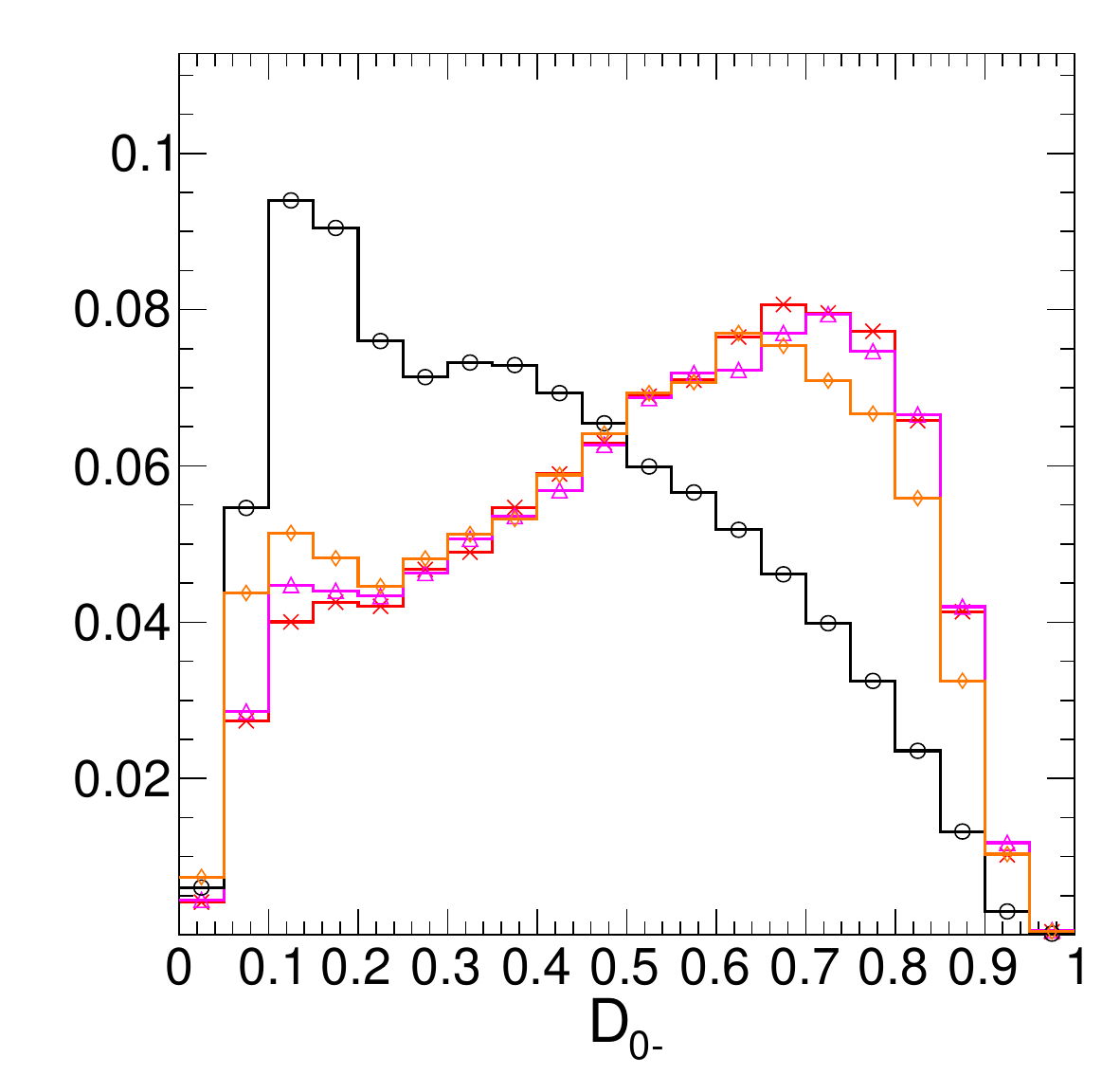}
\includegraphics[width=0.38\linewidth]{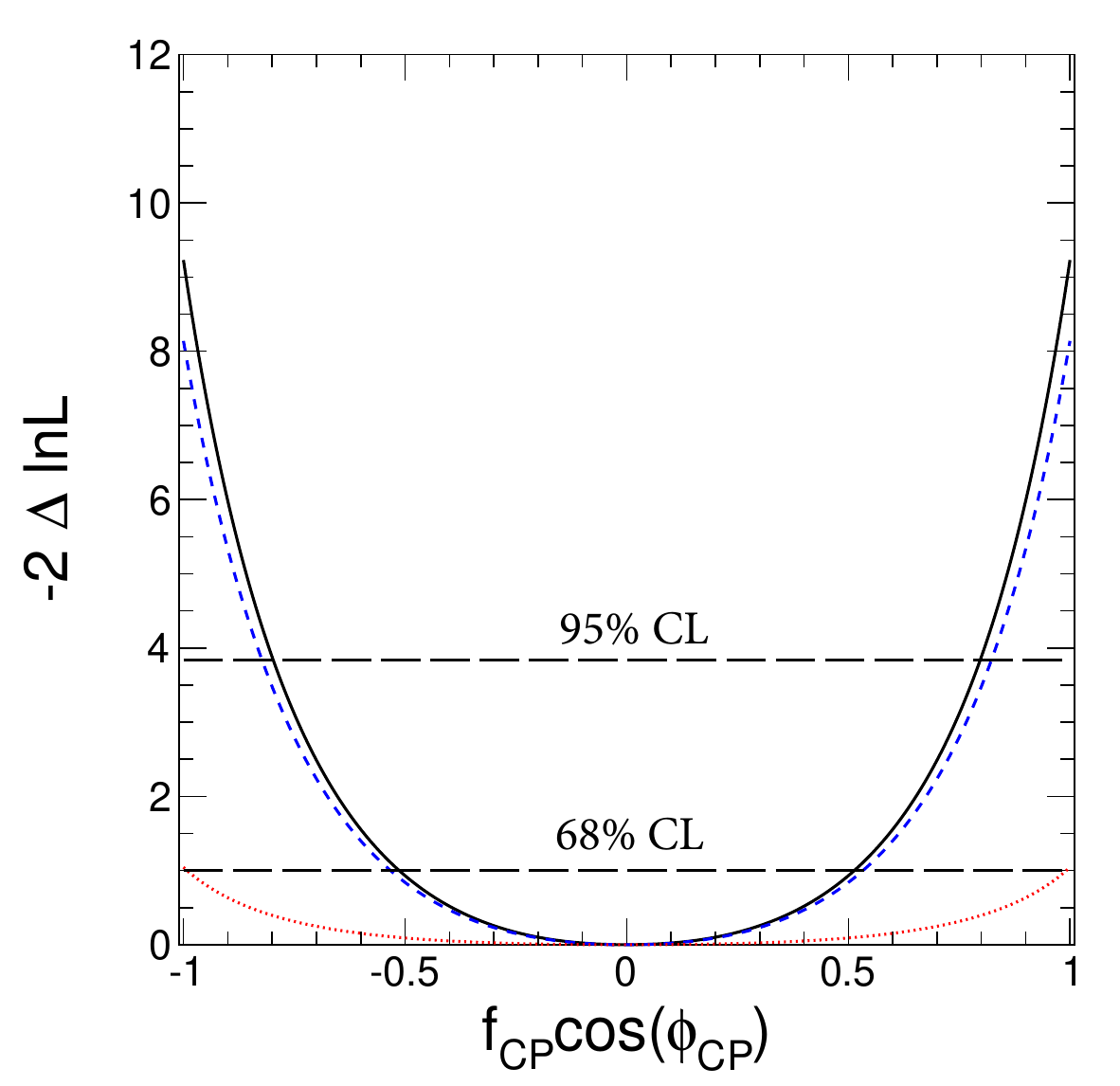}
}
\caption{
Left: the ${\cal D}_{0-}$ discriminant distribution for $\ttH$, $H \to \gamma \gamma$ process after reconstruction
discussed in text. The following distributions are shown: $\Hff$ coupling as $J^P=0^{+}$ signal 
(red crosses) and as $0^{-}$ signal (black circles), and $tt\gamma\gamma$ background (orange diamonds).
Also shown is the SM $J^P=0^{+}$ signal generated at NLO in QCD with POWHEG (magenta triangles). 
Right: the likelihood scan of $f_{\CP}\cos(\phi_{\CP})$, where $\phi_{\CP}=0$ or $\pi$, in the $\ttH$ process 
with both $H \to\gamma \gamma$ (blue dashed) and $H \to 4\ell$ (red dotted) decays, 
and the combined result (solid black) expected with 300\,fb$^{-1}$ at LHC.
}
\label{fig:D0m_reco_ttH}
\end{figure}

The analysis of the $\ttH$ process uses the ${\cal D}_{0-}$ discriminant, where decay of the top quarks is not considered in the 
matrix element. Consideration of the top quark decays is important in the calculation of the ${\cal D}_{\CP}$ or ${\cal D}_{\CP}^{\perp}$ 
discriminants, but only when the up and down flavors of the quarks in the decay chain is known. The latter is difficult to determine with
the jet reconstruction techniques and therefore the $\CP$ discriminants are not used in this analysis. 
In the $H \to \gamma \gamma$ channel, we use the invariant mass $m_{\gamma\gamma}$ to separate the signal and background. 
Figure~\ref{fig:D0m_reco_ttH} shows the ${\cal D}_{0-}$ distribution in the $H \to\gamma \gamma$ channel, 
where the $J^P=0^+, 0^-$ and background distributions are shown. 

In Fig.~\ref{fig:D0m_reco_ttH}, simulation of the $0^+$ process is also shown with the POWHEG generator at NLO in QCD. 
In all cases, parton showering is performed with PYTHIA8. Similar to the study presented in Section~\ref{sect:tth_nlo},
the NLO QCD effects are found to have a small effect on accuracy of ${\cal D}_{0-}$ simulation, especially after parton 
showering is included in simulation. Any residual effects are consistent with systematics also arising from PDF and
QCD scale variations. 

The expected precision of the $f_{\CP}$ measurement in the $\ttH$ process with both $H \to\gamma \gamma$
and $H \to 4\ell$ decays, and their combination, is shown in Fig.~\ref{fig:D0m_reco_ttH} for integrated luminosity of 300\,fb$^{-1}$.
The maximum likelihood fit is based on the probability density functions following Eq.~(\ref{eq:fractions-P}) parameterized
with template distributions filled with generated events as discussed above. 
About 3$\sigma$ exclusion of the pure pseudoscalar state is expected in such a scenario, which is comparable to the
current precision with the $HVV$ measurements, but provides a fundamentally different approach through fermion couplings. 
Scenarios with a sizable $C\!P$ mixture,  $| f_{\CP} \cos \phi_{\CP} | \gtrsim 0.8$, are excluded at 2$\sigma$.  

\subsection{Study of the $\tqH$ process} \label{sect:study_tqh}

\begin{figure}[ht]
\centerline{
\includegraphics[width=0.32\linewidth]{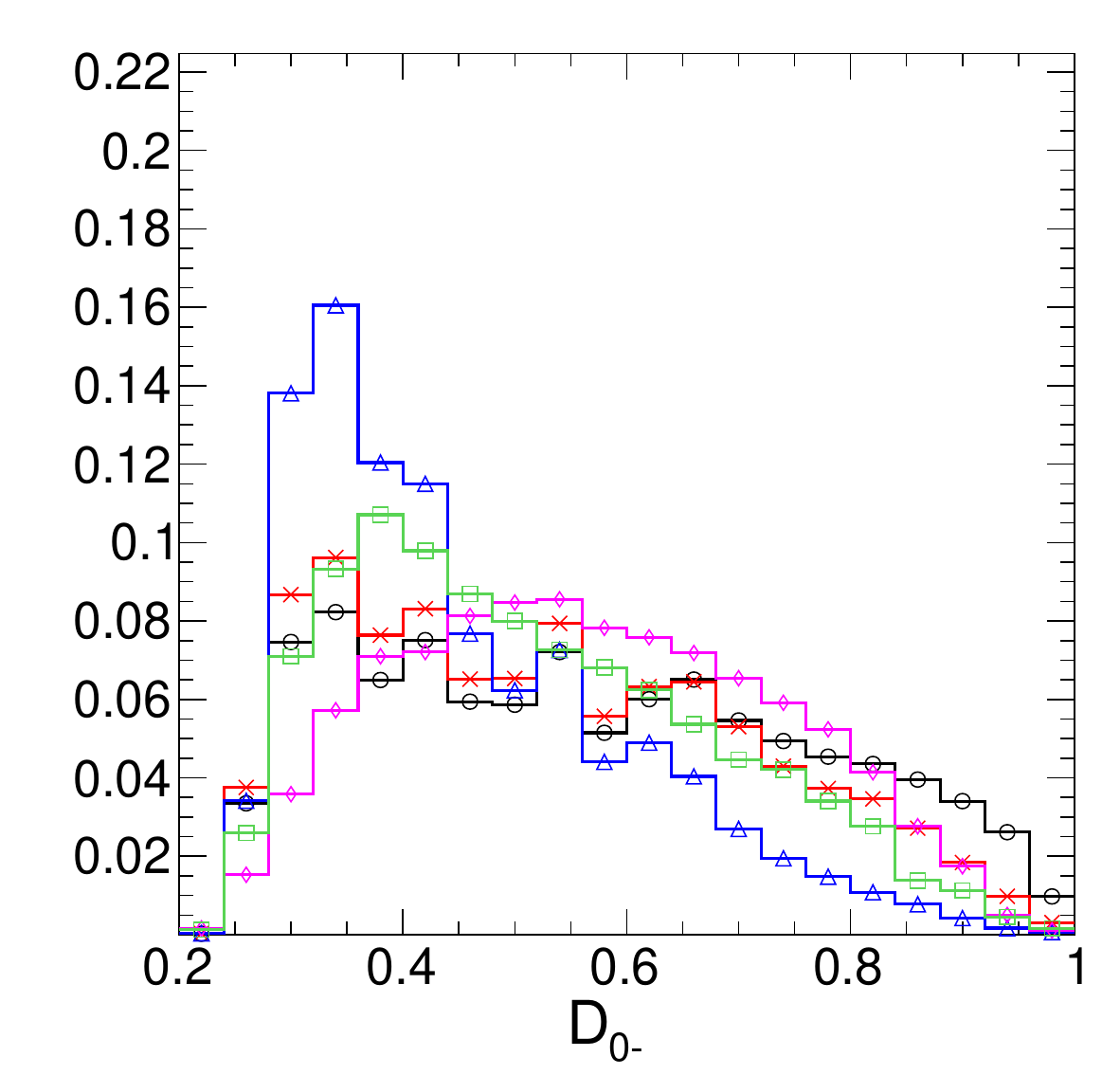}
\includegraphics[width=0.32\linewidth]{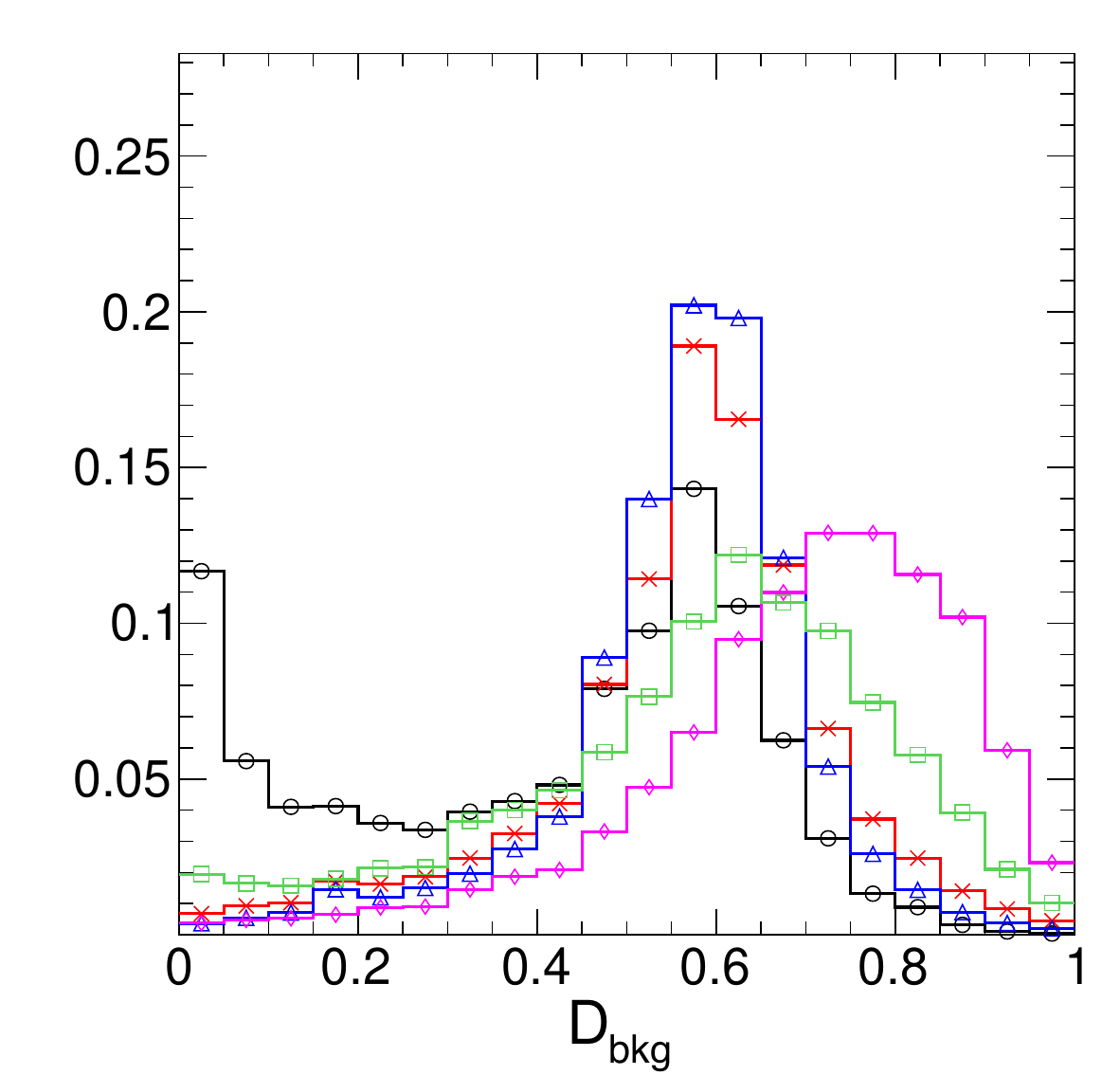}
\includegraphics[width=0.32\linewidth]{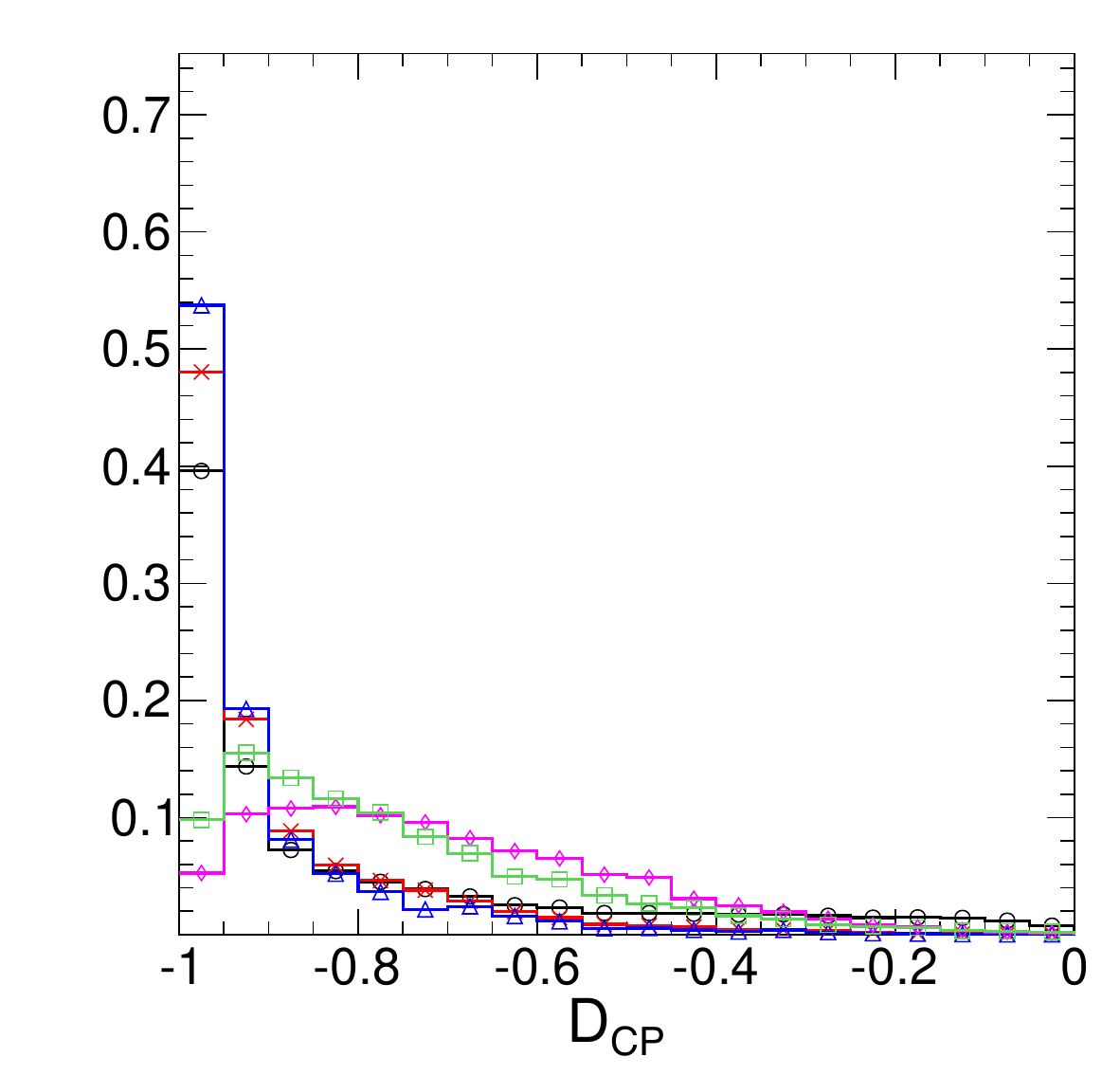}
}
\caption{
Distributions of ${\cal D}_{0-}$ (left), ${\cal D}_{\rm bkg}$ (middle), and  ${\cal D}^{\rm int}_{\rm bkg}$ (right) discriminants 
in the $\tqH$ study where the $\Hff$-induced process is considered as signal and the $HVV$-induced process 
is considered as background. The following three contributions are considered:  $J^P=0^{+}$ signal (red crosses), 
$0^{-}$ signal (blue triangles), $HVV$-induced process as background (black circles). Also shown are distributions of 
mis-reconstructed $\ttH$ signal with $J^P=0^{+}$ (magenta diamonds) and $0^{-}$ signal (green squares).
All distributions appear after simulation and reconstruction discussed in text. 
}
\label{fig:D0m_reco_tHq}
\end{figure}

The analysis of the $\tqH$ process uses the ${\cal D}_{0-}$, ${\cal D}_{\rm bkg}$, ${\cal D}^{\rm int}_{\rm bkg}$ discriminants,
shown in Fig.~\ref{fig:D0m_reco_tHq}. In this study, the $\Hff$-induced process is considered as signal and the 
$HVV$-induced process is considered as background. Similar to the $\ttH$ study, the decay of the top quarks is not considered 
in the matrix element and the ${\cal D}_{\CP}$ discriminants provide little information and therefore are not used. 
There is a sizable contribution of the $\ttH$ events misreconstructed as $\tqH$ and they carry information on $f_{\CP}$.
The above observables provide sufficient information to differentiate between ${\CP}$ components of the $\ttH$ process as well. 

All event contributions in this study can be parameterized with three couplings, $\kappa$, $\tilde\kappa$, and $a_1$,
which are assumed to be real, as 
\begin{eqnarray}
\label{eq:yields_thq}
&& N^{\tqH}_{\rm tot} = \mathcal{L} (a_1^2 \sigma^{\tqH}_{\rm bkg} 
+ \kappa^2 \sigma^{\tqH}_{0+} 
+ \tilde\kappa^2 \sigma^{\tqH}_{0-} 
+ a_1 \kappa \sigma^{\tqH}_{\rm int({\rm bkg},0+)} 
+ a_1 \tilde\kappa \sigma^{\tqH}_{\rm int({\rm bkg},0-)} 
+ \kappa \tilde\kappa\sigma^{\tqH}_{\rm int({0-},{0+})} ) \,, \\ 
&& N^{\ttH}_{\rm tot} = \mathcal{L} (\kappa^2 \sigma^{\ttH}_{0+} 
+ \tilde\kappa^2 \sigma^{ttH}_{0-} 
+ \kappa \tilde\kappa\sigma^{ttH}_{\rm int({0-},{0+})}) \,,
\end{eqnarray}
where ${\cal L}$ is integrated luminosity and $\sigma$ is the product of cross section and reconstruction efficiency 
of a particular process corresponding to the unit value of the coupling ($\kappa$, $\tilde\kappa$, or $a_1$).
The interference cross section can be negative, as it is for interference between the $\kappa$ and $a_1$ terms in the $\tqH$ process. 
In the $\tqH$ process, we express $\kappa$, $\tilde\kappa$, and $a_1$ in terms of two effective cross section fractions with phases 
and the overall normalization as follows
\begin{eqnarray} 
\label{eq:fkappa_thq}
&& f_{\kappa} = \frac{\kappa^2 \sigma^{\tqH}_{0+} }
{ (a_1^2 \sigma^{\tqH}_{\rm bkg} + \kappa^2 \sigma^{\tqH}_{{0+}} + \tilde\kappa^2 \sigma^{\tqH}_{{0-}})} 
\,,\, ~~~\phi_{\kappa}={\rm arg}({\kappa}/a_1) = {\rm 0~or~\pi} \,,
\\
&& f_{\tilde\kappa} = \frac{{\tilde\kappa} ^2 \sigma^{\tqH}_{0-} }
{ (a_1^2 \sigma^{\tqH}_{\rm bkg} + \kappa^2 \sigma^{\tqH}_{{0+}} + \tilde\kappa^2 \sigma^{\tqH}_{{0-}})} 
\,,\, ~~~\phi_{\tilde\kappa}={\rm arg}({\tilde\kappa}/a_1) = {\rm 0~or~\pi}
\,.
\end{eqnarray}
In the SM, $f_\kappa=0.46$, $f_{\tilde\kappa}=0$, and $\phi_{\kappa}=0$. 
The ratios of cross section is $\sigma^{\tqH}_{0+}/\sigma^{\tqH}_{\rm bkg}=0.86$. 

\begin{figure}[ht]
\centerline{
\includegraphics[width=0.38\linewidth]{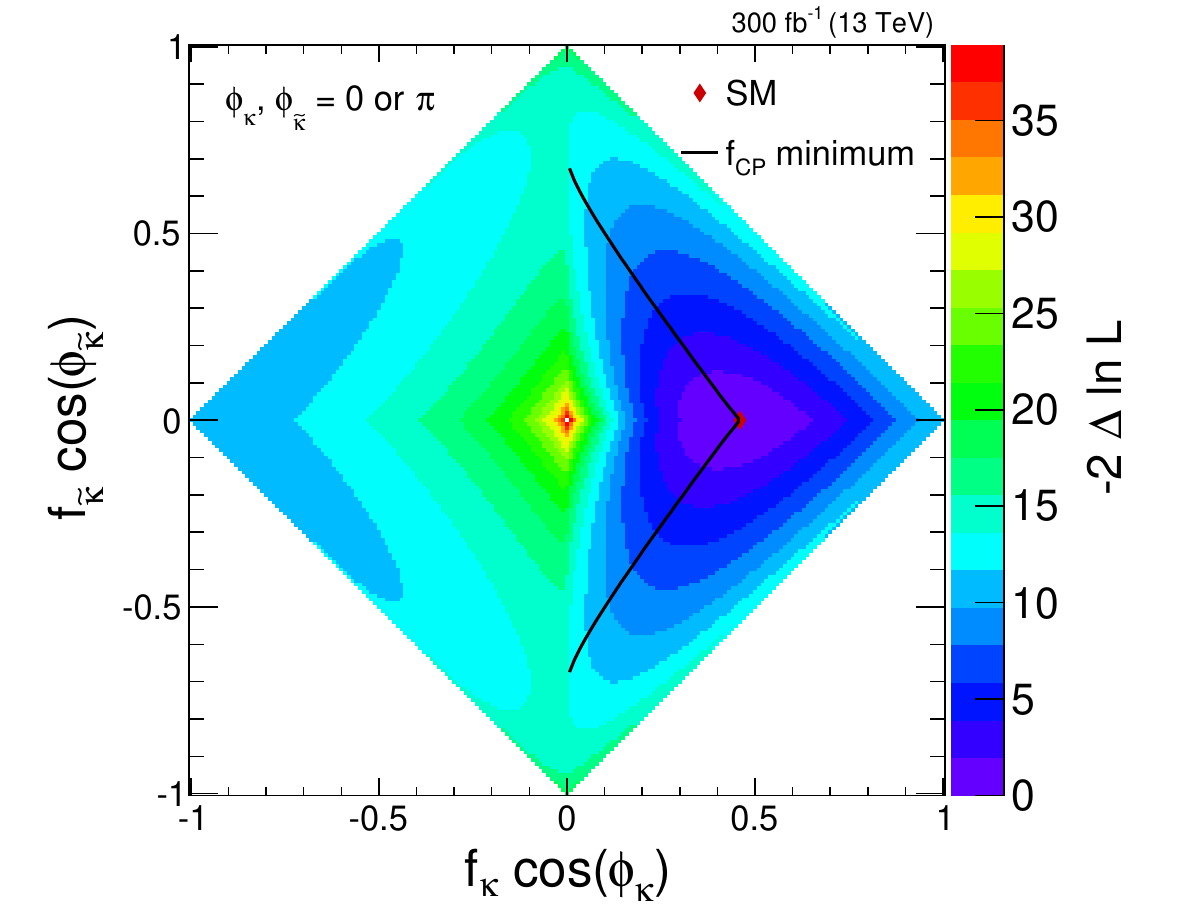}
\includegraphics[width=0.3\linewidth]{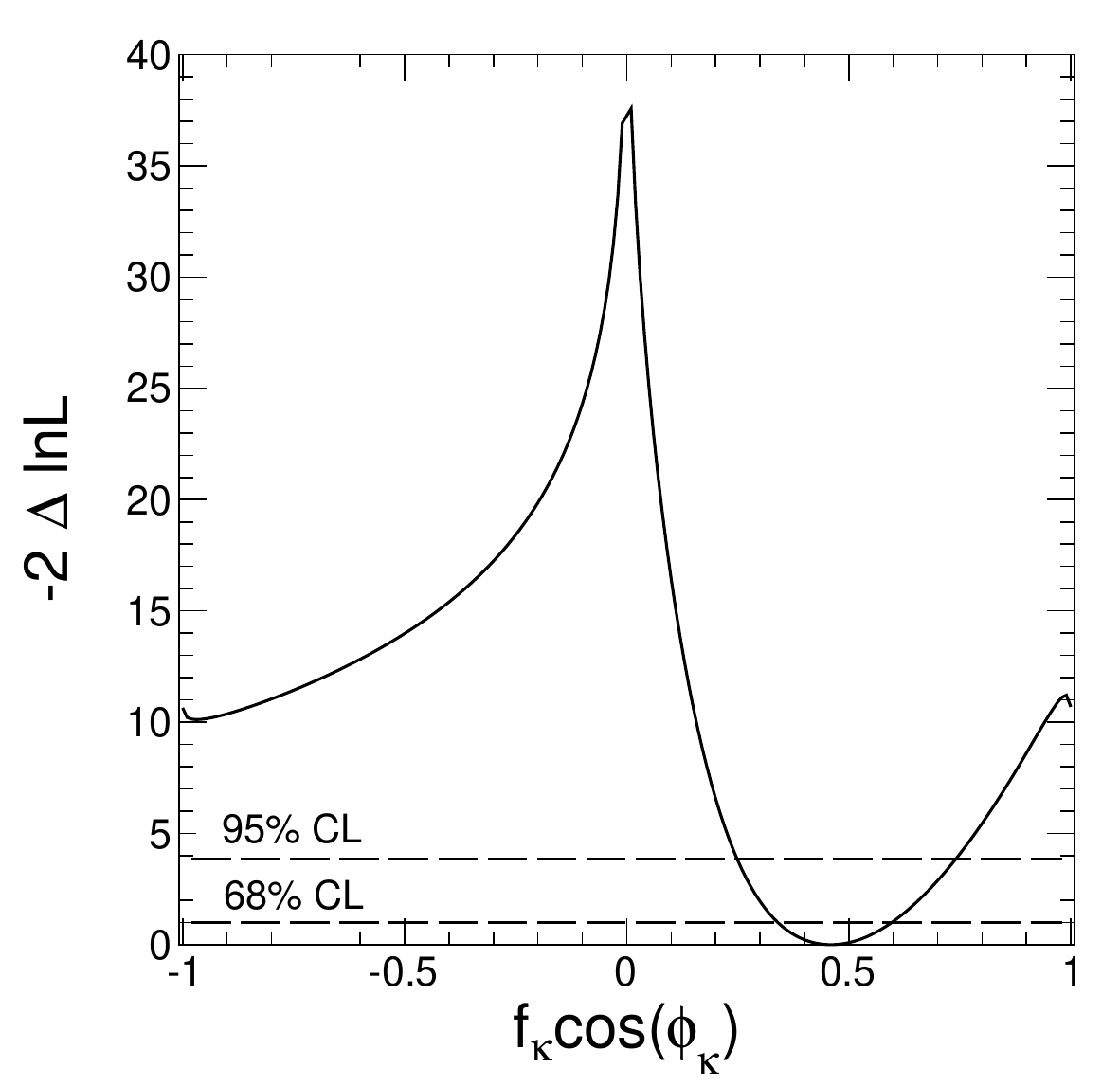}
\includegraphics[width=0.3\linewidth]{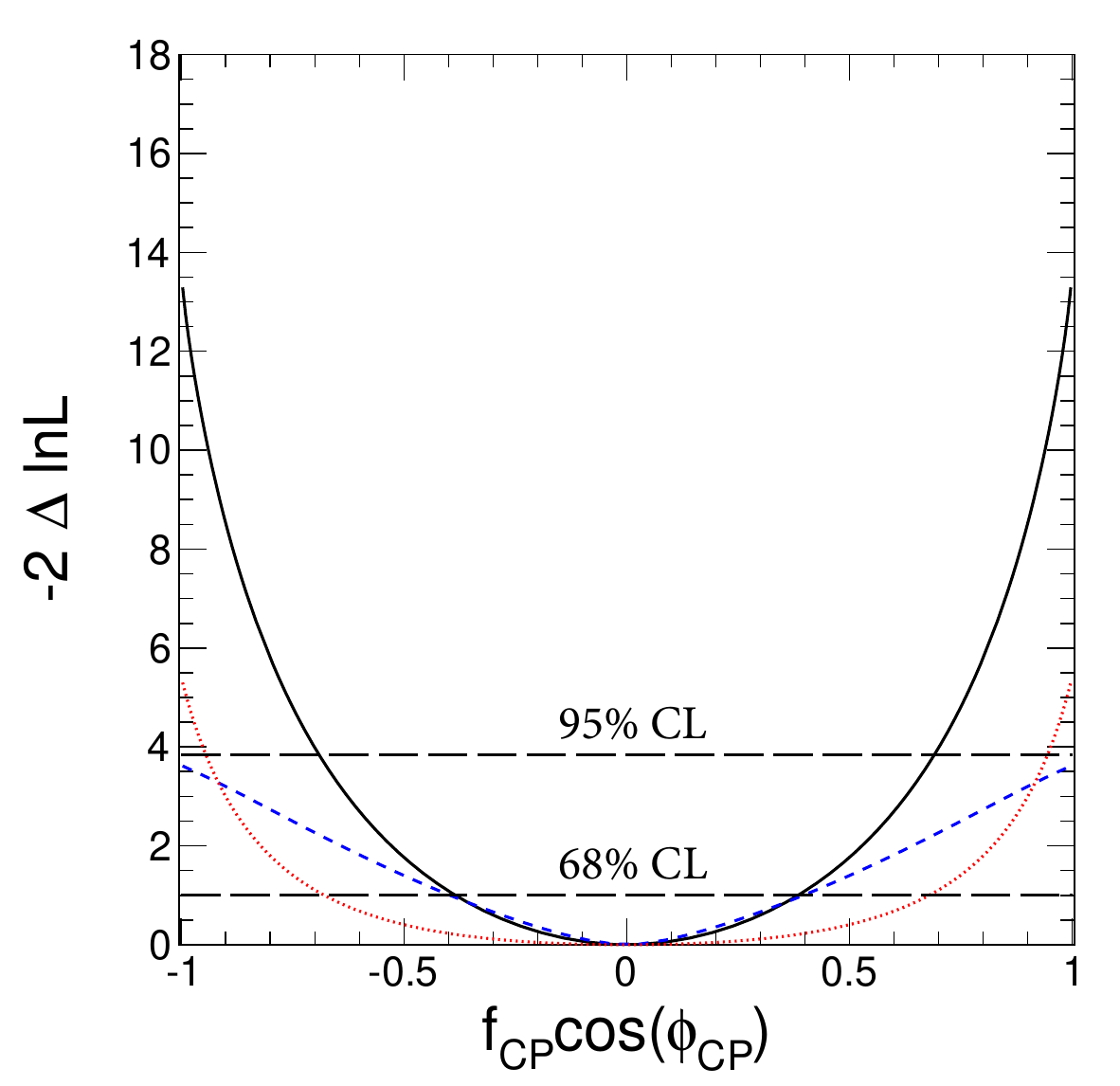}
}
\caption{
The likelihood scan of parameters of interest in the $\tqH$ study expected with 300\,fb$^{-1}$ at LHC.
Left: $f_{\tilde\kappa}\cos(\phi_{\tilde\kappa})$ vs. $f_{\kappa}\cos(\phi_\kappa)$, 
where the SM expectation is $f_\kappa=0.46$ and $f_{\tilde\kappa}=0$, 
and where $\phi_{\tilde\kappa}$ or $\phi_{\CP}=0$ or $\pi$.
Middle: $f_{\kappa}\cos(\phi_\kappa)$ with the constraint of no anomalous couplings, $f_{\tilde\kappa}=0$.
Right: $f_{\CP}\cos(\phi_{\CP})$ corresponding to the curved line on the left plot going from 
$f_\kappa=0$ ($f_{\CP}=1$) to $f_{\tilde\kappa}=0$ ($f_{\CP}=0$) following the minimum of $-2\Delta\!\ln{\cal L}$.
The solid line shows expectation considering all contributions, while 
the red dotted line assumes that only $\ttH$ mis-reconstructed events are present in the sample, 
and the blue dashed line assumes no contribution of $\ttH$ events. 
}
\label{fig:tHq_scan}
\end{figure}

The maximum likelihood fit, similar to the $\ttH$ analysis, uses a 3D template approach 
of three observables ${\cal D}_{0-}$, ${\cal D}_{\rm bkg}$, ${\cal D}^{\rm int}_{\rm bkg}$,
and with $f_{\kappa}$, $f_{\tilde\kappa}$, and total event yield as free parameters. 
The expected precision of the fit is shown in Fig.~\ref{fig:tHq_scan} (left plot).
This approach allows simultaneous measurement of both the relative fraction of $HVV$ and $\Hff$ induced processes
and of the anomalous contribution in the $\Hff$ coupling, with proper accounting for all interference effects. 
This measurement can be reduced either to the measurement of $f_{\kappa}$ with the constraint $f_{\tilde\kappa}=f_{\CP}=0$
(middle plot), or to the measurement of $f_{\CP}$ (right plot). Precision on the $\Hff$ couplings is driven by both 
$\tqH$ and $\ttH$ processes in this analysis, as illustrated with the likelihood scans separated for the two samples of events
in the right plot of Fig.~\ref{fig:tHq_scan}. More than 3$\sigma$ exclusion of the pure pseudoscalar state is expected in such 
a scenario, which is a measurement independent from that discussed in Section~\ref{sect:study_tth} since $\ttH$ events have
little overlap. It is important to note that in this scenario it will be possible to determine the relative sign of the $\Hff$ and 
$HVV$-induced contributions and exclude both extreme scenarios of either pure $\Hff$ or pure $HVV$ processes,
assuming events follow SM expectation. 

\section{Summary and conclusions}
\label{sect:tth_summary}

We have developed the Monte Carlo simulation and matrix element analysis tools and investigated prospects 
for measurement of anomalous interactions in the $H$ boson production in association with top or bottom 
quarks at the LHC, as well as its decay in two tau leptons.
The study is based on the JHU generator framework and the matrix element MELA analysis technique. 
We find that it is difficult to measure anomalous couplings in the $\bbH$ process, while in both 
$\ttH$ and $\tqH$ analyses it is possible to have more than 3$\sigma$ separation of the pseudoscalar
hypothesis from the scalar with 300\,fb$^{-1}$ of integrated luminosity at LHC at 13 TeV. 
It is also possible to separate the $\Hff$ and $HVV$ processes and determine their relative phase
in the $\tqH$ production, where in the SM the two processes interfere strongly and destructively. 
This feasibility study considers only representative decay channels of the top quark (hadronic decay
of one top) and $H$ boson (diboson decay) and inclusion of other final states would only enhance 
expected precision. Systematic uncertainties from QCD effects, such as PDF, scale, parton showering, 
and higher order corrections, are shown to be relatively small compared to expected statistical precision. 
The tools and techniques presented should facilitate measurements of SM and anomalous $\Hff$ couplings.


\bigskip

\noindent
{\bf Acknowledgments}:
We acknowledge contribution of CMS collaboration colleagues to the MELA project development
and thank Roberto Covarelli, Chris Martin, Candice You for help with the generator validation. 
We are grateful to our co-authors of the JHU generator project for valuable contributions, 
and in particular to Ulascan Sarica and Heshy Roskes for the generator package support
and Fabrizio Caola for providing useful comments on the manuscript. 
This research is partially supported by U.S. NSF under grant PHY-1404302 and 
by the German Federal Ministry for Education and Research (BMBF) under grant 05H15VKCCA.
Calculations reported in this paper were performed on the Homewood High Performance Cluster 
(HHPC) of the Johns Hopkins University
and the Maryland Advanced Research Computing Center (MARCC). 
%


\appendix 

\section{Supplemental information on kinematics}
\label{sec:appendixA} 

\begin{figure}[ht]
{\includegraphics[width=0.32\linewidth]{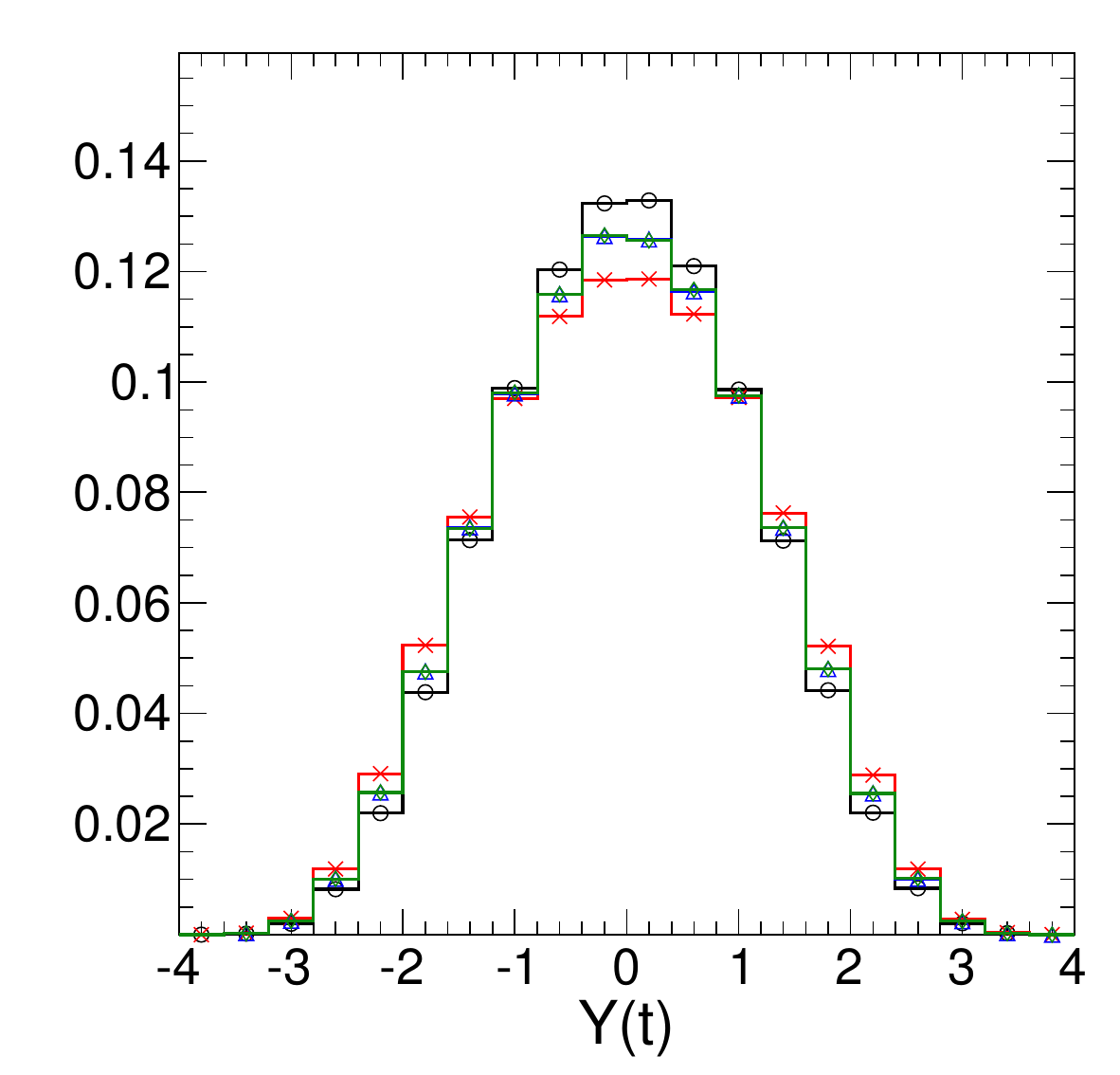}}
{\includegraphics[width=0.32\linewidth]{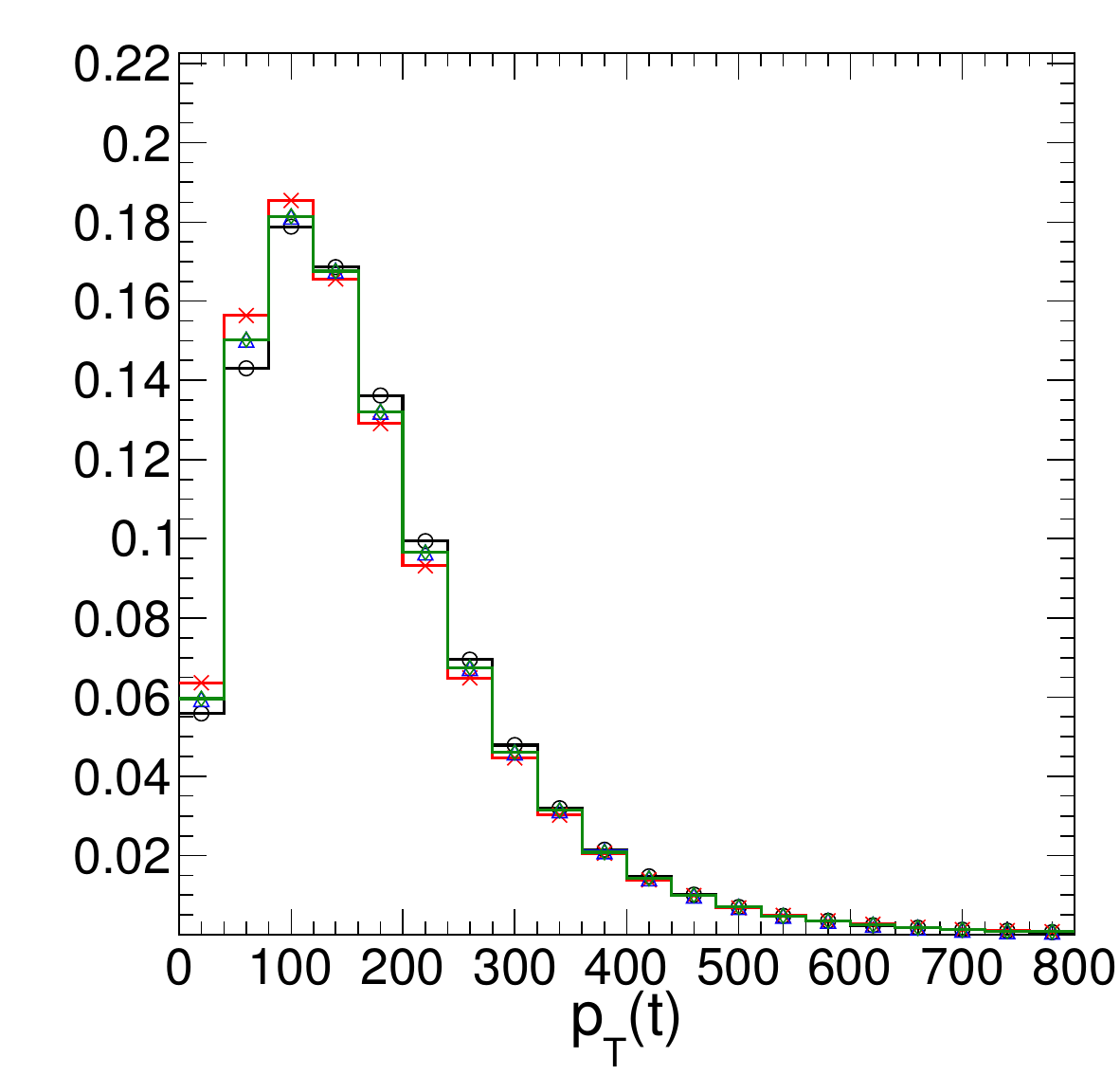}}
{\includegraphics[width=0.32\linewidth]{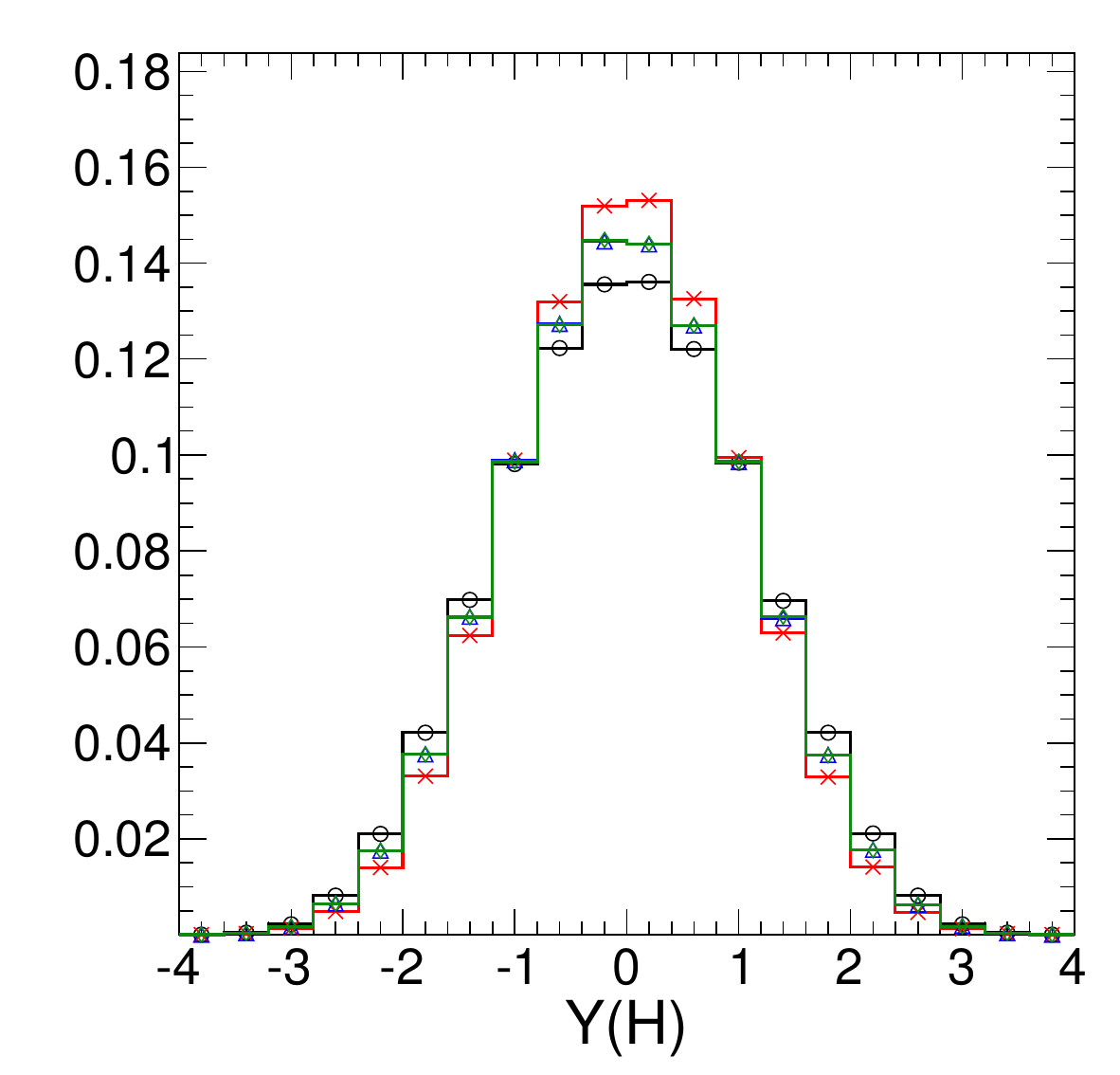}}
{\includegraphics[width=0.32\linewidth]{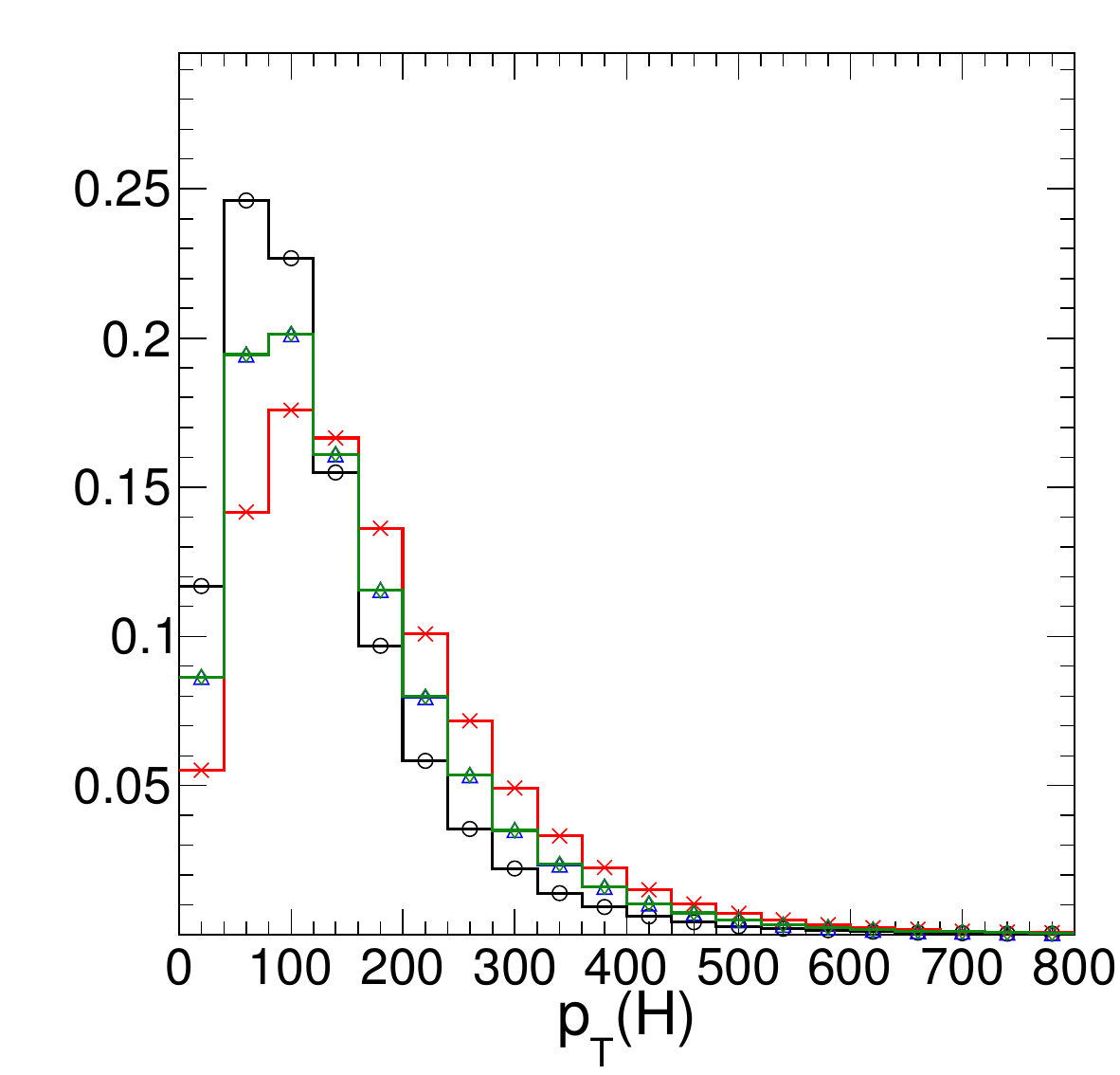}}
{\includegraphics[width=0.32\linewidth]{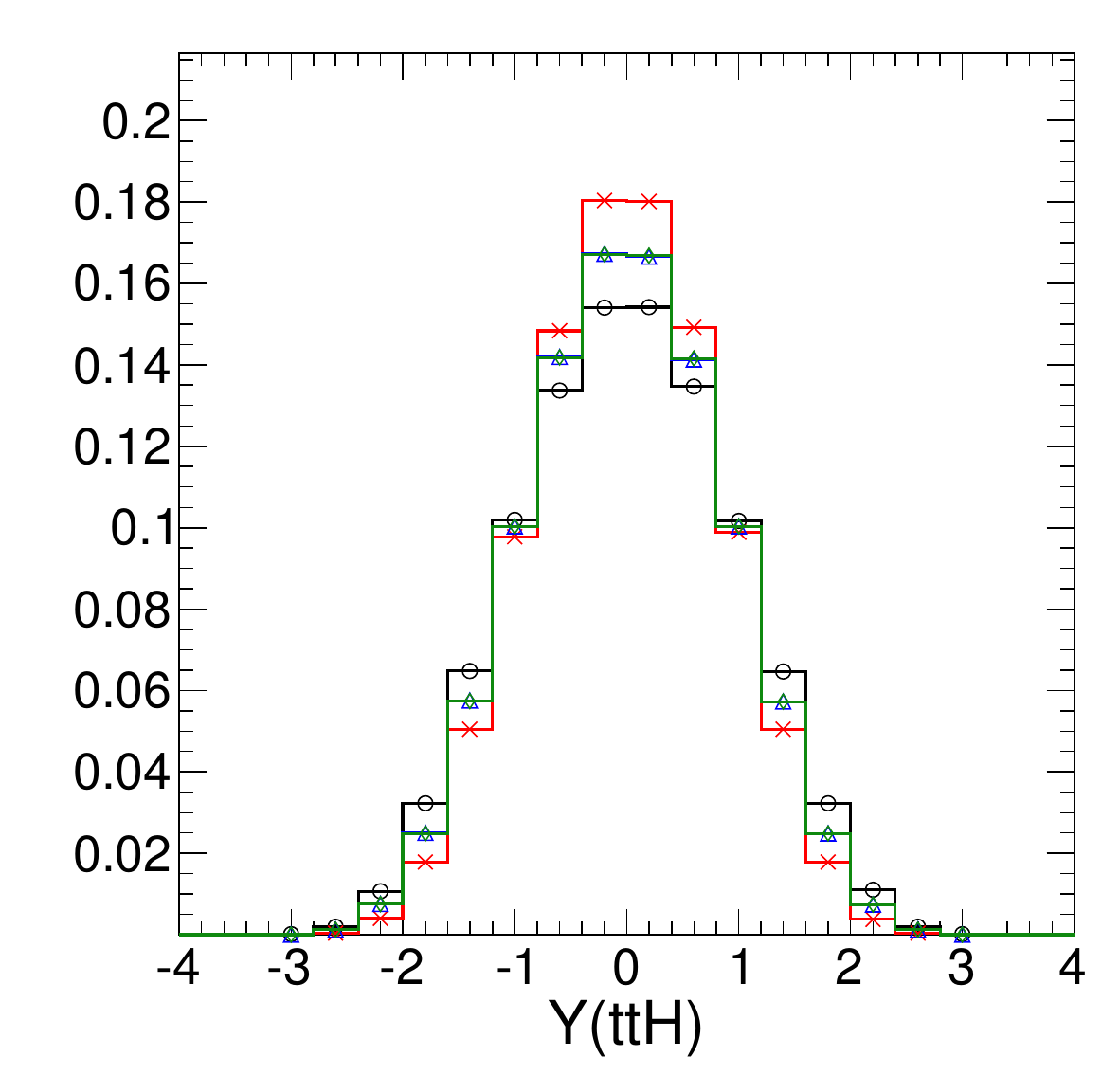}}
{\includegraphics[width=0.32\linewidth]{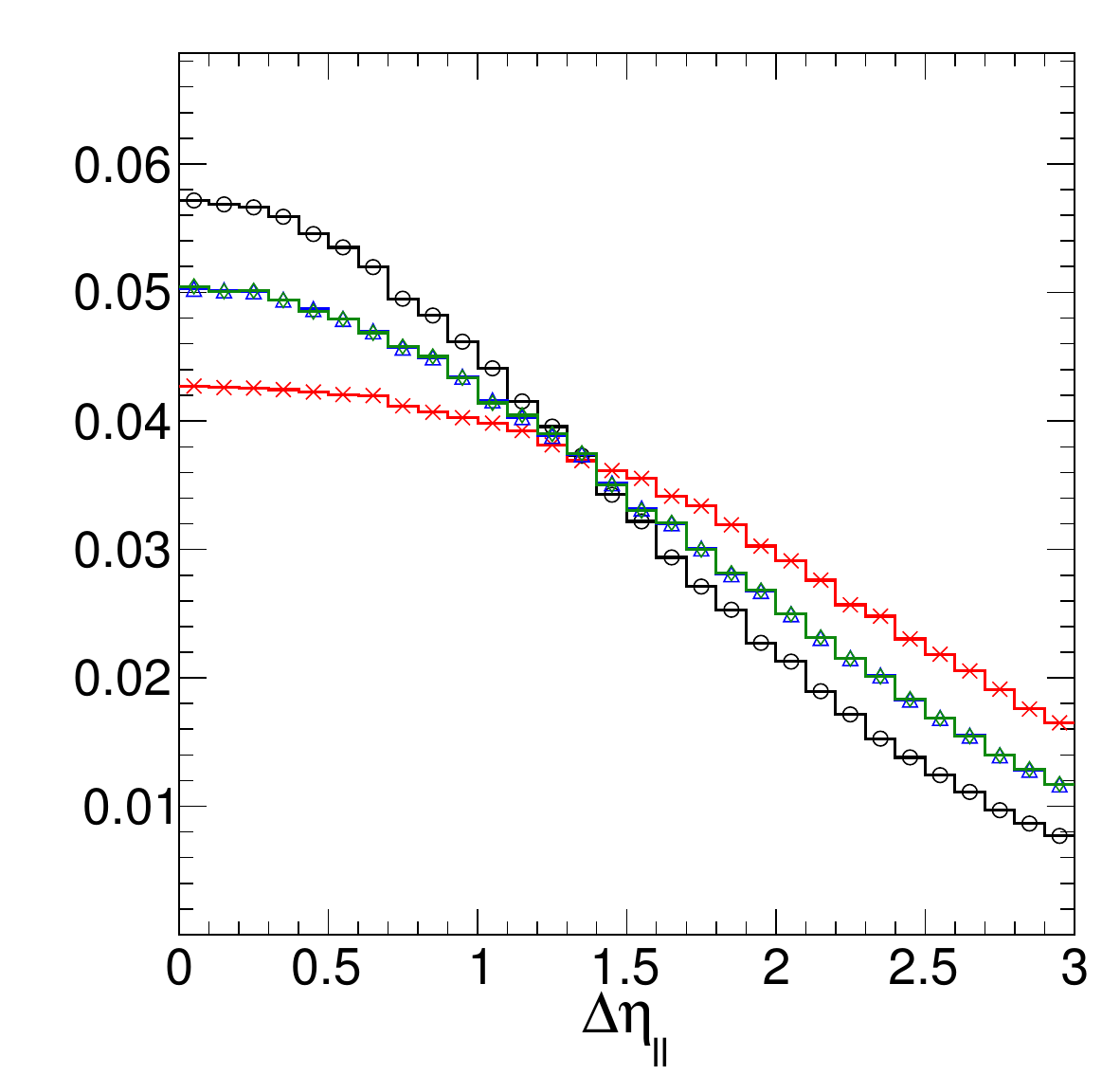}}
{\includegraphics[width=0.32\linewidth]{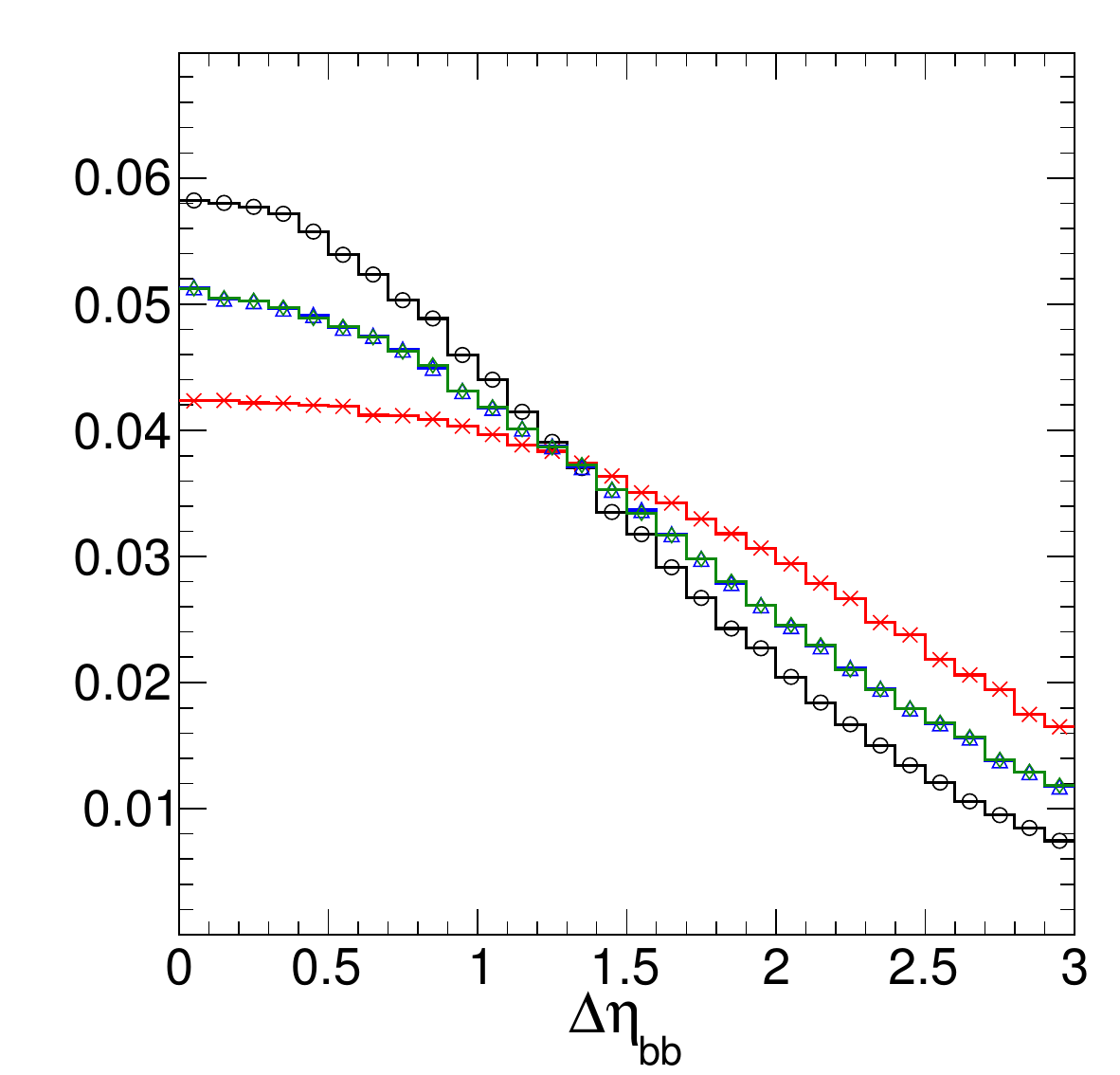}}
{\includegraphics[width=0.32\linewidth]{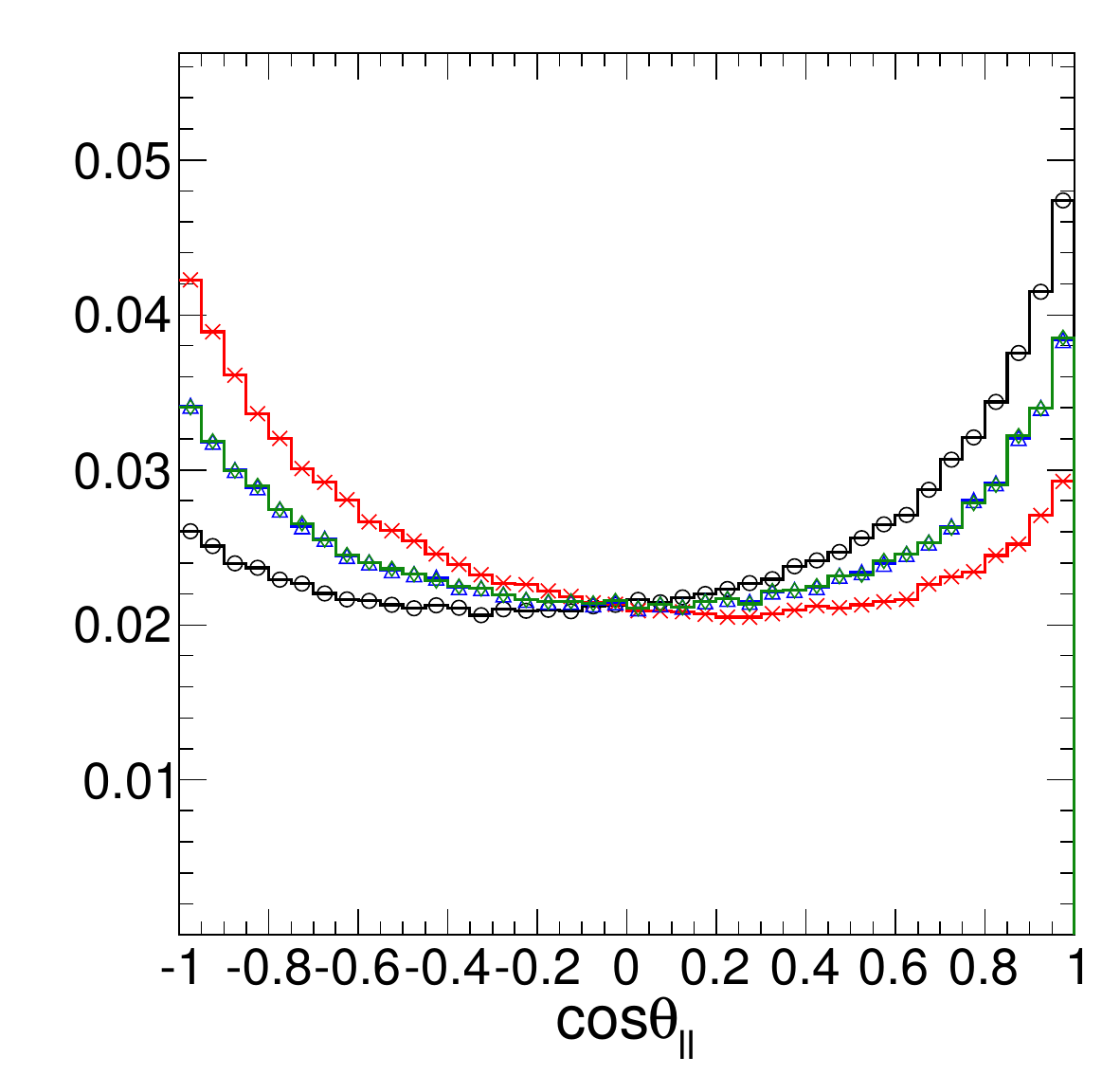}}
{\includegraphics[width=0.32\linewidth]{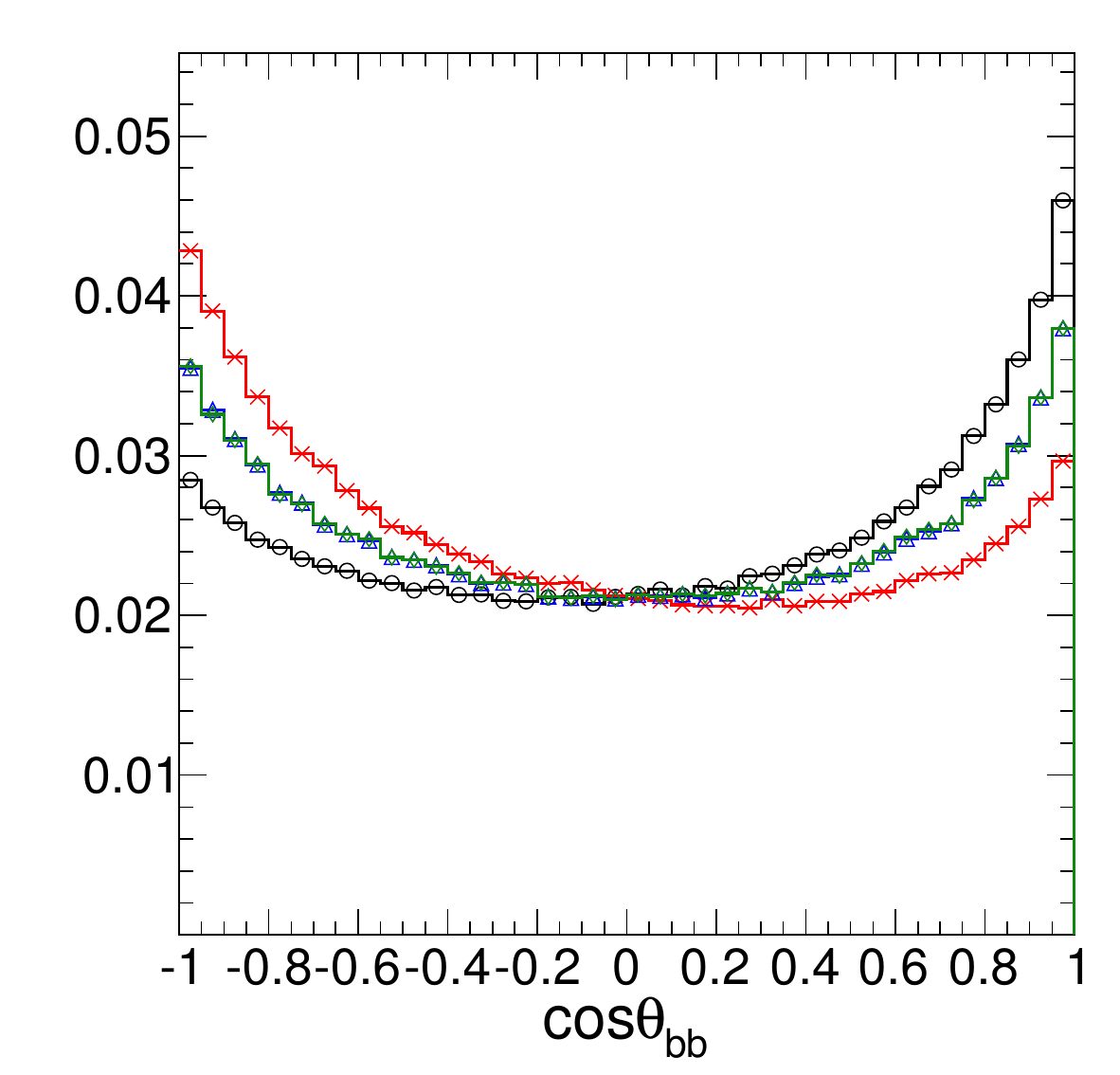}}
\caption{
The kinematic distributions in the process $gg$ and $\qqbar\to\ttH$ defined in the laboratory frame:
top quark rapidity ($Y(t)$), 
transverse momentum of the top quark ($p_T(t)$),
$H$ boson rapidity ($Y(t)$), 
transverse momentum of the $H$ boson ($p_H(t)$),
$\ttH$ system rapidity ($Y(\ttH)$), 
pseudorapidity difference 
between the two down type fermions decayed from top and anti-top ($\Delta\eta_{ll}$) 
and between two bottom quarks ($\Delta\eta_{bb}$),
$\cos\theta_{ll}$ between the two down type fermions, and
$\cos\theta_{bb}$ between the two bottom quarks. 
Four scenarios of anomalous $\ttH$ couplings are shown:
$f_{\CP}=0$ (SM $0^{+}$, black circles),
$f_{\CP}=1$ (pseudoscalar $0^{-}$, red crosses),
$f_{\CP}=0.28$ with $\phi_{\CP}=0$ (blue triangles)
and $\phi_{\CP}=\pi/2$ (green diamonds). 
The LHC pp energy of 13 TeV and $H$ boson mass of 125 GeV are used in simulation.
}
\label{fig:kinematic}
\end{figure}

Figure~\ref{fig:kinematic} shows the kinematic observables defined in the laboratory frame 
in the SM process $gg$ and $\qqbar\to\ttH$, corresponding to four scenarios of anomalous $\ttH$ couplings.
These observables can be derived from those shown in Fig.~\ref{fig:angular} and defined in Section~\ref{sect:tth_mela_angles}. 

\providecommand{\href}[2]{#2}\begingroup\raggedright\endgroup

\end{document}